\documentclass[12pt]{iopart}

\usepackage{iopams}  

\expandafter\let\csname equation*\endcsname\relax
\expandafter\let\csname endequation*\endcsname\relax

\usepackage{amsmath}
\usepackage{graphicx}

\usepackage{appendix}

\usepackage{color}

\usepackage{centernot}

\newcommand{\nn}{{\nonumber}}
\newcommand{\bfs}{\mathbf{s}}
\newcommand{\pr}{\mathrm{P}}
\newcommand{\bfxi}{\boldsymbol{\xi}}
\newcommand{\bfJ}{\mathbf{J}}
\newcommand{\pa}{\partial}
\newcommand{\Wr}{\mathrm{W}}

\newcommand{\balign}{\begin{align}}
\newcommand{\ealign}{\end{align}}

\begin{document}

\title[Sparse Boolean networks with multi-node and self-interactions]{Dynamics of sparse Boolean networks with multi-node and self-interactions}

\author{Christian John Hurry\textsuperscript{1}, Alexander Mozeika\textsuperscript{2}, Alessia Annibale\textsuperscript{1,3}}

\address{\textsuperscript{1}Department of Mathematics, King's College London, Strand, WC2R 2LS \\
\textsuperscript{2}School of Cancer and Pharmaceutical Sciences, King’s College London, London, London, UK\\
\textsuperscript{3}Institute for Mathematical \& Molecular Biomedicine, Hodgkin Building, Guy's Campus, London, SE1 1UL}
\ead{christian.hurry@kcl.ac.uk \\
alexander.mozeika@kcl.ac.uk\\
alessia.annibale@kcl.ac.uk}
\vspace{10pt}
\begin{indented}
\item[]Date: June 2022
\end{indented}

\begin{abstract}
We analyse the equilibrium behaviour and non-equilibrium dynamics of sparse Boolean networks with self-interactions that evolve according to synchronous Glauber dynamics. 
Equilibrium analysis is achieved via a novel 
application of the cavity method to the temperature-dependent pseudo-Hamiltonian that characterizes the equilibrium state of systems with parallel dynamics.
Similarly, the non-equilibrium dynamics can be analysed by using
the dynamical version of the cavity method. It is well known, however, that when self-interactions are present, direct application of the dynamical cavity method is cumbersome, due to the presence of strong memory effects, which prevent explicit analysis of the dynamics beyond a few time steps.
To overcome this difficulty, we show that it is possible to map a system of $N$ variables to an equivalent bipartite system of $2N$ variables, for which the dynamical cavity method can be used under the usual one time approximation scheme. This substantial technical advancement allows for the study of transient and long-time behaviour of systems with self-interactions. Finally, we study the dynamics of systems with multi-node interactions, recently used to model gene regulatory networks, 
by mapping this to a bipartite system of Boolean variables with 2-body interactions. We show that when interactions have a degree of bidirectionality such systems are able to support a multiplicity of diverse attractors,
an important requirement for a gene-regulatory network to sustain multi-cellular 
life.

\end{abstract}

%
\vspace{2pc}
\submitto{\JPA}

%

\maketitle

\section{Introduction}

Sparse Boolean networks are popular models to analyse the operation of a broad range of complex systems, ranging from credit contagion 
\cite{hatchett2009credit,gu2013modeling,liang2014construction} 
to epidemic spreading \cite{zhu2018stochastic} and opinion dynamics \cite{green2007emergence,li2017boolean}. Originally introduced, 
and broadly used, to study the dynamics of  gene-regulatory networks (GRNs) \cite{kauffman1969metabolic,kauffman1969homeostasis},
in these models each node of the network represents a gene, that is assumed to be in one of two states (expressed/not expressed), and its state is described 
by a 
Boolean variable (1 or 0). Genes are assumed to evolve synchronously in time so that all nodes of the network update together based on the state of their neighbouring  nodes at the previous time step \cite{liang2012stochastic}.

It is well known that gene expression is biologically regulated 
by transcription factors (TFs). These are single (or small complexes of) proteins, synthesised by genes, which can bind to certain portions of DNA, and selectively promote or inhibit the expression of genes.
Recently, {\it bipartite} Boolean networks have been introduced to incorporate the role of TFs 
in the dynamics of gene expression
\cite{hannam2019percolation,torrisi2020percolation,torrisi2022uncovering}. In these models, genes and TFs 
are modelled by two sets of Boolean variables which interact with each other via directed links. A directed link from a gene to a TF indicates that the gene codes for a protein that 
constitutes the TF. Conversely, a directed link from a TF to a gene indicates that the TF regulates the expression of that gene. In these models, links were assumed to be sparse, directed and drawn randomly and independently from given distributions, so that links were typically unidirectional, 
i.e. the probability to have a bidirectional link vanished in the thermodynamic limit, where the number of nodes in the network is infinitely large, due to sparsity and directionality.
Unidirectional interactions meant that genes did not contribute to the synthesis of TFs that would regulate them, a process we refer to as self-regulation. It is known that self-regulation is a common feature of GRNs,
and feedback loops, where a gene is directly or indirectly involved in the regulation of its own expression, are important. It is the objective of this work to relax the assumptions made in earlier works, and allow interactions to be correlated, such that we can study the affect of self-regulation on the dynamics of 
GRNs. 

Past models of GRNs have been shown to support only a single attractor and this was hypothesised to be due to the lack of bidirectional links in the networks that were studied \cite{hannam2019percolation, torrisi2020percolation,torrisi2022uncovering}. This is consistent with recent numerical work which suggests that sparse, fully asymmetric networks tend toward supporting a single attractor as dilution is increased \cite{folli2018effect}. A study of sparsely connected, \textit{partially} symmetric networks, to the best of our knowledge, is absent from the literature. In this work we show that partially symmetric networks can support multiple attractors, a requirement for multi-cellular life, at low, but finite, temperature. In order to do so we apply the dynamical cavity method to the study of bipartite systems, with two distinct sets of variables that evolve according to different sets of linear threshold functions. By suitable parameterisation of such systems one can study models of gene expression, with pairwise and multi-node interactions, previously studied in \cite{hannam2019percolation,torrisi2020percolation,torrisi2022uncovering}.   
By mapping these models to suitably defined bipartite systems, we find that we can solve for the dynamics of systems with two-body, arbitrarily asymmetric interactions, even in the presence of {\it self-interactions}. This represents a major technical advance, as self-interactions are well known to introduce long-time correlations which would normally make direct application of the dynamical cavity method (or equivalent methods based on generating functionals)  
cumbersome. 
By a similar mapping to bipartite networks, we are also able to investigate 
the dynamics of systems with multi-node (arbitrarily 
asymmetric) interactions, which can be regarded as Boolean equivalents of mixed p-spin models.
We find that both models support, at low temperature, a multiplicity of (cyclic) attractors, as soon as a degree of bidirectionality is introduced in the links. 
Interestingly, multi-node interactions increase the diversity of attractors.

The remainder of this work is split into several sections. In the following section we define our model of GRNs, and include two different models of the dynamics of gene expression with pairwise and multi-node interactions, leading to linear and nonlinear dynamics, respectively. We show that both models are described by a bipartite Boolean system, and provide a general model from which the linear or nonlinear model can be recovered by suitable choice of parameters. In Sec. \ref{sec: cavity method}  we describe how the dynamical cavity method may be extended to bipartite spin systems, and present a closed set of equations which may be solved numerically within a one time approximation (OTA) scheme, originally developed for monopartite systems \cite{neri2009cavity,aurell2011message,aurell2012dynamic,zhang2012inference}. In Sec. \ref{sec: linear } we solve for the dynamics of the linear model of gene expression with self-interactions. Additionally, we analyse this model in equilibrium  and show that is in agreement with the dynamical cavity in the steady-state. In Sec. \ref{sec: nonlinear } we study the nonlinear model of gene expression. We provide an assessment of the OTA scheme in the presence of nonlinear dynamics and then go on to demonstrate the existence of multiple attractors caused by bidirectional links and the impact of multi-node interactions on their diversity. In Sec. \ref{sec: thermodynamic limit } we demonstrate how the dynamical cavity method may derive an efficient set of equations to solve for the dynamics of systems in the thermodynamic limit. We end with a discussion of our results, and posit further ideas for the study of multiple attractors in GRNs from a statistical mechanics perspective.  Technical details are contained in the appendices.

\section{Model definitions}\label{sec: model defintions}

\begin{figure*}[t]
\centering
  \includegraphics[width=0.6\textwidth]{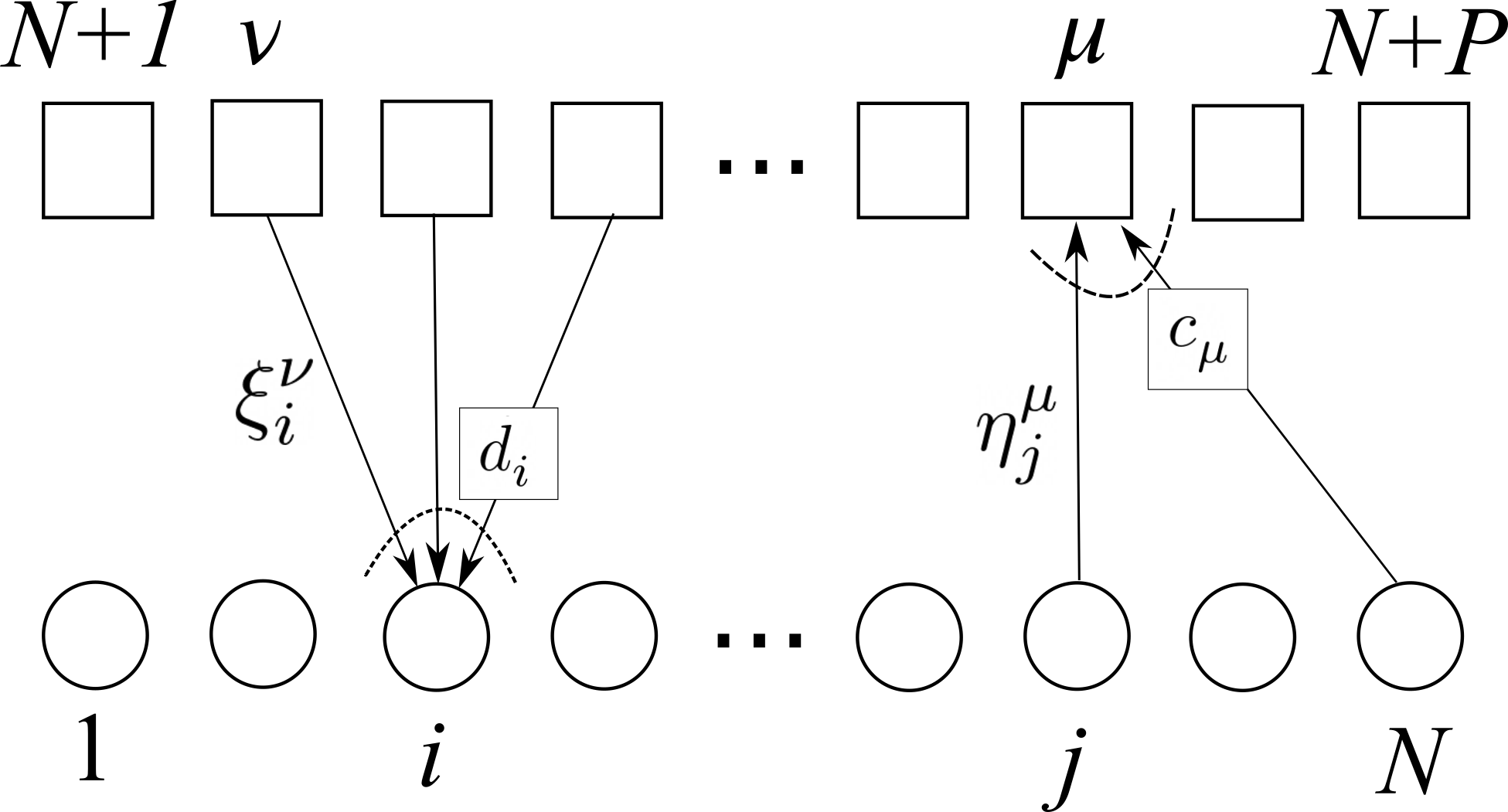}
  \caption{A sketch of the bipartite gene-TF network. Circles indicate genes, and squares indicate TFs. }
  \label{fig: bipartite sketch}
\end{figure*}

We consider a bipartite model of gene-regulatory networks (GRNs), recently introduced in \cite{hannam2019percolation},
which comprises $N$ genes and $P = \alpha N$ transcription factors (TFs), with directed interactions described by 
two matrices, $\bfeta$ and $\bfxi$,
giving unweighted links from genes to TFs and 
weighted links from TFs to genes, respectively.
Entry
$\eta_i^\mu\in\{0,1\}$ denotes whether ($1$) or not ($0$) gene $i$ contributes to the synthesis of TF $\mu$, whereas entry $\xi_i^\mu \in \mathbb{R}$ describes the regulatory effect of TF $\mu$ on gene $i$, which can be excitatory ($\xi_i^\mu >0$), inhibitory ($\xi^\mu_i< 0 $) or null ($\xi^\mu_i=0$).
We will denote with $\partial_i=\{\mu: \xi_i^\mu\neq 0\}$ and $\partial_\mu=\{i: \eta_i^\mu=1\}$ the in-neighbourhoods of gene $i$ and TF $\mu$, respectively, and with $d_i=|\partial_i|$ and $c_\mu=|\partial_\mu|$ their in-degrees, respectively, as sketched in figure \ref{fig: bipartite sketch}. Interactions are assumed sparse, 
as in real GRNs, 
i.e. the average values of the local degrees $\langle d \rangle= N^{-1}\sum_i d_i$, $\langle c\rangle = (\alpha N)^{-1}\sum_\mu c_\mu$ are $\mathcal{O}(N^{0})$.

Earlier analysis of this model \cite{hannam2019percolation, torrisi2020percolation, torrisi2022uncovering} regarded $\bfxi$ and $\bfeta$ as quenched random variables, drawn independently from given
probability distributions $\pr(\bfxi)$ and $\pr(\bfeta)$, respectively. For this choice, the probability to observe a bidirectional link $\xi_i^\mu \eta_i^\mu\neq 0$ is $\mathcal{O}(N^{-1})$, due to the sparsity of the links, and thus vanishes in the thermodynamic limit $N\to \infty$. This assumption is however not justified by biological observations, suggesting that TFs which regulate genes contributing to their own synthesis are commonplace. In this work, we will regard $\bfxi$ and $\bfeta$ as 
drawn from a joint probability distribution $\pr(\bfxi,\bfeta)=\pr(\bfxi|\bfeta)\pr(\bfeta)$, 
such that there is a finite probability $\pr(\xi_i^\mu\neq 0|\eta_i^\mu=1)=p$ to observe a bidirectional link between a TF $\mu$ and a gene $i$.

Following earlier studies \cite{hannam2019percolation,torrisi2020percolation,torrisi2022uncovering}, 
we model 
the state of each gene, labelled by $i=1,\dots,N$, with a Boolean variable $s_i\in \{0,1\}$, indicating whether gene $i$ is expressed ($1$) or not ($0$). Each TF, labelled by $\mu=N,\dots, N+P$, is associated with a variable $\tau_\mu \in[0,1]$, which describes its concentration. Genes update their state at regular time intervals of duration $\Delta$ according to a linear threshold model, 
\begin{align}
    s_i(t + \Delta) = \theta\left( \sum_\mu \xi_i^\mu \tau_\mu(t) + \vartheta_i -  z_i(t) \right),
    \label{eq:LTB}
\end{align}
where $\theta(x)$ is the Heaviside step function, defined such that $\theta(x) =1$ for $x>0$ and $\theta(x) =0 $ for $x<0$,
$\vartheta_i$ can be thought of as a local external field or threshold and the $z_i(t)$'s are independent identically 
distributed zero-averaged random variables mimicking biological noise. We denote their cumulative distribution function (c.d.f.) with ${\rm Prob}[z_i(t) < x]=\phi_\beta(x)$, for all $i, t$, where $ \beta^{-1}=T$ is a parameter that characterises the 
noise strength. We shall assume, throughout the paper, a ``thermal'' (or Glauber) noise distribution with 
\begin{equation}
\phi_\beta(x)=\frac{1}{2}[1+\tanh \frac{\beta x}{2}]
\label{eq:thermal}
\end{equation}

Two types of dynamics for the synthesis of TFs were considered in \cite{hannam2019percolation,torrisi2020percolation,torrisi2022uncovering}, following different logic operations, respectively
\begin{itemize}
    \item `OR' logic
    \begin{equation}
        \tau_{\mu}(t ) = \frac{1}{c}\sum_{j} \eta_{j}^{\mu} s_{j}(t)
        \label{eq:OR}
    \end{equation} 
    where at least one gene in the in-neighbourhood of $\mu$ must be expressed for the synthesis of that TF to occur, 
    \item `AND' logic 
    \begin{equation}
        \tau_\mu(t ) = \prod_{j \in \pa_\mu} s_j(t)
        \label{eq:AND}
    \end{equation} 
    where all genes in the in-neighbourhood of $\mu$ must be expressed for TF $\mu$ to be synthesised (in this case $\tau_\mu$ takes values in $\{0,1\}$ rather than $[0,1]$). 
\end{itemize}
If we choose that TFs evolve according to an `OR' logic, 
upon eliminating the TFs from the description, 
we obtain a system of genes with their evolution given by a
linear threshold model 
\begin{align}
    s_i(t + \Delta) = \theta\left( \sum_j J_{ij} s_j(t)+  \vartheta_i - z_i(t) \right) \label{eq: LTMM update rule}
\end{align}
with effective two-body interactions $J_{ij} = \sum_\mu  \frac{\xi_i^\mu \eta_j^\mu}{c_\mu}$.
Conversely, if TFs perform an `AND' logic, the evolution of genes is given by \textit{nonlinear} threshold dynamics
\begin{align}
    s_i(t + \Delta) = \theta\left( \sum_\mu \xi_i^\mu \prod_{j \in \pa_\mu}s_j(t)+  \vartheta_i - z_i(t) \right), \label{eq: NLTM update rule}
\end{align}
with multi-body interactions between genes.
For later convenience, it is useful to note that 
the `AND' logic \eqref{eq:AND} can be expressed 
in terms of a linear threshold function with suitably chosen threshold
\begin{align}
\tau_\mu(t) &= \theta\left( \sum_j \eta_j^\mu s_j(t) -c_\mu +\epsilon \right)  
\label{eq:AND-alt}
\end{align}
where $0<\epsilon\ll 1$ ensures that the argument of the step function is greater than zero when {\it all} of the $c_\mu$ genes contributing to TF $\mu$ are expressed and negative otherwise.  
Thus, an equivalent description of \eqref{eq: NLTM update rule} is given by the system of equations 
\eqref{eq:LTB} and \eqref{eq:AND-alt}.
Out of mathematical convenience we will define a closely related model, where TFs are not updated instantaneously, but on the same timescale as 
the genes 
\begin{align}
    s_i(t + \tilde{\Delta}) &= \theta\left( \sum_\mu \xi_i^\mu \tau_\mu(t) +  \vartheta_i - z_i(t) \right) 
    \nonumber
    \\
   \tau_\mu(t + \tilde{\Delta}) &= \theta\left( \sum_j \eta_j^\mu s_j(t) -c_\mu +\epsilon \right) . 
   \label{eq:convenient}
\end{align}
Upon choosing $\tilde{\Delta}=\Delta/2$ and initial condition $\tau_\mu(0) =\theta\left( \sum_j \eta_j^\mu s_j(0) -c_\mu +\epsilon \right) \forall \mu$, the trajectory $s_i(t)$ resulting from \eqref{eq:convenient}
at times $t=n\Delta/2$, with $n\in \mathbb{N}$, is identical to the trajectory of 
\eqref{eq: NLTM update rule} at 
times $t=n\Delta$. On the other hand, the trajectories of \eqref{eq:convenient} and  
\eqref{eq: NLTM update rule} at $t=n\Delta$ represent two independent thermal histories 
of the same system, 
as noise is drawn at each time step from the same distribution. Hence, the two systems defined in \eqref{eq:convenient} and \eqref{eq: NLTM update rule} are fully equivalent at integer multiples of 
$\Delta$. The advantage of using \eqref{eq:convenient} is that it makes the application of the dynamical cavity method straightforward (see Sec. \ref{sec: cavity method} ).

The dynamics of GRNs evolving according to (\ref{eq: LTMM update rule}) has previously been studied  
for random interactions $J_{ij}$ with an arbitrary degree of symmetry, in the absence of self-interactions, i.e. $J_{ii}=0$ \cite{torrisi2021overcoming}. This corresponds, in the context of our bipartite network, to a lack of bidirectional links i.e. $\xi_i^\mu \eta_i^\mu = 0$. Similarly, the dynamics of GRNs evolving according to (\ref{eq: NLTM update rule}) has been studied under the assumption that $\pr(\bfxi,\bfeta) = \pr(\bfxi)\pr(\bfeta)$, such that the probability of observing a bidirectional link is zero in the limit of large system size \cite{hannam2019percolation, torrisi2020percolation, torrisi2022uncovering}. 
It was shown in \cite{hannam2019percolation} that in the absence of bidirectional links this model has no multiplicity of attractors, i.e. of stable gene expression profiles, even in the absence of noise, $T=0$. Thus, the fully asymmetric version of this model does not fulfill an important requirement for 
a GRN to sustain multiple cell types, as required in multi-cellular life. 

In this work we solve the dynamics  
(\ref{eq: LTMM update rule}) in the presence of self-interactions, i.e. $J_{ii} \neq 0$, and the dynamics \eqref{eq:convenient} in the presence of bidirectional links, i.e. $\pr(\xi_i^\mu\neq 0|\eta_i^\mu=1)=\mathcal{O}(N^0)$. Previous work has shown that dynamical analysis of \eqref{eq: LTMM update rule} via generating functionals is cumbersome in the presence of self-interactions due to the strong memory effects that they produce \cite{mozeika2012phase}, which make the complexity of the analysis exponential in the time horizon considered. We show that, while the dynamical cavity method faces a similar time-complexity
barrier, when applied to the system \eqref{eq: LTMM update rule} with self-interactions, 
it is possible, by a 
suitable mapping of \eqref{eq: LTMM update rule} to a bipartite system $\hat{\bfeta}, \hat{\bfxi}$ with bidirectional links, to apply the dynamical cavity method and solve the equations explicitly under the so-called
one-time approximation (OTA) scheme, that is effective in reducing the time-complexity of systems with bidirectional links \cite{neri2009cavity,aurell2011message,aurell2012dynamic,zhang2012inference}. 
This allows us to solve the dynamics of 
systems with self-interactions both at short and long times, accessing in particular the non-equilibrium steady-state. To best of our knowledge, this is the first work to do so. 
We will define in Sec. \ref{sec: linear } the bipartite 
model $\hat{\bfeta}, \hat{\bfxi}$ that achieves 
the suitable mapping, here we limit to stress that 
this is different from the links  
$\bfxi, \bfeta$, appearing in \eqref{eq:LTB} and \eqref{eq:OR}, respectively,  
from which the interactions $\{J_{ij}\}$ were derived.
Similarly, by mapping the nonlinear threshold model (\ref{eq: NLTM update rule}) to the bipartite model with two-body interactions \eqref{eq:convenient}, we are able to solve the dynamical cavity equations 
in the presence of bidirectional links, under the OTA scheme, thus relaxing the assumption of fully asymmetric links used in previous work \cite{hannam2019percolation, torrisi2020percolation, torrisi2022uncovering}. We will show that, upon introducing bidirectionality in the links, the nonlinear threshold model \eqref{eq: NLTM update rule} is able to support a multiplicity of attractors at low noise level.

\section{Dynamical cavity method for bipartite systems with parallel dynamics} \label{sec: cavity method}

As explained above, both the linear threshold model with self-interactions (\ref{eq: LTMM update rule}) and the nonlinear model with multi-node interactions (\ref{eq: NLTM update rule}) can be mapped to an equivalent bipartite system. In this section we solve the dynamics of a general bipartite system, from which either the linear or nonlinear model can be recovered by suitable choice of parameters. In this generalised model, two sets of Boolean variables, that we shall refer to as 
genes and TFs, update their state at regular time intervals of duration $\Tilde{\Delta}$ according to the following linear threshold functions
\begin{align}
    s_i(t + \Tilde{\Delta}) &= \theta\left( \sum_\mu \xi_i^\mu \tau_\mu(t) +  \vartheta_i - z_i(t) \right)  &i=1,\dots,N \label{eq: GLTBM update rule s} \\
   \tau_\mu(t + \Tilde{\Delta}) &= \theta\left( \sum_j \eta_j^\mu s_j(t) + \vartheta_\mu - \hat{z}_\mu(t) \right) &\mu=N,\dots,N+P \label{eq: GLTBM update rule tau}
\end{align}
Here, $\vartheta_i$ and $\vartheta_\mu$ are the thresholds of gene $i$ and TF $\mu$, respectively, and $z_i(t)$ and $\hat{z}_\mu(t)$ are random i.i.d random variables with c.d.f. $\phi_\beta(x)$ and $\phi_{\hat{\beta}}(x)$, respectively. 
It is clear that \eqref{eq: GLTBM update rule tau}
reduces to \eqref{eq:convenient} for $\vartheta_\mu=-c_\mu+\epsilon$ and 
$\hat{T}={\hat{\beta}}^{-1}=0$, hence for this choice of parameters, the generalised model defined by  \eqref{eq: GLTBM update rule s} and \eqref{eq: GLTBM update rule tau} recovers model \eqref{eq: NLTM update rule} with $\Delta=2\tilde{\Delta}$ (thanks to its equivalence with
\eqref{eq:convenient}, pointed out earlier).
In Sec. \ref{sec: linear } we will 
show that the system defined in \eqref{eq: GLTBM update rule s} and \eqref{eq: GLTBM update rule tau} 
can also recover 
\eqref{eq: LTMM update rule} for suitable choices of the links $\{\xi_i^\mu\}$ and $\{\eta_j^\mu\}$. 
For the remainder of this manuscript, we shall set $\tilde{\Delta}=1$.

In order to solve the dynamics of the generalised model \eqref{eq: GLTBM update rule s} and \eqref{eq: GLTBM update rule tau} 
we use the dynamical cavity method previously used to study systems with sparse interactions. 
Earlier studies have successfully applied dynamical cavity to the study of monopartite systems of Ising \cite{neri2009cavity, aurell2011message, aurell2012dynamic, zhang2012inference} and Boolean \cite{torrisi2022uncovering} variables with bidirectional links, and bipartite systems of binary variables with unidirectional interactions \cite{torrisi2022uncovering}. Here we apply the method to the study of bipartite systems with (partially) bidirectional links, where each set of variables is subject to a different noise (the method would equally work for two sets of variables evolving via different dynamical rules). 

We can write the probability to observe a trajectory $(\bfs^{0...t_m},\btau^{0...t_m})$,
over time $t=0,\dots, t_m$, for the system defined in (\ref{eq: GLTBM update rule s}) and (\ref{eq: GLTBM update rule tau}), as
\begin{align}
    \mathrm{P}(\bfs^{0...t_m},\btau^{0...t_m}) &= \mathrm{P}(\bfs^{0},\btau^{0})\prod_{t=1}^{t_m}W(\bfs^{t}, \btau^{t}| \bfs^{t-1}, \btau^{t-1}),
    \label{eq:double-trajectory}
\end{align}
where 
$$
W\left(\bfs, \btau| \bfs^\prime, \btau^\prime\right)= \prod_{i=1}^{N} \prod_{\mu=N+1}^{N+P}\Wr(s_i|h_i(\btau'))
\widetilde{\Wr}(\tau_\mu|h_\mu(\bfs'))$$
is the one-time step transition probability and, for the choice of thermal noise, \eqref{eq:thermal}, we have
\begin{align}
    \Wr\left(s_i| h_i(\btau')\right) &= \frac{\rme^{\frac{\beta}{2}\left(2 s_i - 1 \right) h_i(\btau')}}{2 \cosh \frac{\beta}{2} h_i(\btau')} \\
    \widetilde{\Wr}\left(\tau_\mu| h_\mu(\bfs')\right) &=\frac{\rme^{\frac{\beta}{2}\left(2 \tau_\mu - 1 \right) h_\mu(\bfs')}}{2 \cosh \frac{\beta}{2} h_\mu(\bfs')} 
\end{align}
where $h_i(\btau) = \sum_\mu \xi_i^\mu \tau_\mu + \vartheta_i $ and $h_\mu(\bfs) = \sum_j \eta_j^\mu s_j + \vartheta_\mu$ are the local fields acting on $i$ and $\mu$, respectively.
As the local field $h_i$ depends only on the $\tau$'s in the neighbourhood of $i$, $\btau_{\partial_i}=\{\tau_\mu:\xi_i^\mu\neq 0\}$,
we shall write $h_i(\btau_{\partial_i})$. 
Similarly, we shall write $h_\mu(\bfs_{\partial_\mu})$
with $\bfs_{\partial_\mu}=\{s_j:\eta_j^\mu =1\}$. 
As we detail in \ref{app: dyn cav bipartite}, the dynamical cavity method allows us to derive an expression for the probability to observe the trajectory 
$s_i^{0...t_m}$ for the single gene $i$, in the cavity graph where TF $\nu$ has been removed, subject to a time dependent external field with trajectory
$\zeta_i^{(\nu),1\dots t_m}=\xi_i^{\nu} \tau_\nu^{0 \dots t_m-1}$,
\begin{align}
\begin{split}
    \mathrm{P}^{(\nu)}_i(s_i^{0...t_m} | \zeta_i^{(\nu),1\dots t_m}) &=\mathrm{P}(s_i^{0})\sum_{\btau_{\pa_i \setminus \nu}^{0...t_m-1}} \left[\prod_{t=1}^{t_m} \Wr(s_i^{t}| h_i(\btau_{\pa_i}^{t-1})) \right] \\
& \times
\prod_{\mu \in \pa_i\setminus \nu} \mathrm{P}_\mu^{(i)}(\tau_\mu^{0\dots t_m-1}|\zeta_\mu^{(i),1 \dots t_m-1})\label{eq:p_s_cav_exfield}
\end{split}
\end{align}
and similarly for the trajectory of a single TF $\mu$ in the cavity graph where gene $\ell$ has been removed, subject to a time dependent external field $\zeta_\mu^{(\ell),1\dots t_m} = \eta_{\ell}^\mu s_{\ell}^{0 \dots t_m-1}$,
\begin{align}
\begin{split}
    \mathrm{P}^{(\ell)}_\mu(\tau_\mu^{0...t_m}| \zeta_\mu^{(\ell),1\dots t_m}) &=\mathrm{P}(\tau_\mu^{0})\sum_{\bfs_{\pa_\mu \setminus \ell}^{0...t_m-1}} \left[\prod_{t=1}^{t_m} \widetilde{\Wr}(\tau_\mu^{t}| h_\mu(\bfs_{\pa_\mu}^{t-1})) \right] \\
&\times\prod_{j \in \pa_\mu \setminus \ell} \mathrm{P}_j^{(\mu)}(s_j^{0\dots t_m-1}|\zeta_j^{(\mu),1 \dots t_m-1}). \label{eq:p_tau_cav_exfield}
\end{split}
\end{align}
These expressions are exact when the bipartite network is a tree, and are expected to be a good approximation for sparse (bipartite) networks, where in the limit $N \to \infty$ the length of loops grows as $\log N$. 
In principle, they may be solved recursively for the probability of a trajectory up to a given time $t_m$. However, in expression (\ref{eq:p_s_cav_exfield}) this requires a sum over $|\btau_{\pa_i\setminus \nu}^{0,\dots,t_m-1}| = (2^{d_i-1})^{t_m}$ variables, which grows exponentially with time, and similarly, expression (\ref{eq:p_tau_cav_exfield}) requires a sum over $|\bfs_{\pa_\mu\setminus \ell}^{0,\dots,t_m-1}| = (2^{c_\mu-1})^{t_m}$ variables. In practice, this means that these equations can only be solved for very short times, making them unsuitable for the study of long time dynamics and stationary states. The exception is for systems with unidirectional interactions, where these equations drastically simplify (see \ref{app: dyn cav bipartite} for details), as reported in the literature 
 for monopartite systems \cite{neri2009cavity}. 

When bidirectional interactions are present, an approximation scheme, known as the One Time Approximation (OTA), has been proposed to reduce the computational complexity of the dynamical cavity method \cite{torrisi2022uncovering,neri2009cavity,aurell2011message,aurell2012dynamic,zhang2012inference}, which factorises the distributions of cavity trajectories into products of single time steps. 
Applying this approximation scheme, originally formulated for monopartite systems, to the present bipartite model, we assume
\begin{align}
    \mathrm{P}^{(\ell)}_\mu(\tau_\mu^{0...t_m}| \zeta_\mu^{(\ell),1\dots, t_m}) = \mathrm{P}_\mu(\tau_\mu^{0})\prod_{t=1}^{t_m}  \mathrm{P}^{(\ell)}_\mu(\tau_\mu^{t}| \zeta_\mu^{(\ell),t}) \label{app eq: OTA assumption s}
\end{align}
and
\begin{align}
    \mathrm{P}^{(\nu)}_i(s_i^{0...t_m}| \zeta_i^{(\nu),1\dots t_m}) = \mathrm{P}_i(s_i^{0})\prod_{t=1}^{t_m}  \mathrm{P}^{(\nu)}_i(s_i^{t}| \zeta_i^{(\nu),t}).\label{app eq: OTA assumption tau}
\end{align}
This allows us to derive
an equation for the probability to observe a gene $i$ in a given state $s_i^{t_m}$, at time $t_m$, in the cavity graph where TF $\nu$ has been removed, \textit{given} the external field induced by its state
at the earlier time step $\zeta_i^{(\nu),t_m} = \xi_i^{\nu} \tau_\nu^{t_m-1}$,
\begin{eqnarray}
    \mathrm{P}_i^{(\nu)}(s_i^{t_m}| \zeta_i^{(\nu),t_m})
    &=& \sum_{s_i^{t_m-2}} \sum_{\btau_{\partial_i\setminus \nu}^{t_m-1}} \Wr(s_i^{t_m}| h_i^{(\nu)}(\btau_{\partial_i}^{t_m-1}) + \zeta_i^{(\nu), t_m}) \nonumber \\ 
    &&  \quad \quad \times \left[ \prod_{\mu \in \partial_i\setminus \nu}\mathrm{P}^{(i)}_\mu(\tau_\mu^{t_m-1}| \zeta_\mu^{(i),t_m-1})\right] \mathrm{P}_i(s_i^{t_m-2}).  
    \nonumber\\
    \label{eq: p i without nu}
\end{eqnarray}
This depends on the probability to observe the TF $\mu$ in a given state $\tau_\mu^{t_m-1}$, at time $t_m-1$, in the cavity graph where gene $i$ has been removed, \textit{given} the state of $i$ at the earlier time step $\zeta_i^{(\nu),t_m-1} = \eta_i^\mu s_i^{t_m-2}$, 
\begin{eqnarray}
    \mathrm{P}_\mu^{(\ell)}(\tau_\mu^{t_m}| \zeta_\mu^{(\ell),t_m})
    &=& \sum_{\tau_\mu^{t_m-2}} \sum_{\bfs_{\partial_\mu\setminus \ell}^{t_m-1}} \widetilde{\Wr}(\tau_\mu^{t_m}| h_\mu^{(\ell)}(\bfs_{\partial_\mu}^{t_m-1}) + \zeta_\mu^{(\ell), t_m}) \nonumber \\ 
    && \quad \quad \times \left[ \prod_{j \in \partial_\mu\setminus \ell}\mathrm{P}^{(\mu)}_j(s_j^{t_m-1}| \zeta_j^{(\mu),t_m-1})\right] \mathrm{P}_\mu(\tau_\mu^{t_m-2}).
    \nonumber\\
    \label{eq: p mu without l }
\end{eqnarray}
As we detail in \ref{app: dyn cav bipartite},
and as observed earlier in the literature (see \cite{torrisi2022uncovering,neri2009cavity, zhang2012inference}),  
the assumption of time factorisation is not enough to find a closed set of equations and one must make additional assumptions. In writing equations (\ref{eq: p i without nu}) and (\ref{eq: p mu without l }) we
assume that the cavity distributions, with external field from the removed site, may be approximated by their non-cavity counterparts i.e 
\begin{align}
    \mathrm{P}^{(\nu)}_i(s_i^{T-2} | \zeta_i^{(\nu),T-2}) & \approx\mathrm{P}_i(s_i^{T-2}) \label{eq: closure approx s}
\end{align}
and
\begin{align}
\mathrm{P}^{(\ell)}_\mu(\tau_\mu^{T-2} | \zeta_\mu^{(\ell),T-2}) & \approx\mathrm{P}_\mu(\tau_\mu^{T-2}) \label{eq: closure approx tau},
\end{align} following the 
approach of \cite{zhang2012inference}. This 
has recently been shown to accurately predict single-site marginals in non-equilibrium steady-states   \cite{torrisi2022uncovering}. Equations (\ref{eq: p i without nu}) and (\ref{eq: p mu without l }) then depend upon the marginal probability to observe gene $i$ at time $t_m$, $\mathrm{P}_i(s_i^{t_m})$, and TF $\mu$ at time $t_m$, $\mathrm{P}_\mu(\tau_\mu^{t_m})$, which evolve according to, 
\begin{eqnarray}
    \mathrm{P}_i(s_i^{t_m})
    &= \sum_{s_i^{t_m-2}} \mathrm{P}_i(s_i^{t_m-2}) \sum_{\btau_{\partial_i}^{t_m-1}} \Wr(s_i^{t_m}| h_i(\btau_{\partial_i}^{t_m-1})) \nonumber \\
    & \quad \times \left[ \prod_{\mu \in \partial_i}\mathrm{P}^{(i)}_\mu(\tau_\mu^{t_m-1}| \zeta_\mu^{(i),t_m-1})\right] \label{eq: p i}
\end{eqnarray}
and
\begin{eqnarray}
    \mathrm{P}_\mu(\tau_\mu^{t_m})
    &= \sum_{\tau_\mu^{t_m-2}}\mathrm{P}_\mu(\tau_\mu^{t_m-2}) \sum_{\bfs_{\partial_\mu}^{t_m-1}} \widetilde{\Wr}(\tau_\mu^{t_m}| h_\mu(\bfs_{\partial_\mu}^{t_m-1})) \nonumber \\ 
    &\quad \times \left[ \prod_{j \in \partial_\mu}\mathrm{P}^{(\mu)}_j(s_j^{t_m-1}| \zeta_j^{(\mu),t_m-1})\right]. \label{eq: p mu}
\end{eqnarray}
For a given bipartite network $(\bfeta,\bfxi)$ and initial condition $\mathrm{P}_i(s_i^{0}) ~ \forall i$ and $\mathrm{P}_\mu(\tau_\mu^{0}) ~ \forall ~ \mu$, one can solve these equations by simple iteration. For networks with arbitrary bidirectionality, equations (\ref{eq: p i without nu}),(\ref{eq: p mu without l }),(\ref{eq: p i}) and (\ref{eq: p mu}) provide an efficient numerical scheme to solve for the transient and long-time dynamics of bipartite systems.

\section{Linear threshold model with self-interactions} \label{sec: linear }

\subsection{Equilibrium analysis of monopartite systems with self-interactions}\label{sec: linear statics }

In this section we analyse the equilibrium behaviour of the linear threshold model with symmetric interactions $J_{ij}=J_{ji}$,
evolving via parallel dynamics (\ref{eq: LTMM update rule}) in the presence of self-interactions $J_{ii}$.
It is well-known that for symmetric interactions $J_{ij}=J_{ji}$, linear threshold models evolving via parallel dynamics converge, both in the presence and the absence of self-interactions, to 
an equilibrium state described by the Peretto distribution, which is characterised by a temperature-dependent pseudo-Hamiltonian \cite{peretto1984collective}.
Equilibrium analysis of such systems 
has been carried out for Ising variables via transfer matrices \cite{skantzos2000random,skantzos2001random} and the replica method \cite{fontanari1988information,castillo2004little}. An equilibrium analysis of sparse \textit{Boolean} networks, however, is absent from the literature.  Here we fill this gap by using the cavity method,
originally formulated for the Gibbs-Boltzmann distribution, reached by systems evolving via sequential dynamics in the absence of self-interactions. We then use this method to study the equilibrium of the linear threshold model \eqref{eq: LTMM update rule} with symmetric interactions, evolving by parallel dynamics, in the presence of self-interactions. This is described  
by Peretto distribution $p_{\text{eq}}(\bfs)= Z^{-1}\rme^{ - \beta H_{\beta}(\bfs)}$
with the \textit{pseudo}-Hamiltonian 
\begin{equation}
H_{\beta}(\bfs) =  - \frac{1}{\beta} \sum_i \ln 2 \cosh \frac{\beta}{2}h_i(\bfs_{\pa_i},s_i) - \frac{1}{2}\sum_i h_i(\bfs_{\pa_i},s_i) -\sum_i \vartheta_i s_i
\end{equation} 
where $h_i(\bfs_{\pa_i},s_i) = \sum_{j \in \pa_i} J_{ij} s_j + J_{ii}s_i + \vartheta_i$.
The function $H_{\beta}(\bfs)$ is not a true Hamiltonian due to its dependence on noise level, $T= \beta^{-1}$. 
It is
helpful to introduce a fictitious set of variables $\btau = \{0,1\}^{N}$ such that we extend our system to one of size $2N$, as it was similarly done 
in \cite{fontanari1988information,castillo2004little} for models of Ising spins. This allows us to rewrite the equilibrium distribution as the marginal of the joint distribution of real and fictitious variables, 
\begin{align}
    p_{\text{eq}}(\bfs) &=  \sum_{\btau}p(\bfs,\btau) \\
    p(\bfs,\btau) &= \frac{1}{Z} \rme^{-\beta \mathcal{H}(\bfs,\btau)} \label{eq: statics joint dist}
\end{align}
with $\mathcal{H}(\bfs,\btau)  = -\sum_{\ell,j} s_{\ell} J_{\ell j} \tau_j - \sum_{\ell} s_{\ell} J_{\ell \ell} \tau_{\ell} - \sum_{\ell} \vartheta_{\ell}(s_{\ell} + \tau_{\ell})$. In \ref{app: equib analysis}  we use the equilibrium cavity method to find a closed set of equations for single site marginals,
\begin{align}
  p_i(s_i,\tau_i) &= \frac{1}{Z_i} \rme^{\beta \left(s_i J_{ii}\tau_i + \vartheta_i (s_i + \tau_i)\right)} \prod_{j \in \pa_i} \sum_{s_j} \sum_{\tau_j}\rme^{\beta \left(s_i J_{ij}\tau_j + \tau_iJ_{ij}s_j\right)} p^{(i)}_j(s_j, \tau_j)\label{eq: site marginal }    \\
  p_i^{(\ell)}(s_i,\tau_i) &= \frac{1}{Z_i^{(\ell)}}  \rme^{\beta \left(s_i J_{ii}\tau_i + \vartheta_i (s_i + \tau_i)\right)}\prod_{j \in \pa_i \setminus \ell} \sum_{s_j} \sum_{\tau_j}\rme^{\beta \left(s_i J_{ij}\tau_j + \tau_iJ_{ij}s_j\right)} p^{(i)}_j(s_j, \tau_j)  \label{eq: cavity site marginal }
\end{align}
which can be solved by simple iteration (see \ref{app: equib analysis} for details). These equations are exact when the interaction matrix is a tree, and give a good approximation for sparse graphs where the length of loops grows logarithmically with the system size. From the solution of these equations we can compute the average activation probability of a site $\langle s_i \rangle = \sum_{s_i,\tau_i} s_i p_i(s_i,\tau_i)$.

To verify these equations, we now consider a system with interactions drawn according to the following probability distributions 
\begin{align}
    \pr(J_{ij}) &= \left(1 - \frac{c}{N}\right)\delta_{J_{ij},0} + \frac{c}{ N}\Big[\frac{1+a}{2} \delta_{J_{ij},1} + \frac{1-a}{2}\delta_{J_{ij},-1}\Big] \quad\quad \forall i\neq j \\
    \pr(J_{ii}) &= \left(1 - p\right)\delta_{J_{ii},0} + p\Big[\frac{1+b}{2}\,\delta_{J_{ii},1} + \frac{1-b}{2}\delta_{J_{ii},-1}\Big] \quad\quad \forall i
\end{align}
such that $c$ and $p$ control the density of links, and $a,b \in [-1,1]$ control the sign of the interactions. 
Predictions from the cavity method are compared 
with results from MC simulations via 
\begin{align}
    \left<s_i\right>_{\rm MC} = \frac{1}{t_{l}}\sum_{t=t_{eq}}^{t_{l} + t_{eq}} s_i^{t}  \label{eq: MC site act}
\end{align}
where $t_{eq}$ is a large time where the system has reached equilibrium and $t_{l}$ is a long time that we average the state of the site $i$ over.

\begin{figure*}[t]
  \begin{tabular}{c @{\quad} c }
    \hspace{-3mm} \includegraphics[width=0.5\linewidth]{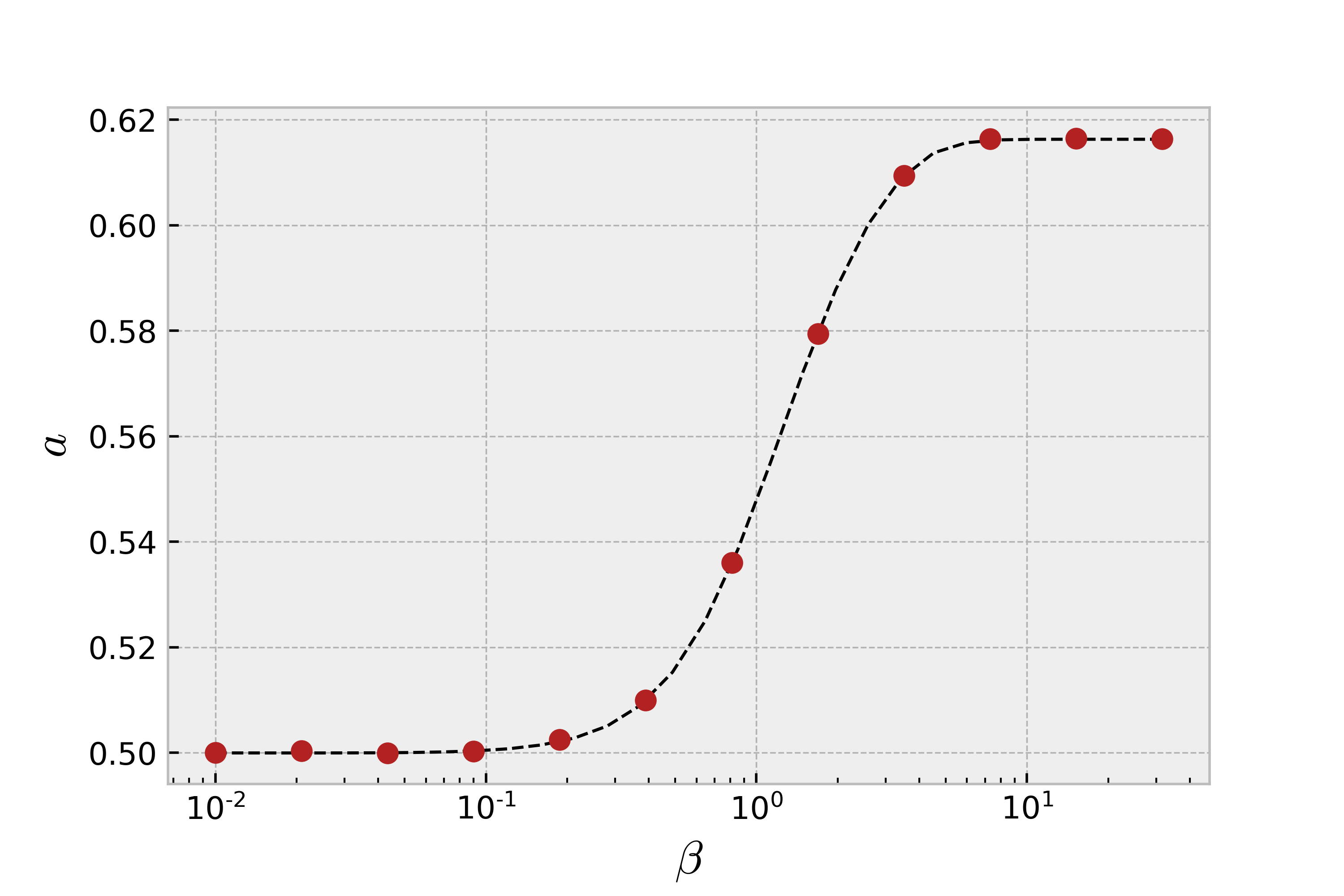}  &  \hspace{-3mm}
    \includegraphics[width=0.5\linewidth]{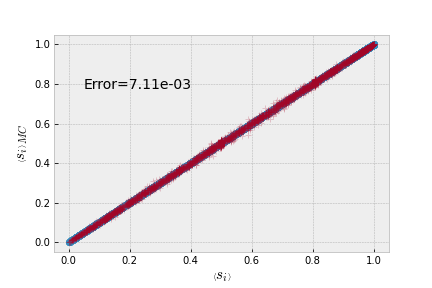} \vspace{-2mm} \\
     \vspace{-3mm}
     \hspace{-60mm} \text{a)} & \hspace{-58mm} \text{b)} 
  \end{tabular}
  \caption{ (a) Average activation probability of sites $a = \frac{1}{N}\sum_i \langle s_i\rangle$ against inverse noise level $\beta$. Symbols indicate the results of MC simulations, dotted line indicates solution from cavity equations (\ref{eq: site marginal }). (b) Scatter plot of site activation probabilities $\langle s_i\rangle$ computed from static cavity method and MC simulations at noise level $\beta=2$. Activation probability from MC simulations is computed by (\ref{eq: MC site act}) with $t_{l}=1000$ in (a) and $t_{l}=5000$ in (b). 
  In (a) and (b), the network size is $N=5000$ with average connectivity $c=2$, density of self-interactions $p=0.5$ and bias $a=b=0$. The external field is $\vartheta_i=0 ~ \forall ~ i$. Annotation in (b) indicates the root mean square error.}
\vspace{-3mm}
\hspace{-60mm} \textbf{a)}  \hspace{-58mm} \textbf{b)} 
  \label{fig: monopartite comparison equib}
\end{figure*}

In figure \ref{fig: monopartite comparison equib}(a), we show results for 
the average activation probability 
$a=N^{-1}\sum_i \langle s_i \rangle$, 
as a function of the noise level, whereas in figure \ref{fig: monopartite comparison equib}(b), we show a scatter plot of the 
equilibrium activation probabilities of individual sites, 
at a given noise level, as 
predicted by the cavity method and computed from MC simulations, finding excellent agreement.
It is well-known that finitely connected systems of Ising spins ($s_i\in\{-1,1\}$) undergo a spin glass transition at low noise, where replica-symmetry is broken and the (replica-symmetric equivalent) cavity equations are expected not to converge to the correct solution \cite{mezard2001bethe}. Results in figure \ref{fig: monopartite comparison equib}(a)
show, however, that in the system of sparsely connected {\it Boolean} variables under study, the cavity equations converge to the correct solution down to low levels of noise.
Since a system of Boolean variables can be mapped to an equivalent Ising spin system by adding a quenched random external field $\vartheta_i \to \vartheta_i + \sum_jJ_{ij}$ 
and the presence of an external field is known to change the noise level at which a finitely connected system will enter a replica-symmetry-broken phase \cite{parisi2014diluted,altieri2017loop,thouless1986spin,carlson1990betheI,carlson1990betheII}, 
this may explain the relatively low noise level at which the cavity equations successfully converge (and  
replica symmetry is observed). 
We may expect, however, that further lowering the temperature our system may undergo ergodicity breaking, as our (non-equilibrium) analysis at zero temperature will reveal for the case of partially symmetric interactions (see Sec. \ref{sec: nonlinear }).

\begin{figure*}[t]
  \begin{tabular}{c @{\quad} c }
    \hspace{-3mm} \includegraphics[width=0.5\linewidth]{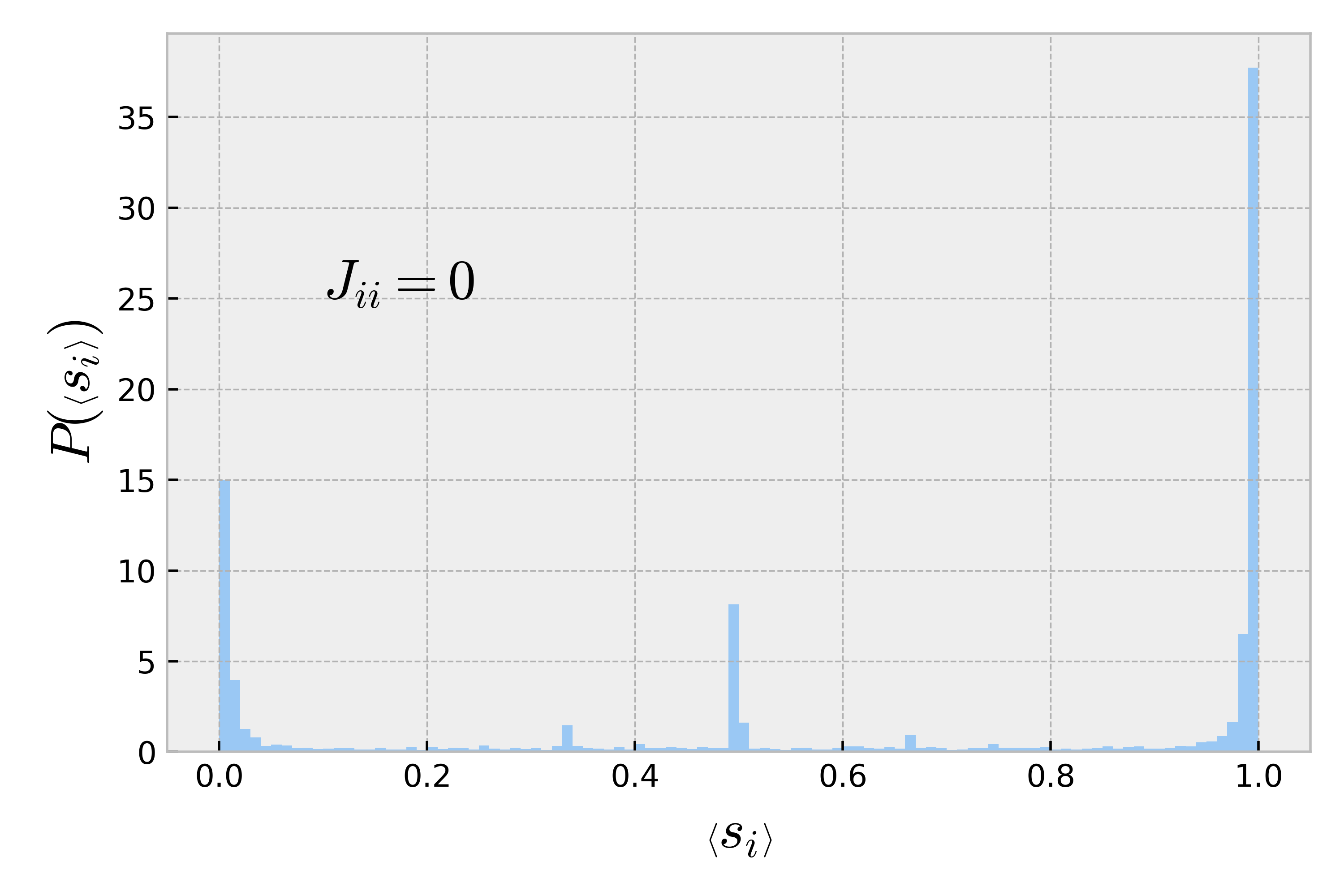}  &  \hspace{-3mm}
    \includegraphics[width=0.5\linewidth]{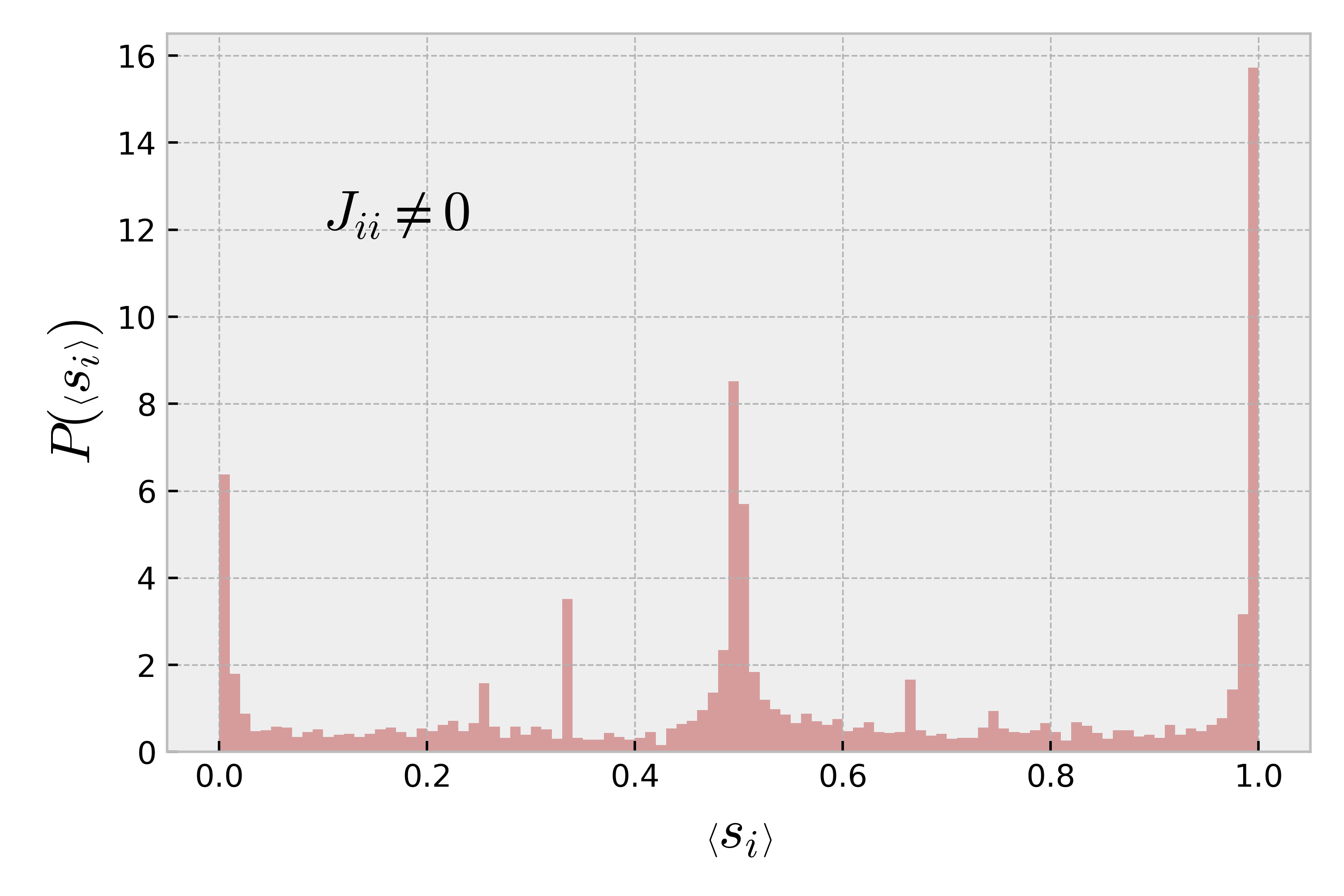} \vspace{-3mm} \\
     \hspace{-60mm} \text{a)} & \hspace{-58mm} \text{b)} \\
     \hspace{-3mm} \includegraphics[width=0.49\textwidth]{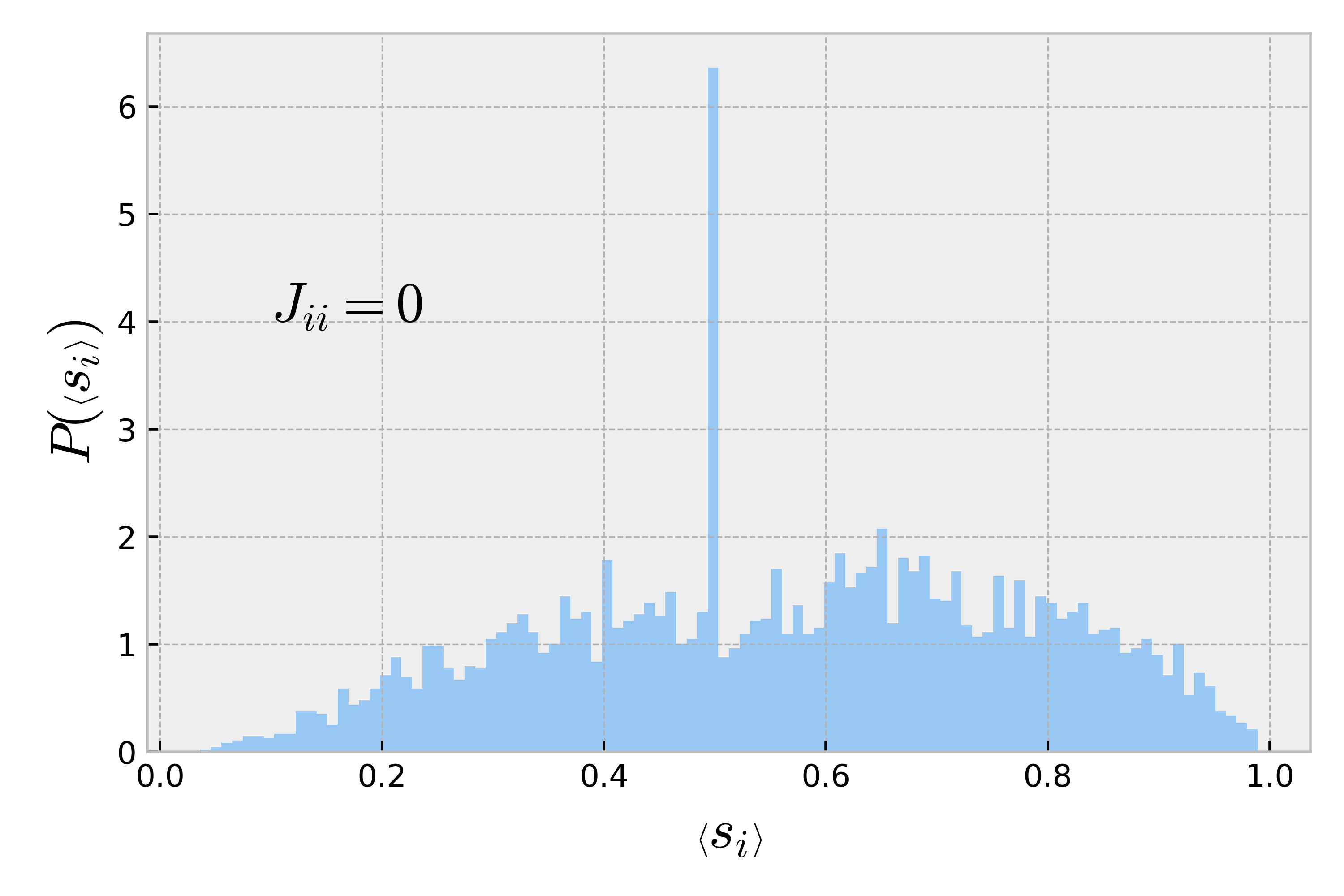} & 
    \hspace{-3mm}
   \includegraphics[width=0.49\textwidth]{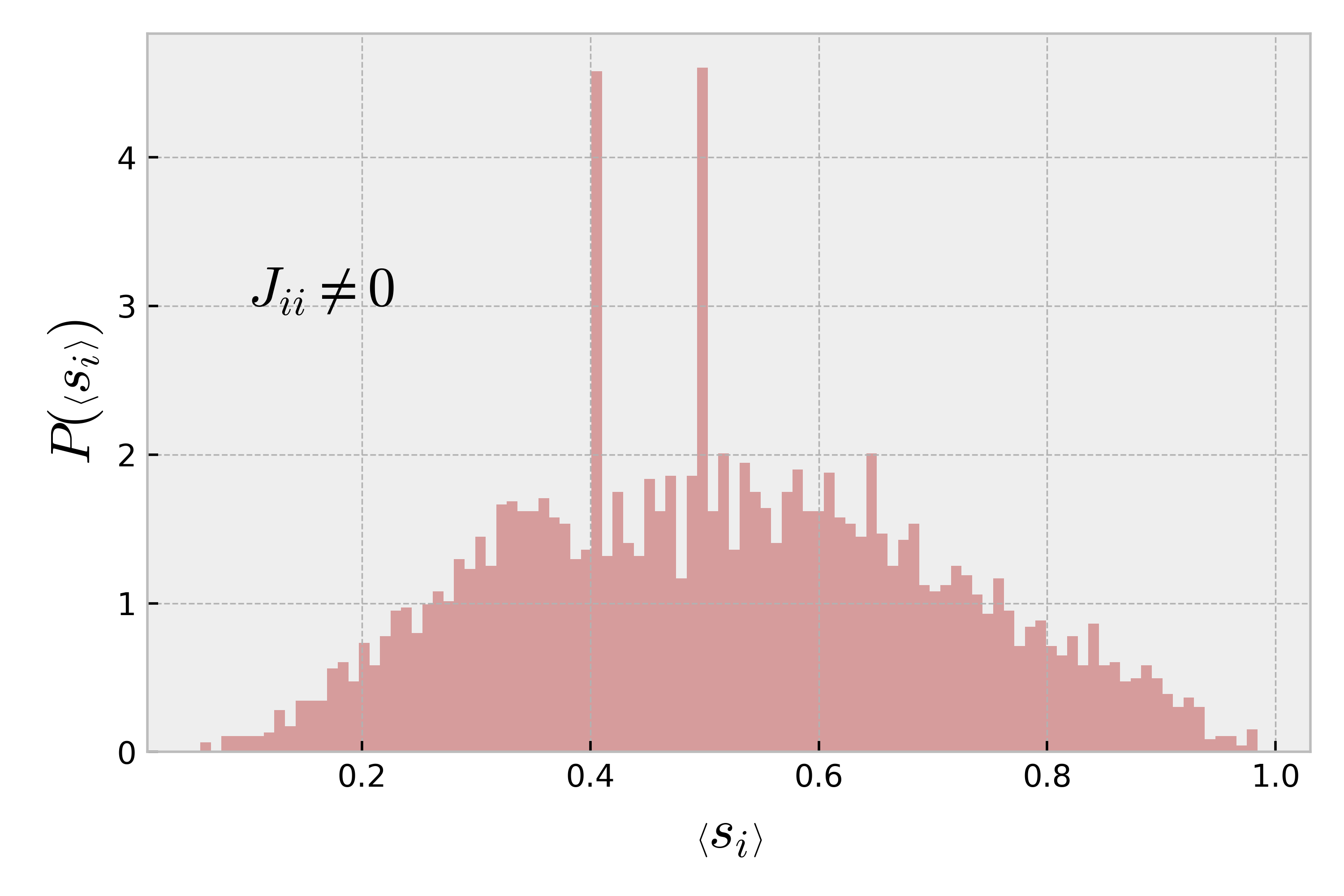} \vspace{-3mm} \\
     \vspace{-3mm}
     \hspace{-60mm} \text{c)} & \hspace{-58mm} \text{d)} 
  \end{tabular}
  \caption{Probability density function for the activation probability of sites $\pr(\langle s_i\rangle )$ computed from the equilibrium cavity equations. Results in each panel are for the same network of size $N=5000$, with connectivity $c=3$, unbiased interactions $a=0$ and external field $\vartheta_i=0 ~ \forall ~ i$. In (b) and (d), self-interactions are chosen with parameters $p=0.5$ and $b=-1$ such that all self-interactions $J_{ii}=-1$. In (a) and (c), $J_{ii}=0 ~ \forall i$. In (a) and (b) inverse noise is set to $\beta=5$, in (c) and (d) $\beta=1$. }
  \label{fig: monopartite equib dist}
\end{figure*}

In figure \ref{fig: monopartite equib dist} we plot the distribution $\pr(\langle s_i \rangle)$ at different temperatures, as predicted by the cavity method at equilibrium, for unbiased interactions i.e. $a=0$. In the absence of self-interactions (figures \ref{fig: monopartite equib dist}(a) and \ref{fig: monopartite equib dist}(c)), 
a bias in the fraction of sites that are activated in the steady state, emerges at low temperature. This is 
consistent with results obtained in  \cite{torrisi2022uncovering} via dynamical cavity approaches (iterated until convergence), in the absence of self-interactions. Figures \ref{fig: monopartite equib dist}(b) and \ref{fig: monopartite equib dist}(d) show the distribution $\pr(\langle s_i \rangle)$, in the presence of self-interactions, where direct application of the dynamical cavity method would be cumbersome (see \ref{app: dyn cav self-interactions}). By choosing that all self-interactions, where present, are negative, we see that the bias in the fraction of sites which are active at low temperature is modified. For example, at relatively high temperature, we see that, in addition to the expected peak at $a_i=1/2$ (corresponding to fluctuating variables with zero field),
there is an extra peak in the presence of self-interactions, due to sites associated with the single peak having their activation probability reduced by their self-interaction.

\subsection{Dynamics of monopartite spin systems with self-interactions} \label{sec: linear dynamics}
Previously, the dynamics of the linear threshold model (\ref{eq: LTMM update rule}) have been studied using the dynamical cavity method in the absence of self-interactions \cite{torrisi2022uncovering}. As we mentioned earlier, for networks with bidirectional links, the computational complexity of the cavity equations is exponential in time, due to memory effects, 
and the one time approximation (OTA) scheme has been proposed to reduce the computational complexity, so that the cavity equations can be solved at long and short times \cite{neri2009cavity,aurell2011message,aurell2012dynamic,zhang2012inference}. However, in \ref{app: dyn cav self-interactions} we show that for systems with self-interactions the OTA scheme fails to provide a closed set of equations, and one is left with the computational complexity being exponential in time, which would prevent solving the equations at long times. To 
overcome this difficulty, 
we map our system to an equivalent bipartite system, which can be solved using the bipartite cavity equations under the OTA scheme detailed in section \ref{sec: cavity method}. 

\begin{figure*}[t]
\centering
 \includegraphics[width=0.79\textwidth]{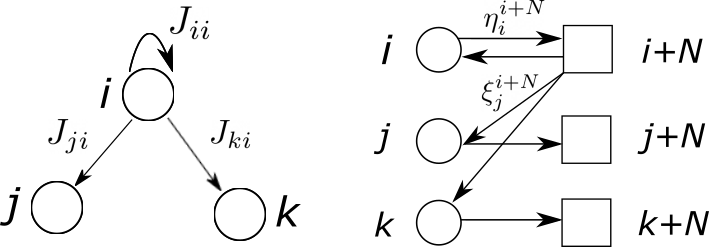}
  \caption{Sketch of how we map a monopartite system with self-interactions (left) to an equivalent bipartite spin system (right). In this mapping we have that the interaction site $i$ has on site $j$ is given by $J_{ji}= \xi_j^{i+N}$, and we have $\eta_i^\mu=\delta_{i,i+N}$.}
  \label{fig: monopartite sketch}
\end{figure*}

\begin{figure*}[t]
  \begin{tabular}{c @{\quad} c }
    \hspace{-3mm} \includegraphics[width=0.5\linewidth]{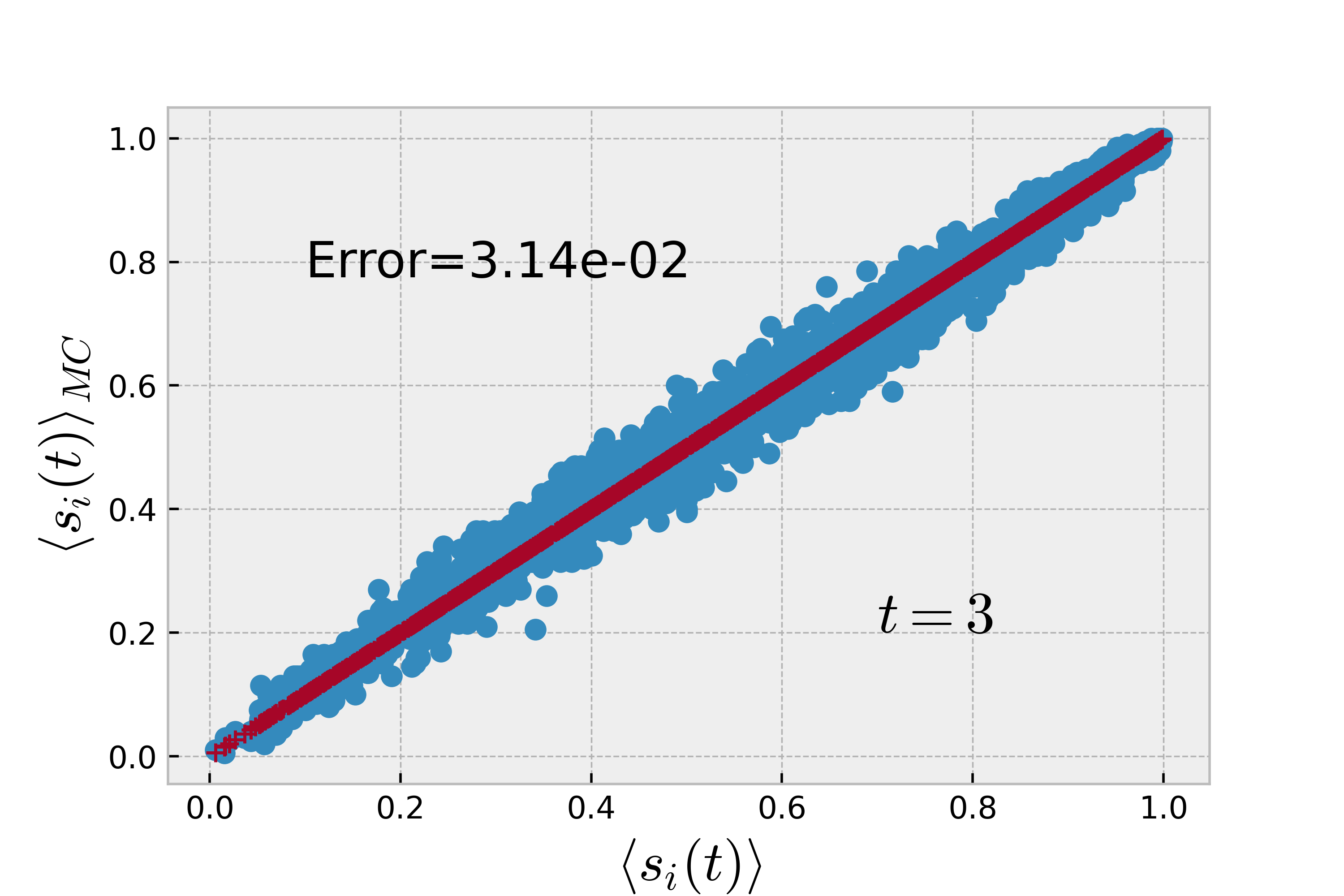}  &  \hspace{-3mm}
    \includegraphics[width=0.5\linewidth]{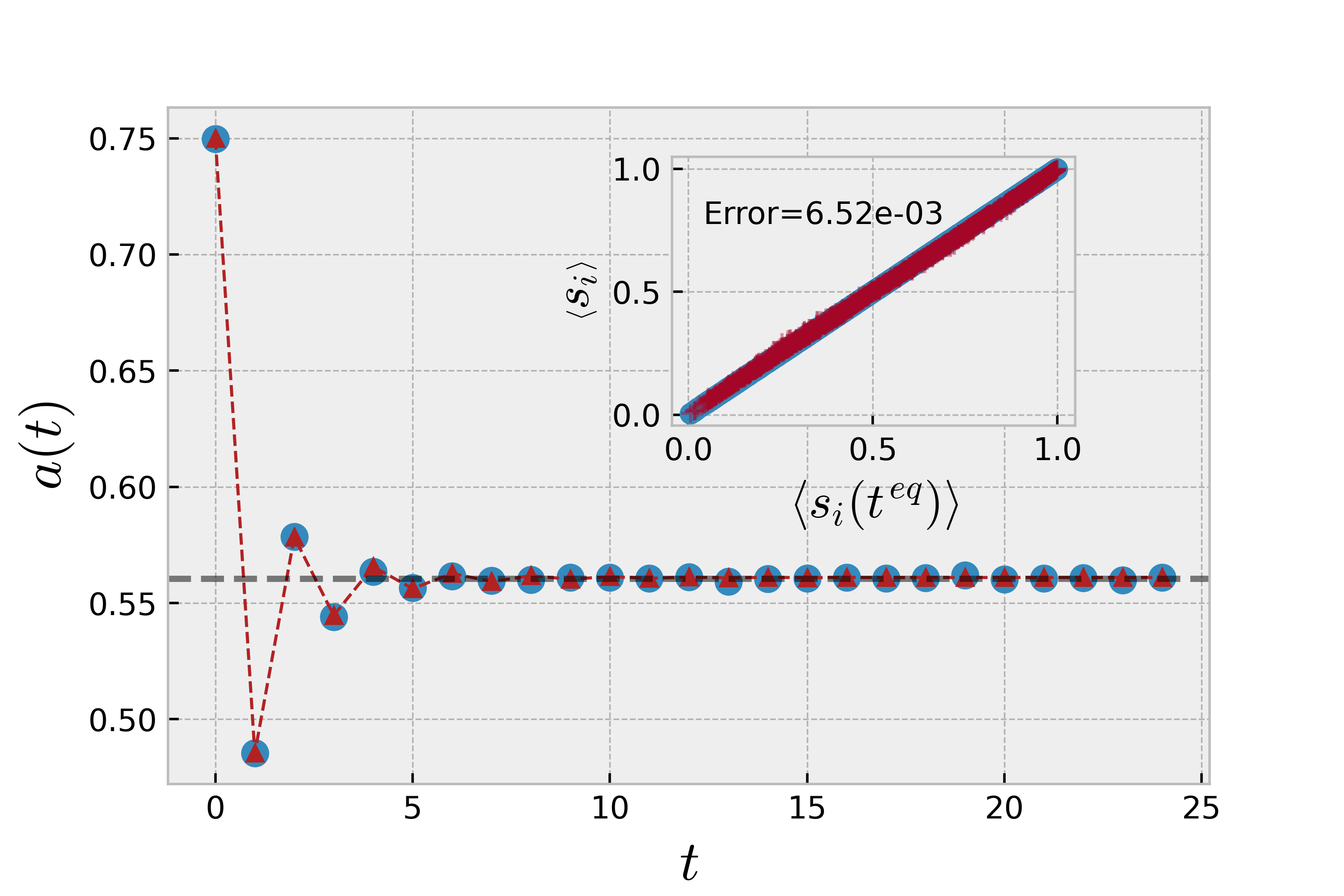} \vspace{-2mm} \\
     \vspace{-3mm}
     \hspace{-60mm} \text{a)} & \hspace{-58mm} \text{b)} 
  \end{tabular}
  \caption{ (a) Scatter plot of site activation probabilities $\langle s_i(t)\rangle$ computed via dynamical cavity and from MC simulations (average over 200 runs with different initial conditions drawn from the same distribution).  (b) Site average of activation probabilities against time. Circles indicate MC simulations, triangles, connected by red dashed line for visual aid, indicate the solution of the bipartite dynamical cavity equations,  (\ref{eq: p i}) and (\ref{eq: p mu}). Dashed horizontal line indicates the solution for average activation found from the cavity equations at equilibrium (\ref{eq: site marginal }).
  Inset of (b) shows scatter plot of activation probability of each spin computed from equilibrium cavity and dynamical cavity method in the steady state. Annotation indicates the root mean square error. In (a) and (b), network size is $N=2500$, connectivity $c=2$, density of self-interactions $p=0.25$, bias $a=b=0$, external field $\vartheta_i=0 ~ \forall ~  i$,
  noise level $\beta=1.5$, and initial conditions $P(s_i^{0}) =P(\tau_\mu^{0}) = 0.75$.}
  \label{fig: monopartite mapping comparison equib}
\end{figure*}

To map our system of $N$ variables with interactions $J_{ij}$ to the model defined in (\ref{eq: GLTBM update rule s}) and  (\ref{eq: GLTBM update rule tau}), we create an additional set of $N$ nodes, such that the system is now of size $2N$. This means that each site in the original system is given a partner site in the new set of variables, as shown in figure \ref{fig: monopartite sketch}. To achieve this we set $\eta_i^\mu = \delta_{\mu,i+N}$ in the bipartite model and
$J_{ij} = \xi_i^{j+N}$. If we now set $\hat{T}=0$ and $\vartheta_\mu=-\epsilon$ we have that (\ref{eq: GLTBM update rule tau}) becomes, 
\begin{eqnarray}
   \tau_\mu(t + \Tilde{\Delta}) &=& \theta\left( s_{\mu-N}(t) -\epsilon \right)\nonumber\\
   &=& s_{\mu-N}(t)
\end{eqnarray}
which when inserted into (\ref{eq: GLTBM update rule s}) we find,
\begin{eqnarray}
    s_i(t + \Tilde{\Delta}) &=& \theta\left( \sum_{\mu=N}^{2N} \xi_i^\mu  s_{\mu-N}(t - \Tilde{\Delta}) +  \vartheta_i - z_i(t) \right). 
    \end{eqnarray}
        Hence,
    \begin{align}
    s_i(t + \Delta)
    &= \theta\left( \sum_{j=1}^{N} J_{ij}  s_j(t) +  \vartheta_i - z_i(t+\Delta/2) \right) \label{eq: monpartite nearly mapped}
\end{align}
with $\Delta=2\tilde{\Delta}$.
We now see that equation (\ref{eq: monpartite nearly mapped}) is of the same form as the linear threshold model with self-interactions (\ref{eq: LTMM update rule}). If we impose the initial conditions $\tau_{i+N}(0) = s_i(0) ~ \forall ~ i$, and assume that $z_i(t)$ are drawn from the same distribution (\ref{eq:thermal}) at each time, these two models are equivalent, up to some random noise, at every $2\Delta$ time points. With this mapping the procedure is now as follows 
\begin{enumerate}
    \item Draw interaction matrix $J_{ij}$
    \item Construct bipartite system with $\xi_i^{j+N} = J_{ij}$ and $\eta_i^\mu = \delta_{\mu,i+N}$ 
    \item Solve the bipartite dynamical cavity equations up to some time $t_m$, with $\Tilde{\Delta}=1$
    \item Compare with MC simulations of the monopartite system (\ref{eq: LTMM update rule}) at $t=0,2,4,\dots,t_m$.
\end{enumerate}
We follow this procedure and compute the activation probability of each site obtained via dynamical cavity applied to the bipartite system and via MC simulations of the equivalent monopartite system. Figure \ref{fig: monopartite mapping comparison equib}(a) shows results for the transient behaviour where the dynamical cavity method accurately predicts the activation probability of each site.  Figure \ref{fig: monopartite mapping comparison equib}(b) shows the full time dependence of the average activation probability and agreement is excellent both during transient and in the steady state. Additionally, the inset of figure \ref{fig: monopartite mapping comparison equib}(b) shows that the stationary state reached by the bipartite dynamical cavity method is in excellent agreement with the equilibrium state predicted by the static cavity equations. Therefore, by mapping a monopartite system with self-interactions to an equivalent bipartite system, we can now study their transient and long-time dynamics.

\section{Nonlinear model with correlated, multi-node interactions} \label{sec: nonlinear }

\subsection{Dynamical analysis of systems with multi-node interactions}

Previously, the dynamics of the nonlinear model (\ref{eq: NLTM update rule}) has been studied under the assumption that interactions are uncorrelated, $\pr(\bfxi,\bfeta) = \pr(\bfxi)\pr(\bfeta)$, and in the absence of noise in TF synthesis, $\hat{T}=0$ \cite{hannam2019percolation, torrisi2022uncovering}. Here we solve the dynamics of the system defined in  (\ref{eq:convenient}), from which the dynamics of  (\ref{eq: NLTM update rule}) can be recovered, with arbitrary noise $\hat{T}$ in TF synthesis and correlated interactions $\pr(\bfxi,\bfeta) = \pr(\bfxi|\bfeta)\pr(\bfeta)$.  
We will assume
$\pr(\bfeta) = \prod_{i, \mu} \pr(\eta_i^\mu)$ 
with 
\begin{eqnarray}
\pr(\eta_i^\mu ) &= \left(1 - \frac{c}{N}\right)\delta_{\eta_i^\mu,0} + \frac{c}{ N}\delta_{\eta_i^\mu,1} \label{eq: partially symmetric eta}
\end{eqnarray}
and
$\pr(\bfxi|\bfeta) = \prod_{i, \mu} \pr(\xi_i^\mu|\eta_i^\mu)$, with the choice  
\begin{eqnarray}
\pr(\xi_i^\mu | \eta_i^\mu\!=\!0) &= \left(1 \!-\! \frac{d}{N}\right)\delta_{\xi_i^\mu,0} + \frac{d}{N}\Big[\frac{1+a}{2}\delta_{\xi_i^\mu,\frac{1}{d}}
+ \frac{1-a}{2}\delta_{\xi_i^\mu,-\frac{1}{d}}
\Big]
\label{eq: partially symmetric xi, 1} \\
\pr(\xi_i^\mu | \eta_i^\mu=1) &= \left(1 - p\right)\delta_{\xi_i^\mu,0} + p\left[\frac{1+b}{2}\delta_{\xi_i^\mu,1}+ \frac{1-b}{2}\delta_{\xi_i^\mu,-1}\right] \label{eq: partially symmetric xi, 2}
\end{eqnarray}
such that $c$ is the average in-degree of a TF and $d +pc$ is the average out-degree of a TF, $d$ being the average number of unidirectional links stemming from a TF and $pc$ the average number of bidirectional links stemming from (and pointing to) a TF. 
In particular, $p \in \left[0,1\right]$ controls the density of bidirectional links, such that the network is unidirectional when $p=0$, and when $p=1$ any gene that contributes to the synthesis of a TF will be regulated by that TF, in which case the number of bidirectional links in the network is maximal and equal to $\alpha c N$. We note that when $p=1$ the network is not fully bidirectional as $\eta_i^\mu=0$ does not imply $\xi_i^\mu=0$ and there are still $\alpha d N$ unidirectional links from TFs to genes.
The bias in positive and negative regulatory couplings is controlled through the parameters $a,b \in \left[-1,1\right]$ such that all regulatory effects are excitatory ($\xi > 0$) if $a,b=1$ and all are inhibitory ($\xi <0$) if $a,b=-1$.

\begin{figure}[t]
   \begin{tabular}{c @{\quad} c }
    \hspace{-3mm} \includegraphics[width=0.5\linewidth]{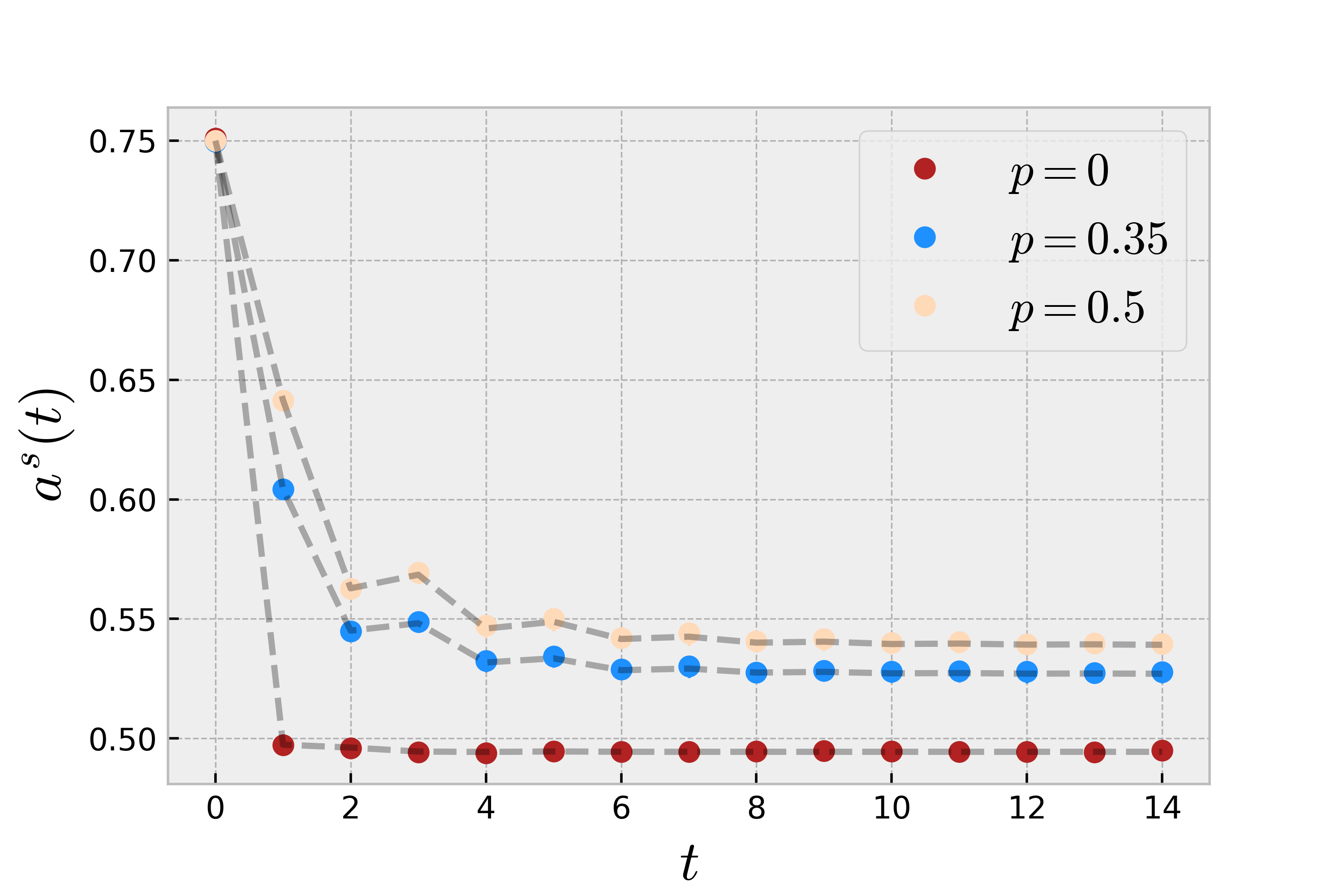}  &  \hspace{-3mm}
    \includegraphics[width=0.5\linewidth]{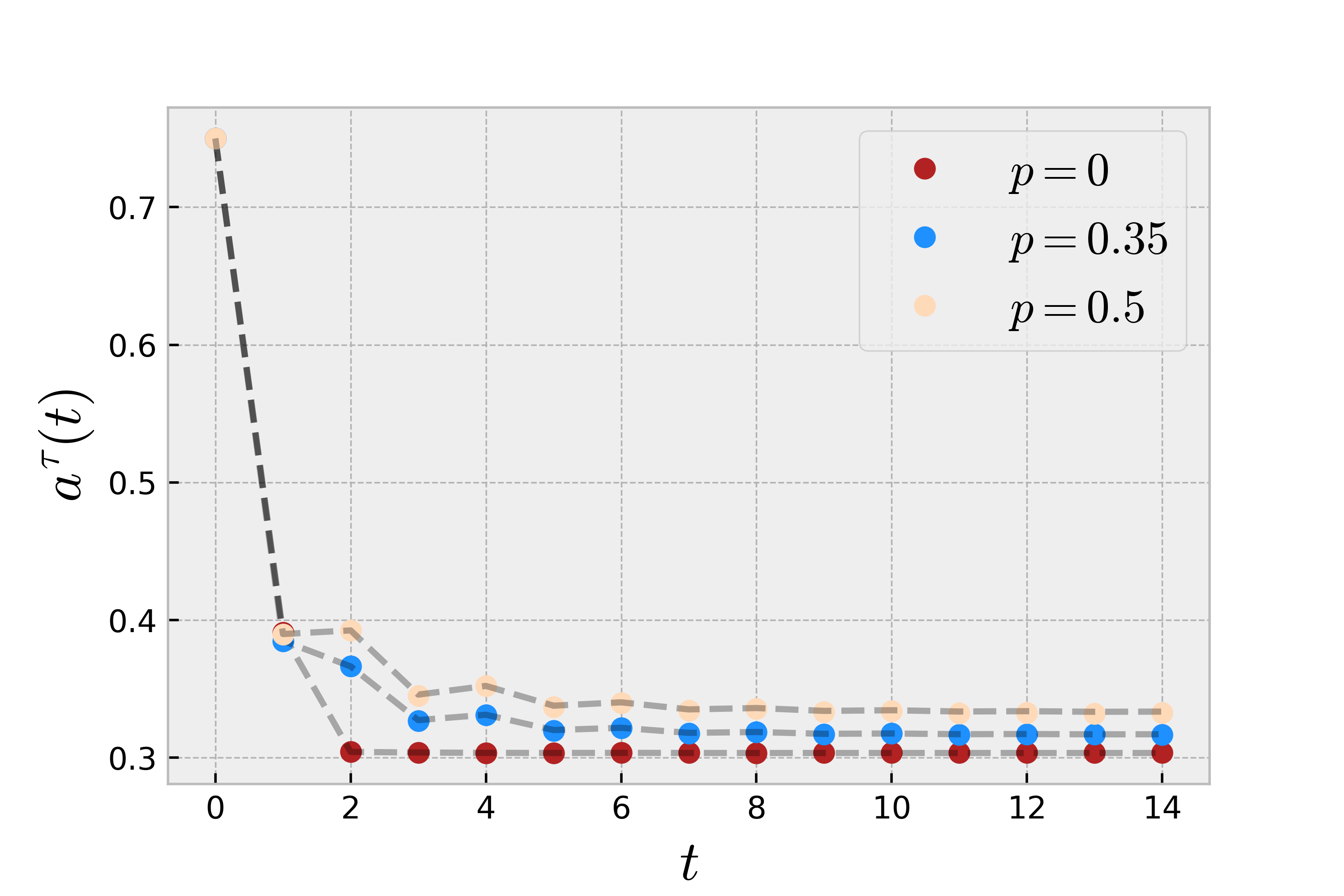}  \vspace{-0.25mm} \\
     \vspace{-3mm}
     \hspace{-60mm} \text{a)} & \hspace{-58mm} \text{b)} 
  \end{tabular}
  \caption{Average activation of (a) genes $a^{s}(t) = N^{-1}\sum_i\langle s_i(t)\rangle$ and (b) TFs, $a^{\tau}(t) = (\alpha N)^{-1}\sum_\mu\langle \tau_\mu(t)\rangle$ against time. Markers indicate MC simulations on networks with size $N=2500$ averaged over $100$ runs with initial conditions drawn from $\pr(s_i^{0}) = \pr(\tau_\mu^{0}) = 0.75 ~ \forall ~ i,\mu $. Noise level is $\beta=\hat{\beta}=10$, network parameters are $c=1$, $d=2$, $a=0$, $b=1$. Results shown for different densities of self-interactions, $p$, as indicated in the legend.}
  \label{fig: partial symmetry DC error finite temp s + tau }
\end{figure}

\begin{figure}[t]
\begin{tabular}{lccccc}
  \includegraphics[width=0.33\textwidth]{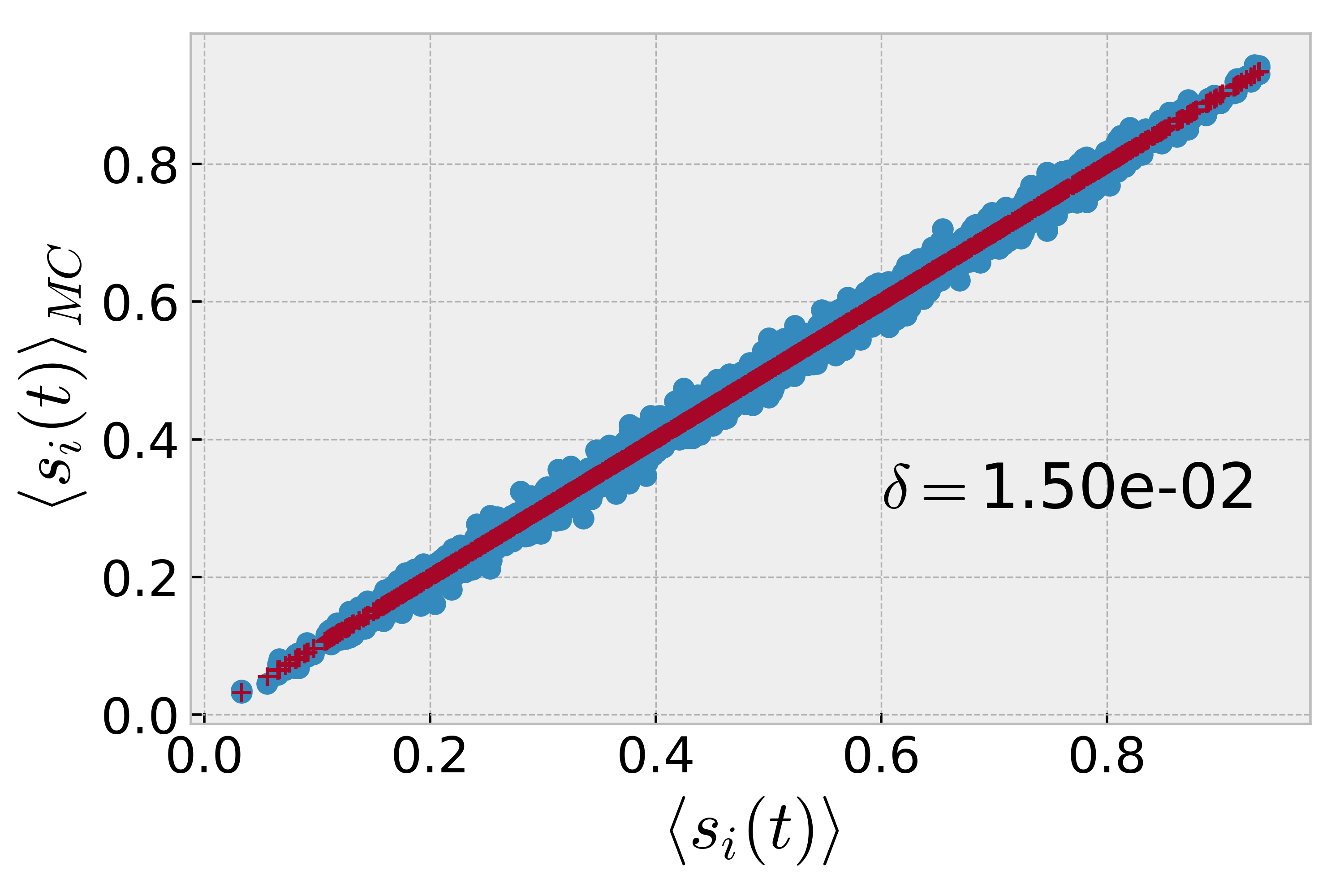}  &  \includegraphics[width=0.33\textwidth]{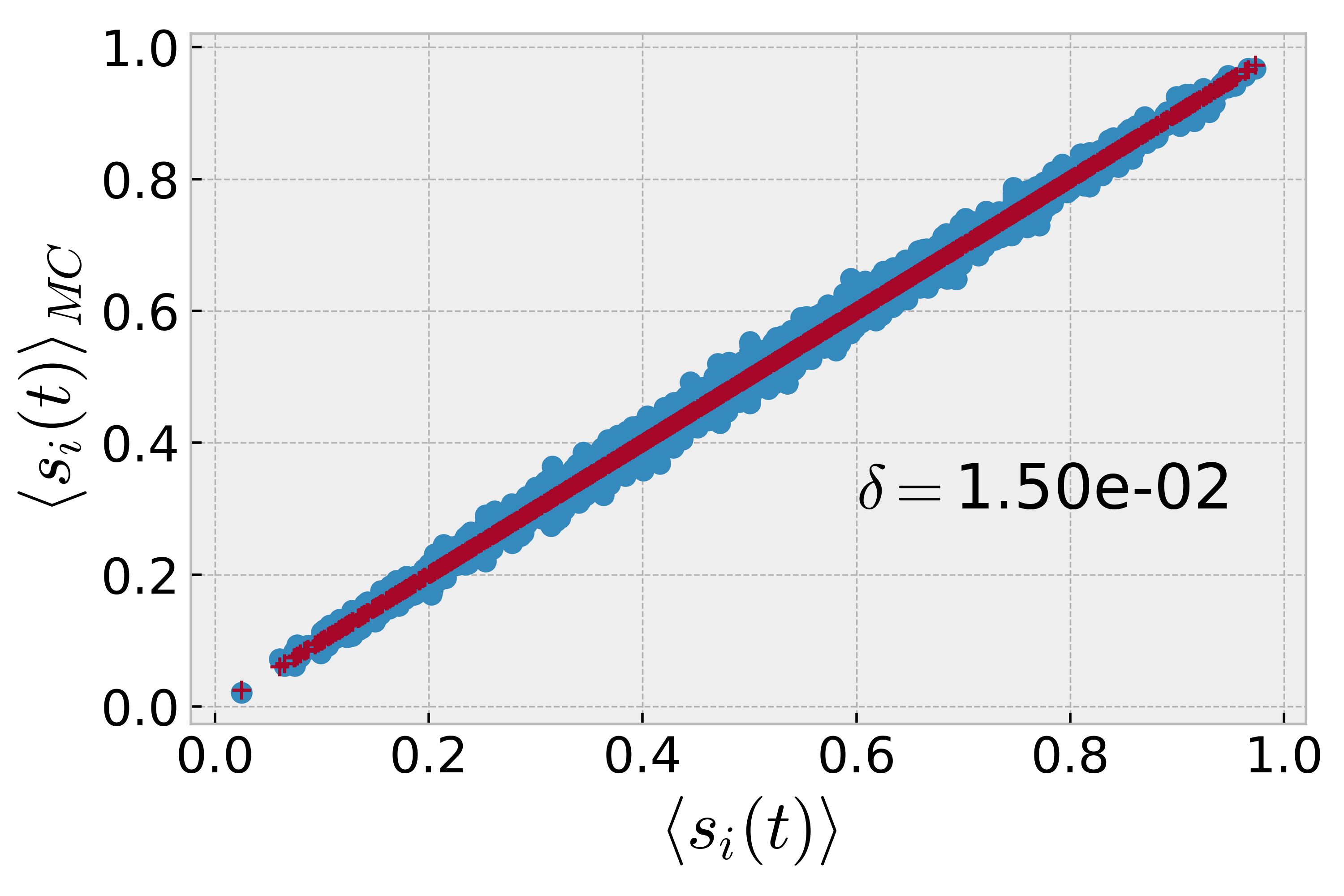} & 
  \includegraphics[width=0.33\textwidth]{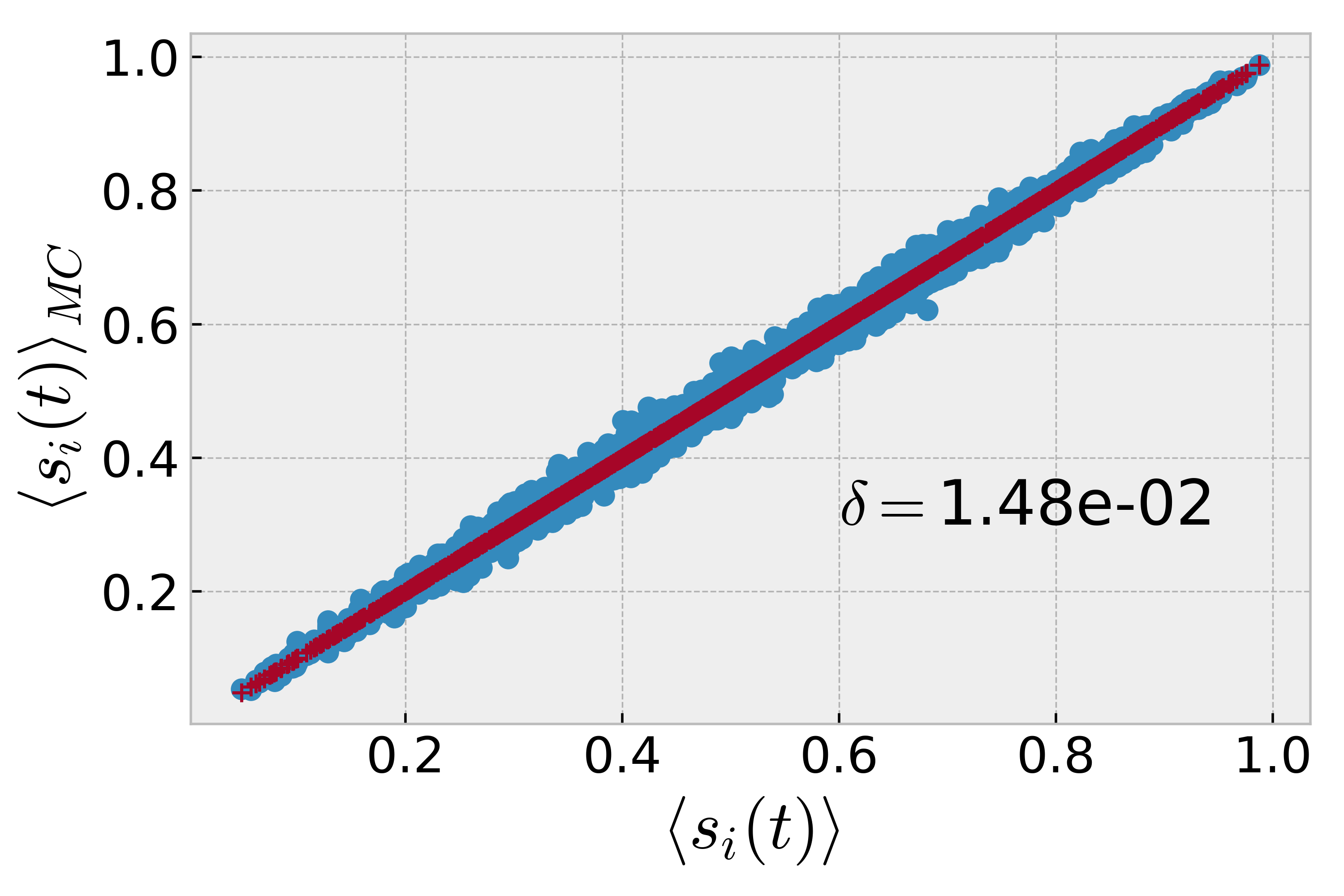} \\
  \text{a)} & \hspace{-35mm} \text{b)} & \hspace{-35mm} \text{c)} \\
 \hspace{-3mm} \includegraphics[width=0.33\textwidth]{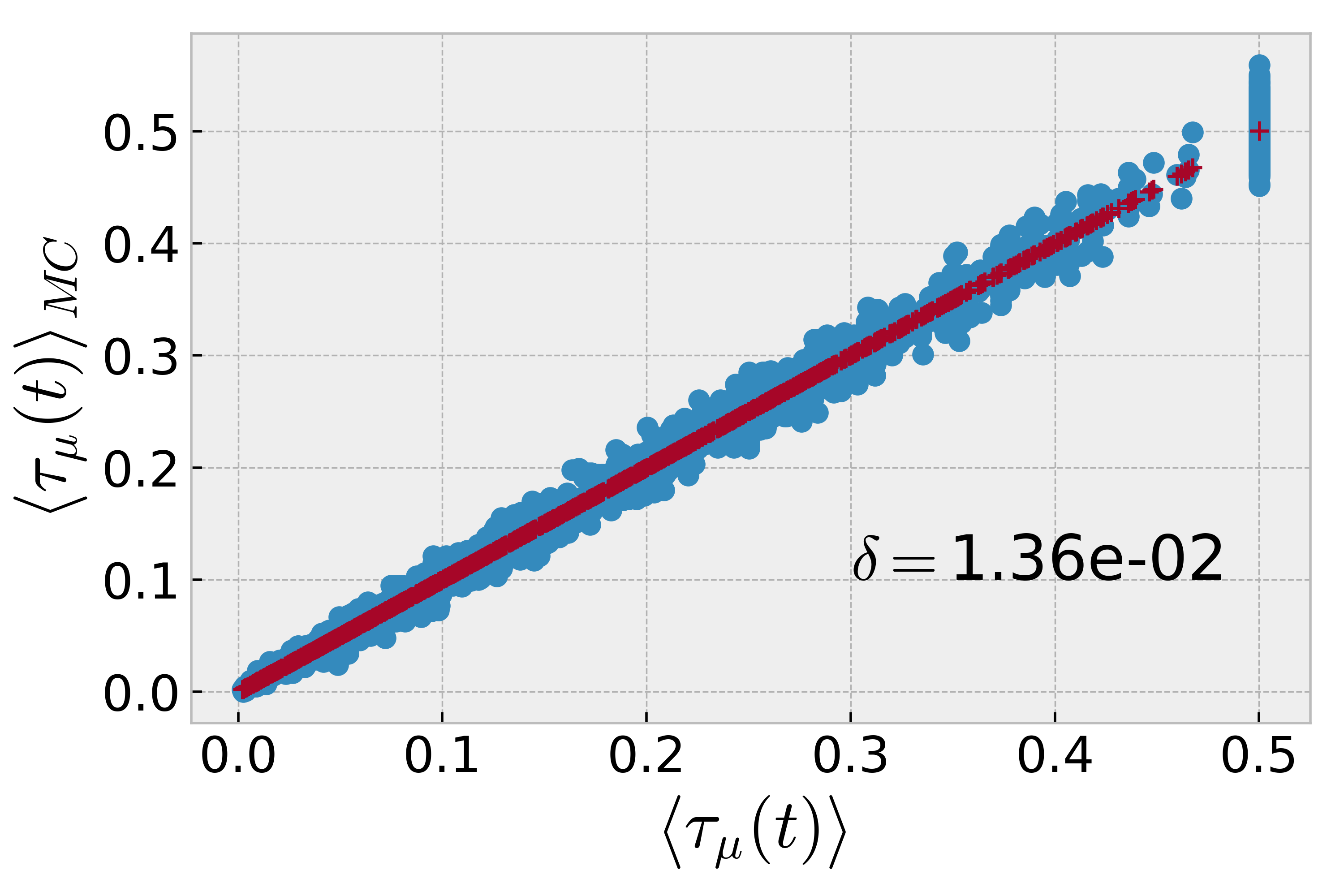} & \hspace{-3mm} \includegraphics[width=0.33\textwidth]{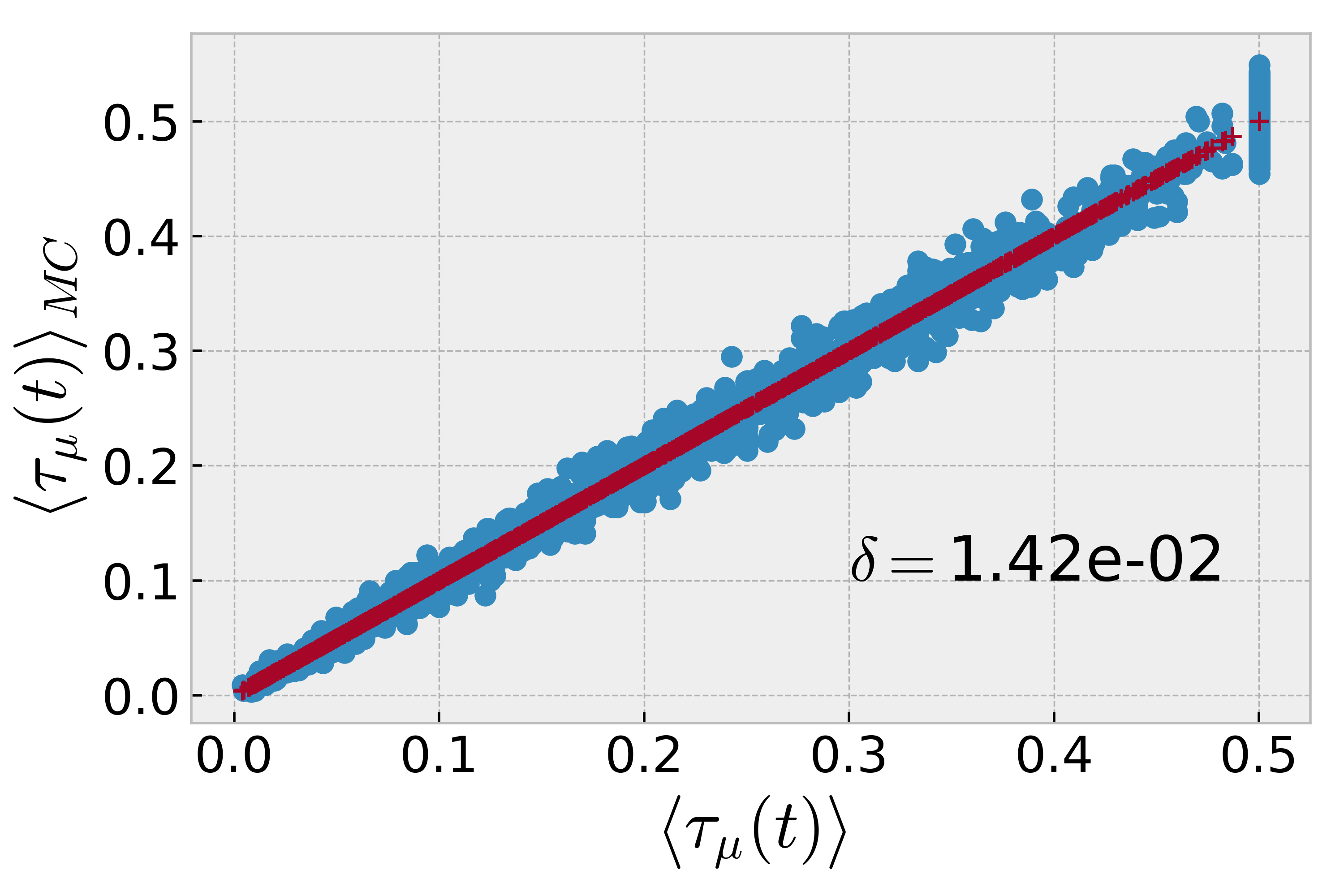}  & \hspace{-3mm}
  \includegraphics[width=0.33\textwidth]{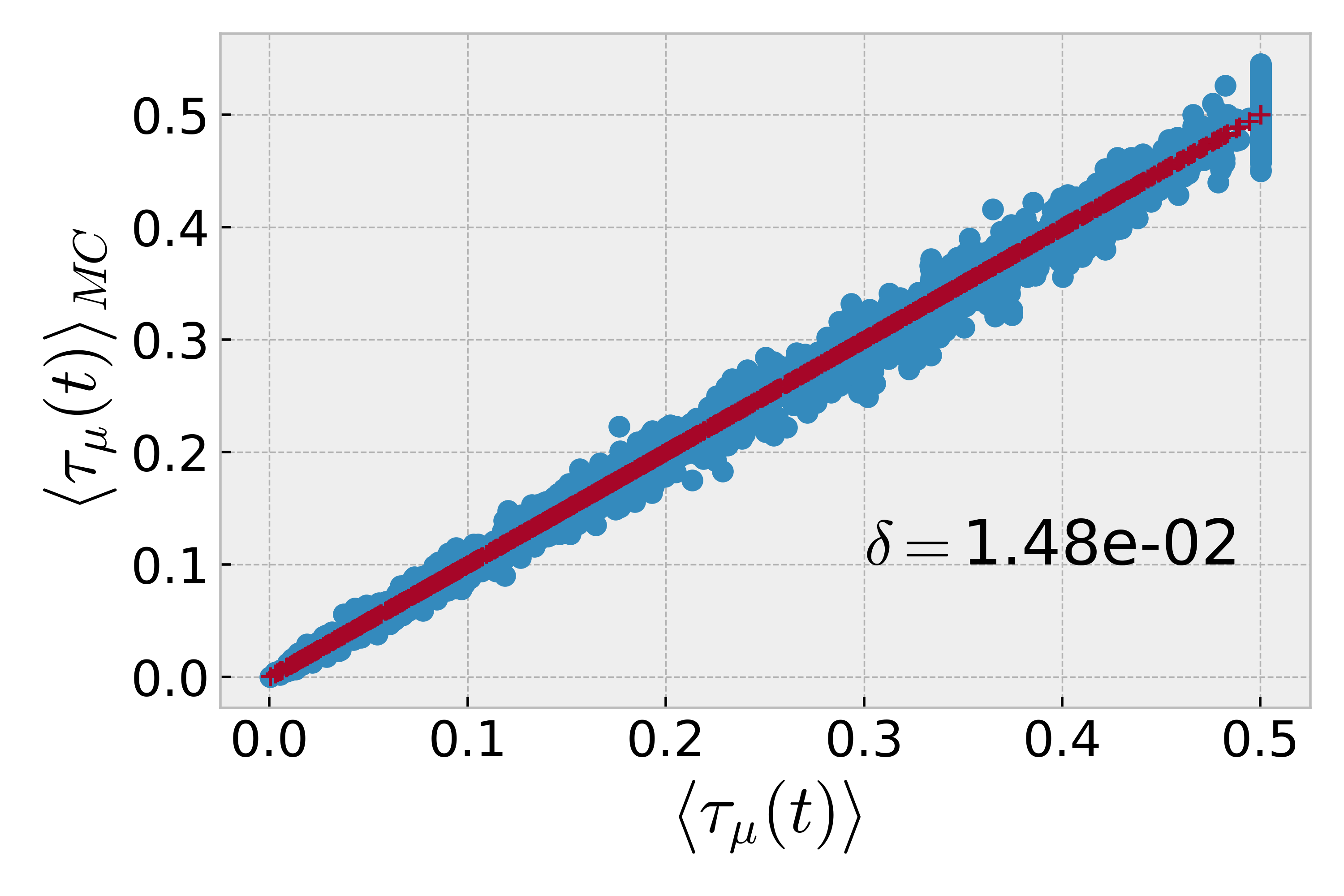} \\
  \vspace{-1.4mm}
   \text{d)} & \hspace{-35mm} \text{e)} & \hspace{-35mm} \text{f)} \\
\end{tabular}
  \caption{ Scatter plot of the activation probability of genes $\langle s_i(t) \rangle$ (a)-(c) and TFs $\langle \tau_\mu(t) \rangle$  (d)-(f) from dynamical cavity method and MC simulations. MC simulations are averaged over 1000 thermal histories. Here we show the system at time $t=15$ where the system has reached its steady-state. Parameters are: $N=2500$, $c=1$, $d=2$, $a=0$, $b=1$, $\beta=\hat{\beta}=10$. The root mean square error $\delta$ for each plot is annotated. Columns correspond to different densities of bidirectional links, from left to right $p=0,0.35,0.5$. }
  \label{fig: partial symmetry scatter plots}
\end{figure}

To assess the accuracy of the dynamical cavity method, we compute the time-dependent average activation probability of genes and TFs, given by $\langle s_i(t) \rangle = \sum_{s_i} s_i^{t} \pr_i(s_i^{t})$ and $\langle \tau_\mu(t) \rangle = \sum_{\tau_\mu} \tau_\mu^{t} \pr_\mu(\tau_\mu^{t})$, respectively. To compute this from MC simulations, we run many thermal realisations of a trajectory and take the average over these thermal realisations, 
\begin{align}
    \langle s_i(t) \rangle_{\text{MC}} &= \frac{1}{n} \sum_{\rho=1}^{n} s_i^{t,\rho} \\
    \langle \tau_\mu(t) \rangle_{\text{MC}} &= \frac{1}{n} \sum_{\rho=1}^{n} \tau_\mu^{t,\rho}
\end{align}
where $n$ is the number of thermal realisations of the trajectory we average over, and $s_i^{t,\rho},\tau_\mu^{t\rho}$ denote the states of gene $i$ and TF $\mu$ at time $t$ in the $\rho^{\text{th}}$ realisation of the trajectory, respectively. In figure \ref{fig: partial symmetry DC error finite temp s + tau } we show the average activation probability of genes and TFs against time, and show that there is good agreement between the dynamical cavity method and MC simulations at all times, for different densities of bidirectional links and relatively low temperature. Additionally, the stationary values of the activation probabilities of individual sites 
are plotted in 
figure \ref{fig: partial symmetry scatter plots} for the same temperature and 
and are found in good agreement with MC simulations. 

\begin{figure}[t]
\begin{tabular}{c @{\quad} c }
\hspace{-2mm}
  \includegraphics[width=0.49\textwidth]{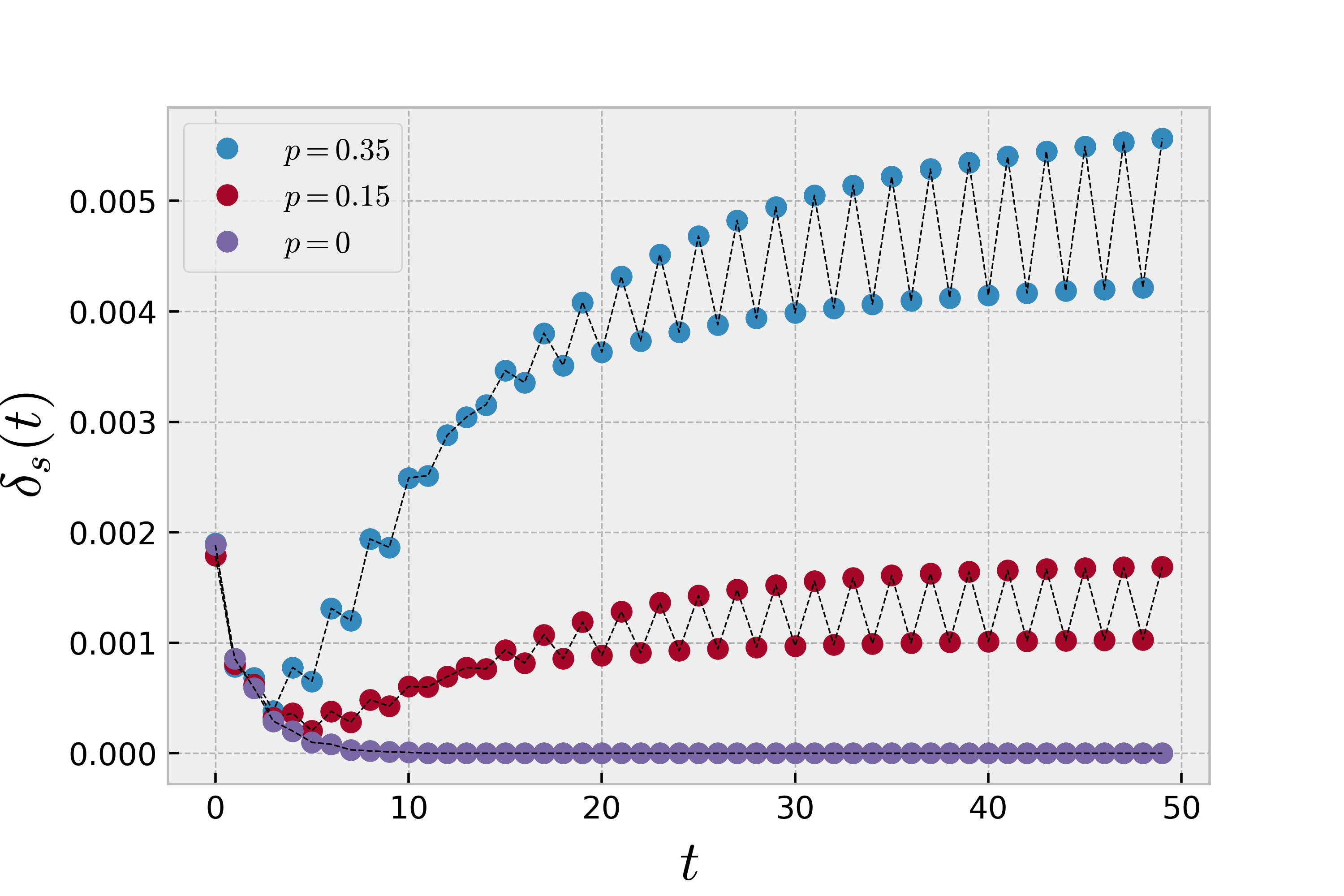} & \hspace{-3mm} 
  \includegraphics[width=0.49\textwidth]{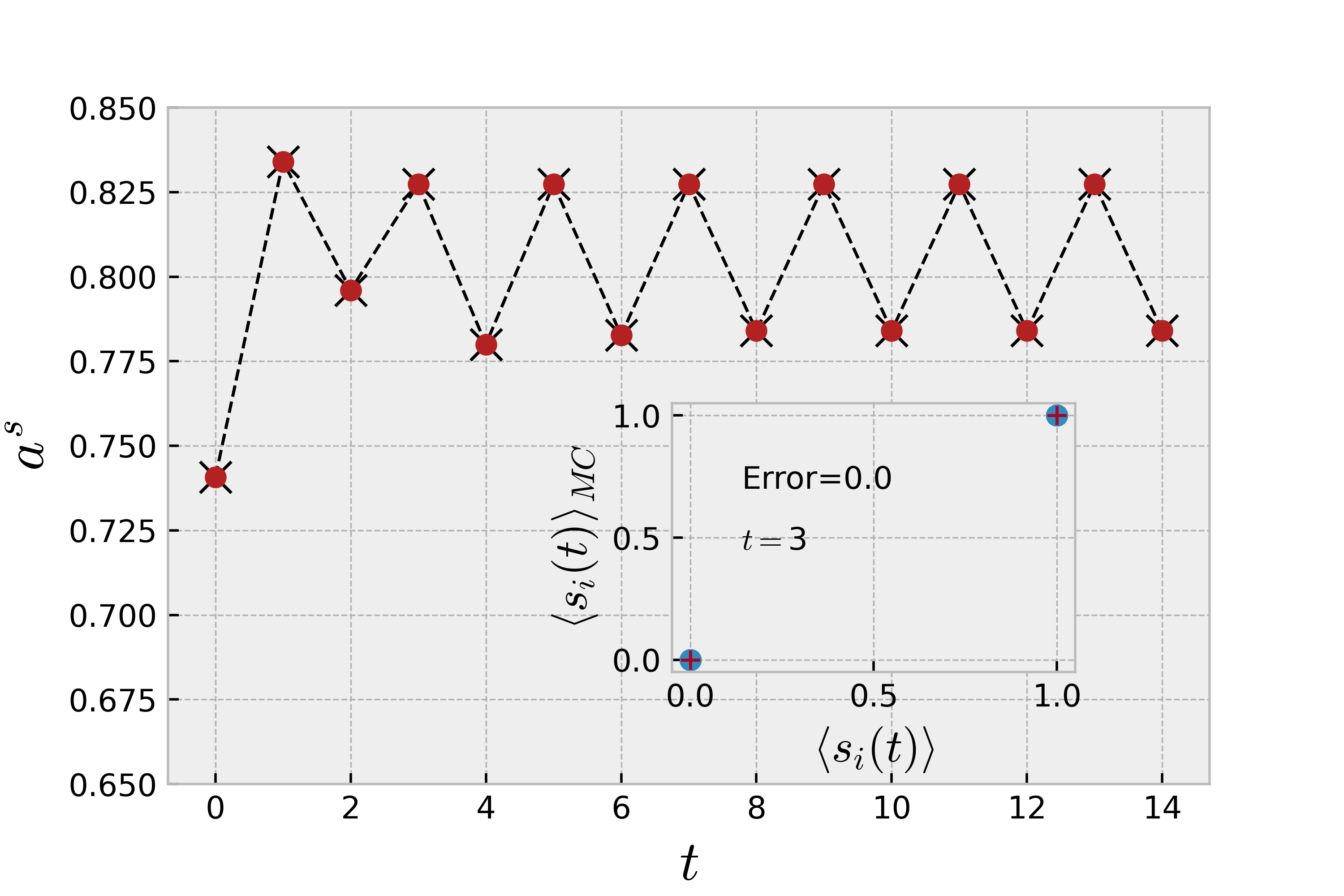}  \\
     \vspace{-3mm}
     \hspace{-60mm} \text{a)} & \hspace{-58mm} \text{b)} 
  \end{tabular}
  \caption{ (a) Mean square error $\delta_{s}(t)$ between MC simulations (averages over 100 runs) and dynamical cavity, for the activation probabilities of genes, against time. Network size is $N=2500$, connectivities are $c=1$, $d=2$, bias are $a=0$, $b=1$, and initial conditions are set to $P(s_i^{0}) =P(\tau_\mu^{0}) = 0.75$. Results shown for different densities of self-interactions, $p$, as indicated in the legend. (b) Average activation probability of genes $a^{s}(t) = N^{-1}\sum_i\langle s_i(t) \rangle$ against time, computed by dynamical cavity (crosses) and MC simulations (circles). The dynamical cavity method is initialised with $P_i(s_i^{0}) = \delta_{s_i^{0}, s_i^{MC}(0)}$ and $P_\mu(\tau_\mu^{0}) = \delta_{\tau_\mu^{0}, \tau_\mu^{MC}(0)}$. Network parameters are: $N=1500$, $c=1$, $d=3$, $a=-1.5$, $b=1$. Inset shows scatter plot of site activation probabilities of genes $\langle s_i(t) \rangle$, at time $t=3$, computed by dynamical cavity and MC simulations. Annotation indicates root mean square error. In (a) and (b) external fields are set to $\vartheta_i=\epsilon$ and $\vartheta_\mu = -c_\mu +\epsilon$, with $\epsilon=10^{-4}$, and the noise level is $\beta=\hat{\beta}=\infty$. }
  \label{fig: partial symmetry DC error zero temp }
\end{figure}

It is well known that the error in the predictions from the dynamical cavity method under the OTA scheme increases as the noise level is decreased. Previous work has demonstrated this looking at the error of macroscopic quantities in Ising spin models \cite{aurell2011message,aurell2012dynamic} whereas for Boolean systems with pairwise interactions, such as the linear threshold model \eqref{eq: LTMM update rule}, predictions within a OTA scheme were 
found to remain accurate for microscopic quantities, even at relatively low temperatures. 
Here, we assess the accuracy of the OTA scheme in predicting microscopic quantities in 
Boolean systems with multi-node interactions, 
like the non-linear threshold model defined in \eqref{eq: NLTM update rule}, and we find that the accuracy of OTA decreases when the temperature is lowered, similarly to 
what was observed in Ising spin models.  In figure \ref{fig: partial symmetry DC error zero temp }(a) we show the mean square error in activation probabilities, $\delta(t) = N^{-1}\sum_i \left(\left<s_i(t)\right> - \left<s_i(t)\right>_{\text{MC}}\right)^{2}$, at zero noise. For networks with an absence of bidirectional links, the OTA scheme is in perfect agreement with the MC simulations at long times, 
when the system has reached equilibrium.
The accuracy of the OTA scheme at zero noise, however, decreases as the density of bidirectional links increases. There are two potential sources of error. %
Firstly, the OTA scheme is based on two explicit assumptions, i.e. the Markovian factorization of cavity trajectories, (\ref{app eq: OTA assumption s}) and  (\ref{app eq: OTA assumption tau}), and the closure assumption, (\ref{eq: closure approx s}) and (\ref{eq: closure approx tau}), which may 
break down at lower temperature due to stronger memory effects. 
Secondly, the dynamical cavity method implicitly assumes that there is no ergodicity breaking. Practically, the assumption that the system is ergodic in the cavity approach means that when we provide the dynamical cavity method with the initial conditions $\pr_{0}(\bfs^{0}) = \prod_iP_{0}(s_i^{0}) $ and $\pr_{0}(\btau^{0})= \prod_\mu\pr_{0}(\tau_\mu^{0})$, it is implied that each configuration drawn from this distribution belongs to the same \textit{ergodic sector}, and 
will hence reach the same attractor, which is untrue when ergodicity is broken. To deduce whether the loss of accuracy at zero temperature is due to the assumptions made in the OTA scheme, or by ergodicity breaking, we provide the dynamical cavity equations with a \textit{specific configuration} $(\bfs^{0},\btau^{0})$ as the initial condition, and then run zero temperature dynamics. In figure \ref{fig: partial symmetry DC error zero temp }(b) we show that 
by initiaizing dynamical cavity in the same configuration (and thus in the same ergodic sector) as MC simulations, 
the OTA scheme predicts the trajectory without error. This suggests that the loss of accuracy observed in figure \ref{fig: partial symmetry DC error zero temp }(a) is due to ergodicity breaking.

\subsection{Multiple attractors induced by self-regulation}

\begin{figure}[t]
\begin{tabular}{c @{\quad} c } 
  \includegraphics[width=0.49\textwidth]{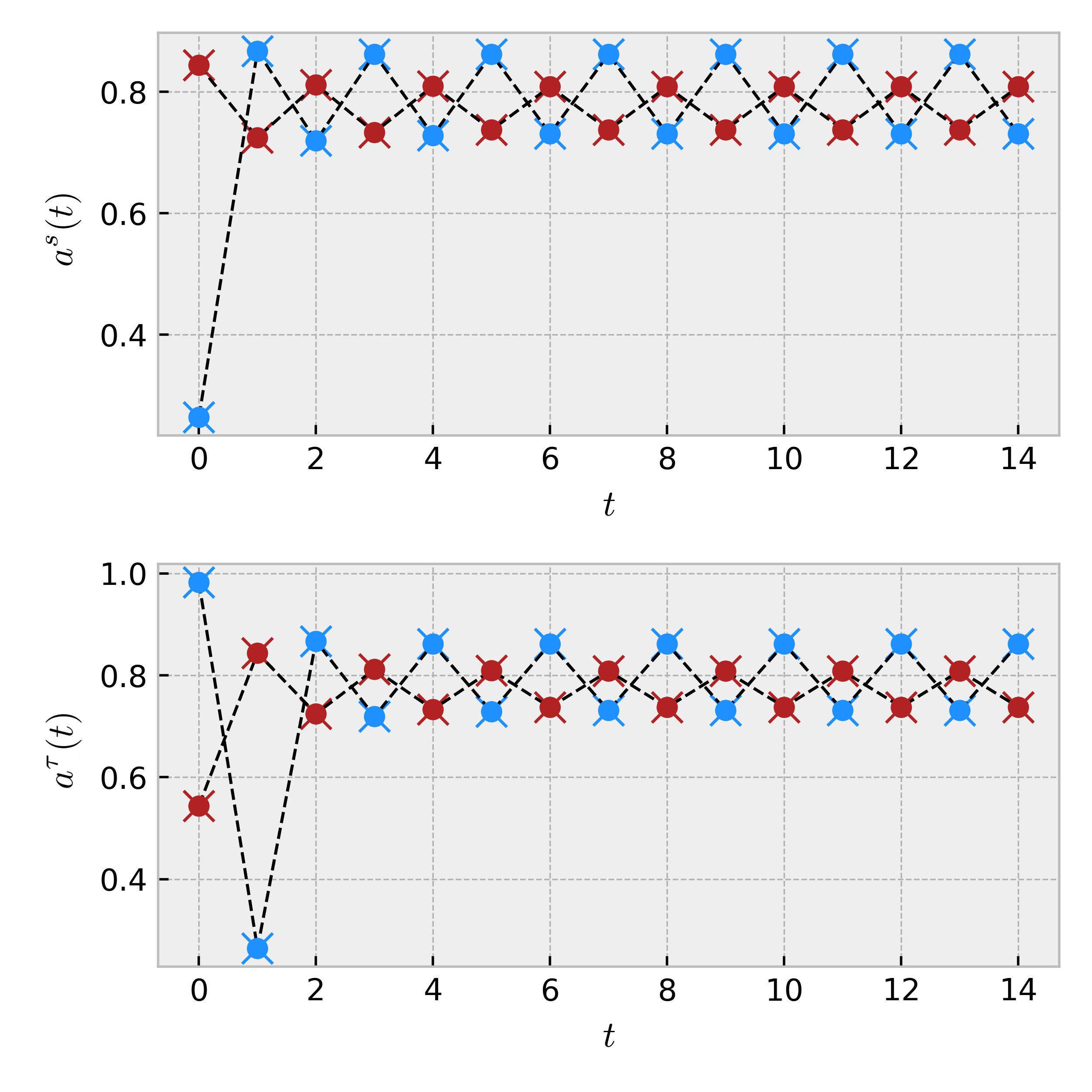}  & 
  \includegraphics[width=0.49\textwidth]{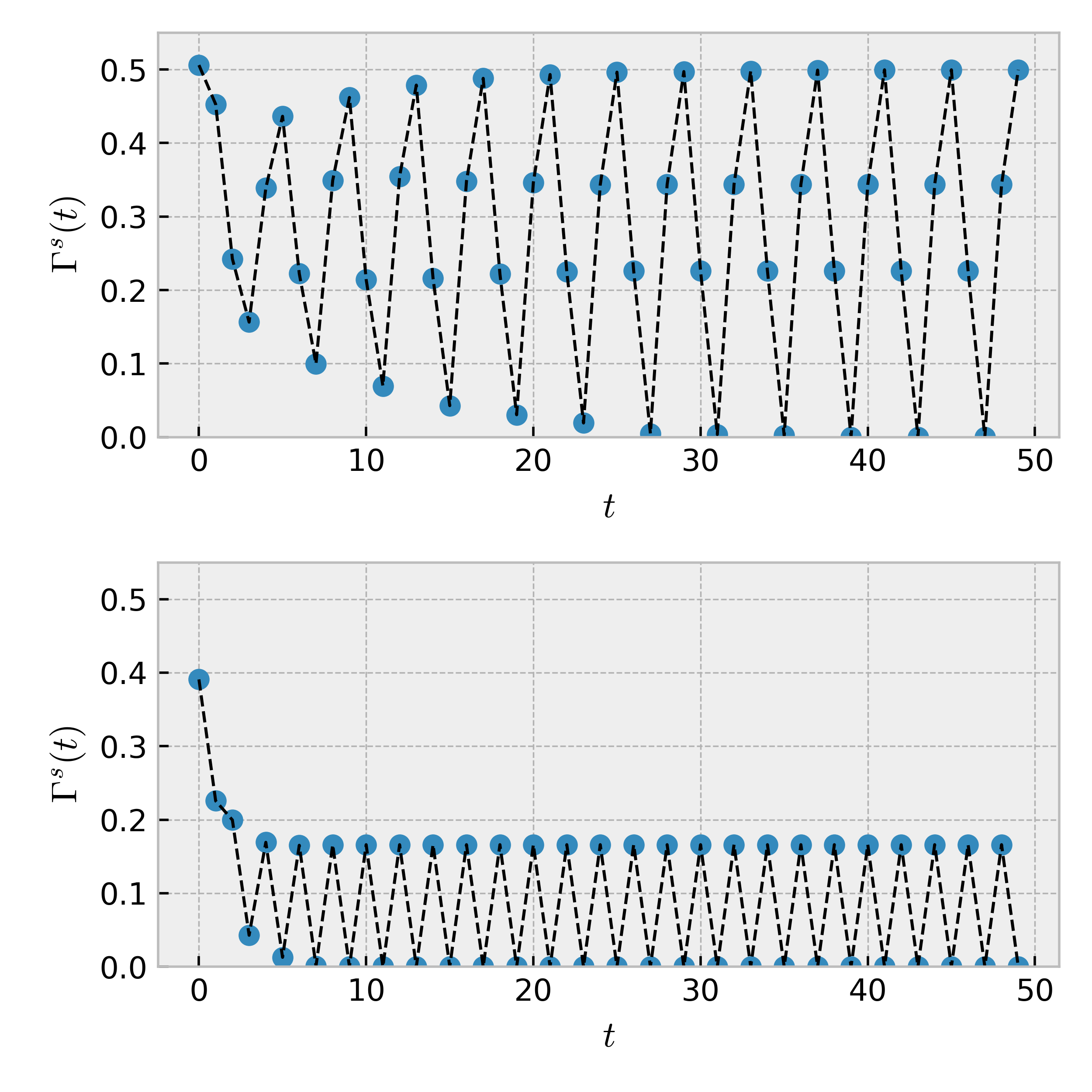} \vspace{-45mm} \\ 
  \hspace{-58mm} a) & \hspace{-60mm} b) \vspace{32mm} \\
  \vspace{-2mm}
  \hspace{-58mm} c) & \hspace{-60mm} d) 
 \end{tabular}
  \caption{ (a) and (c): Average activation of genes (a) and TFs (c) with system size $N=1500$, $c=1$, $d=3$, $a=-1.5$, $b=1$ (i.e. all self-interactions are positive) and $\beta=\hat{\beta}=\infty$. Results for two different initial conditions are shown. Circles indicate MC simulations, crosses indicate dynamical cavity. The dynamical cavity method is initialised with $P_i(s_i^{0}) = \delta_{s_i^{0}, s_i^{MC}(0)}$ and $P_\mu(\tau_\mu^{0}) = \delta_{\tau_\mu^{0}, \tau_\mu^{MC}(0)}$. (b) and (d): Self-overlap $\Gamma^{s}(t,T)$ against time $t$ for genes. Results are computed from MC simulations with $N=2500$, $c=1$, $d=15$, $a=0$, for cases where (b) all self-interactions are inhibitory (i.e. $b=-1$)
or (d) excitatory (i.e. $b=1$). Similar plot can be obtained for the self-overlaps of the TFs, $\Gamma^{\tau}(t,T)$ but are not shown here. }
  \label{fig: multiple attractors }
\end{figure}

The evidence of ergodicity breaking suggests that our system supports multiple attractors. To confirm this we show in figures \ref{fig: multiple attractors }(a) and \ref{fig: multiple attractors }(c) the dynamics of genes and TFs starting from two different initial conditions for the same network. In the network we consider, a TF has an excitatory effect on every gene that is required for its own synthesis, $p=1$ and $b=1$, such that this network has a high density of bidirectional links. We again initialise the dynamical cavity method with the same initial configuration as the MC simulations, such that there is zero error in the dynamical cavity predictions. We see that the dynamics of the same network, starting from two different initial configurations, converges to two different 2-cycles, which shows that, at least at zero temperature, our model supports multiple attractors.

To better assess the type of attractors these systems exhibit, we look at the self-overlap, which we define as,
\begin{align}
    \Gamma^{s}(t,t') &= \frac{1}{N}\sum_i (s_i^{t} - s_i^{t'})^{2} \\
    \Gamma^{\tau}(t,t') &= \frac{1}{N}\sum_\mu (\tau_\mu^{t} - \tau_\mu^{t'})^{2}.
\end{align}
By running MC simulations at zero temperature for some long time $t_{w}$, and computing the self-overlap $\Gamma^{s}(t,t_{w})$, with all other times $t\leq t_w$,  we find, as shown in figures \ref{fig: multiple attractors }(b) and \ref{fig: multiple attractors }(d), that for systems with all positive self-regulatory effects the system will exhibits a 2-cycle, but the same network where the self-regulatory effects are negative will exhibit a 4-cycle. We find these results consistent for different realisations of the network, and for networks with different values of TF in-degree, $d+pc$. This suggests that the type of attractor that the dynamics converges to is determined by the type of self-regulatory effects present, alone.

\begin{figure}[t]
\begin{tabular}{lccccc}
  \includegraphics[width=0.33\textwidth]{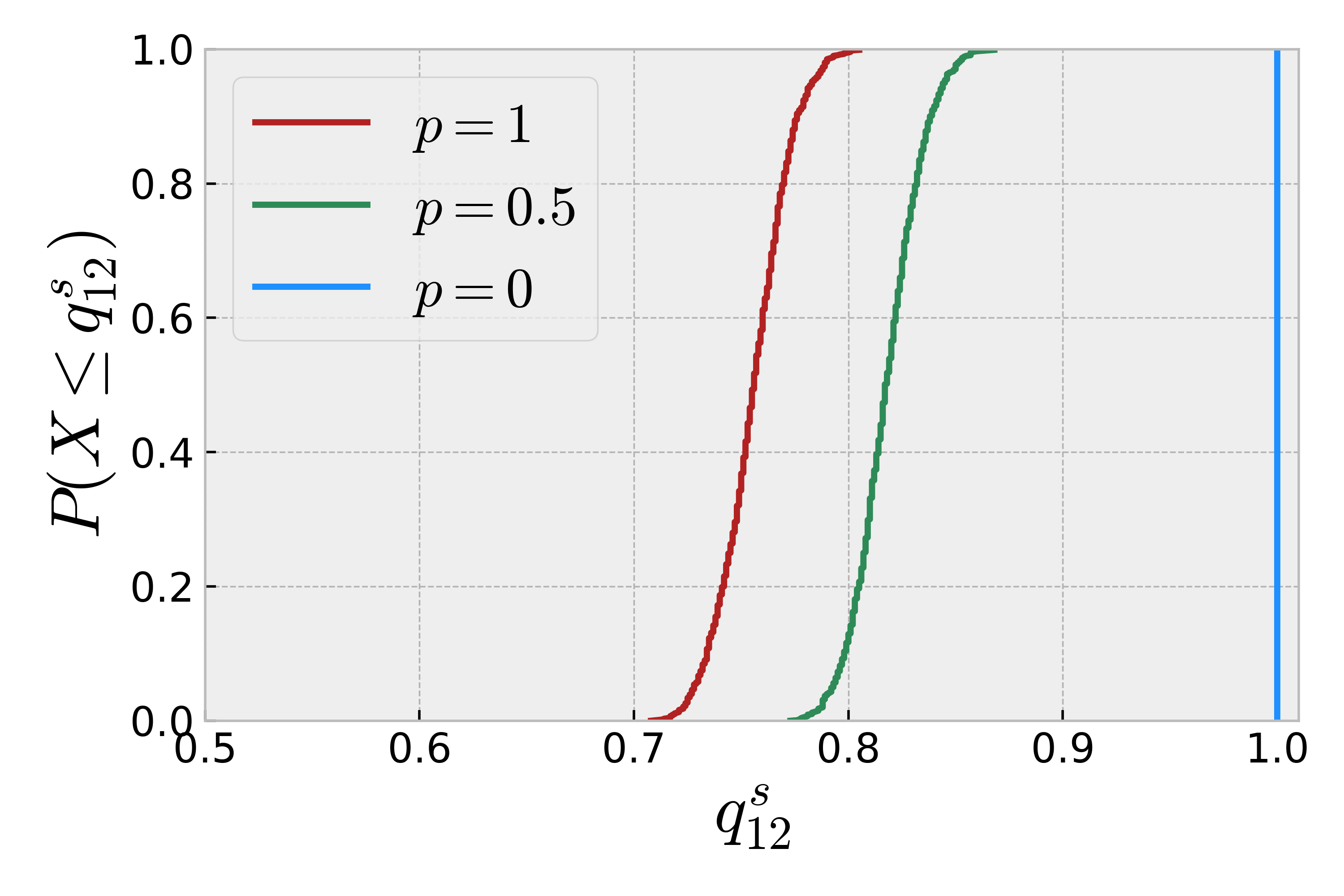}&
  \includegraphics[width=0.33\textwidth]{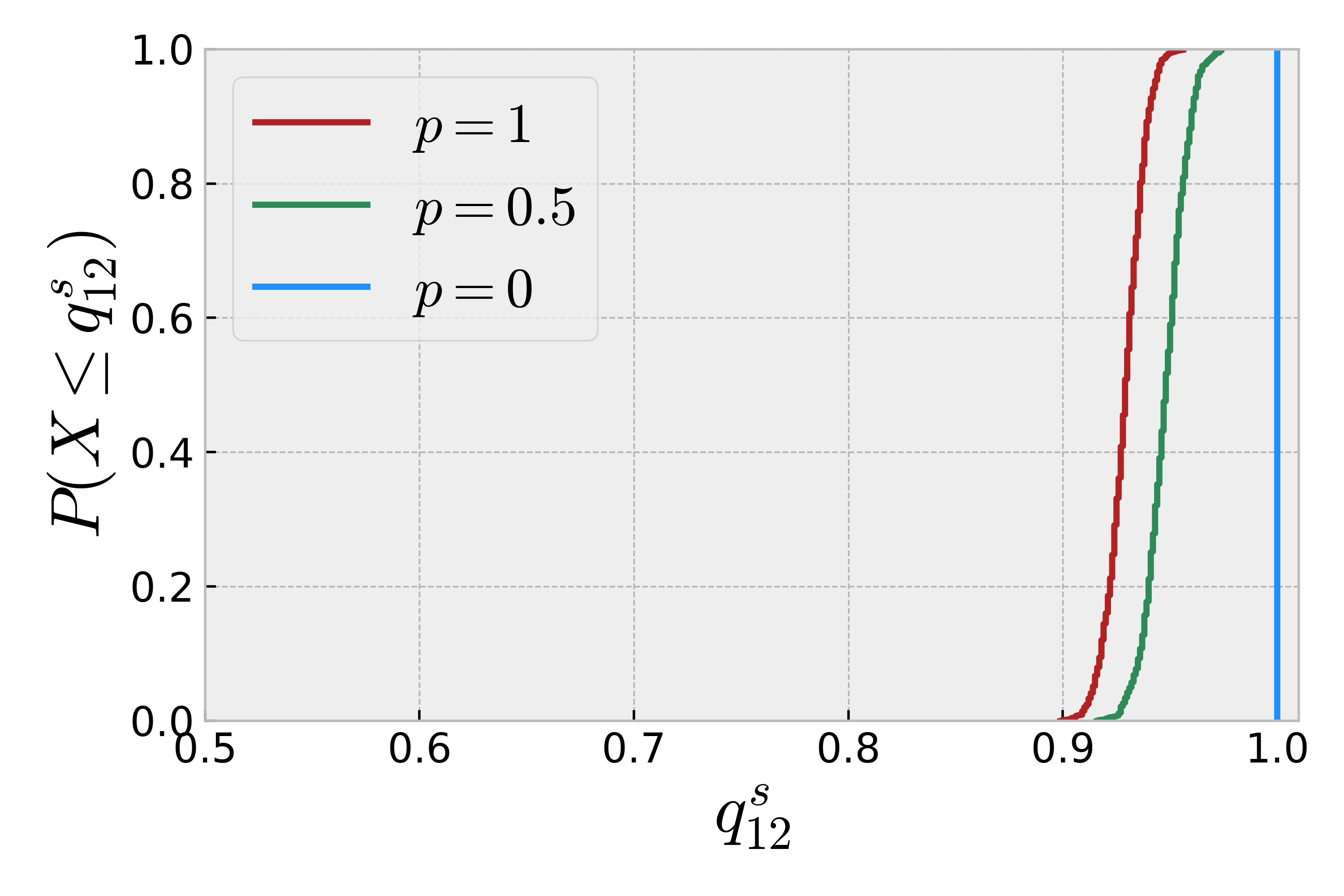}&
  \includegraphics[width=0.33\textwidth]{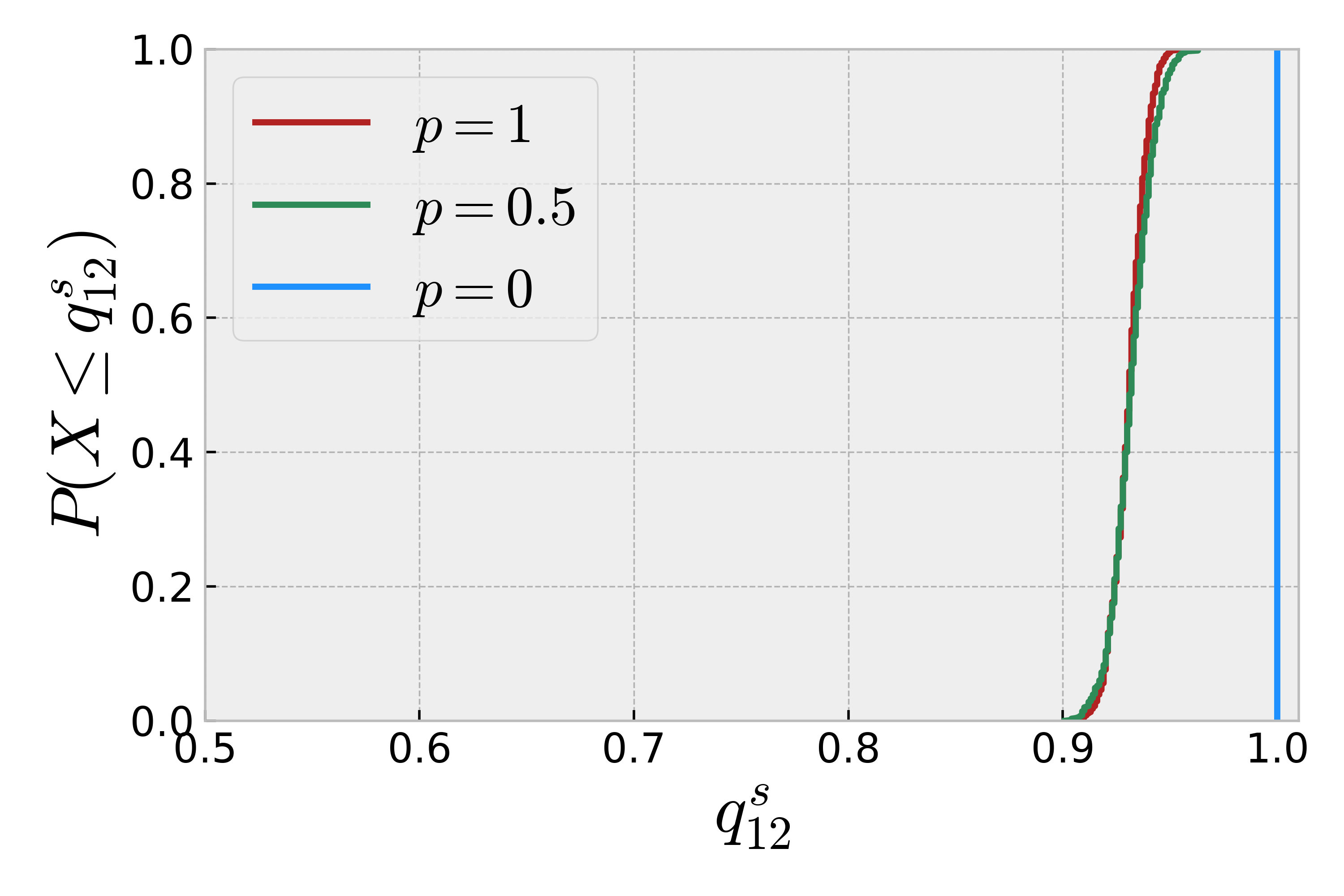} \vspace{-3mm} \\
    \text{a)} & \hspace{-35mm} \text{b)} & \hspace{-35mm} \text{c)} \\
    \includegraphics[width=0.33\textwidth]{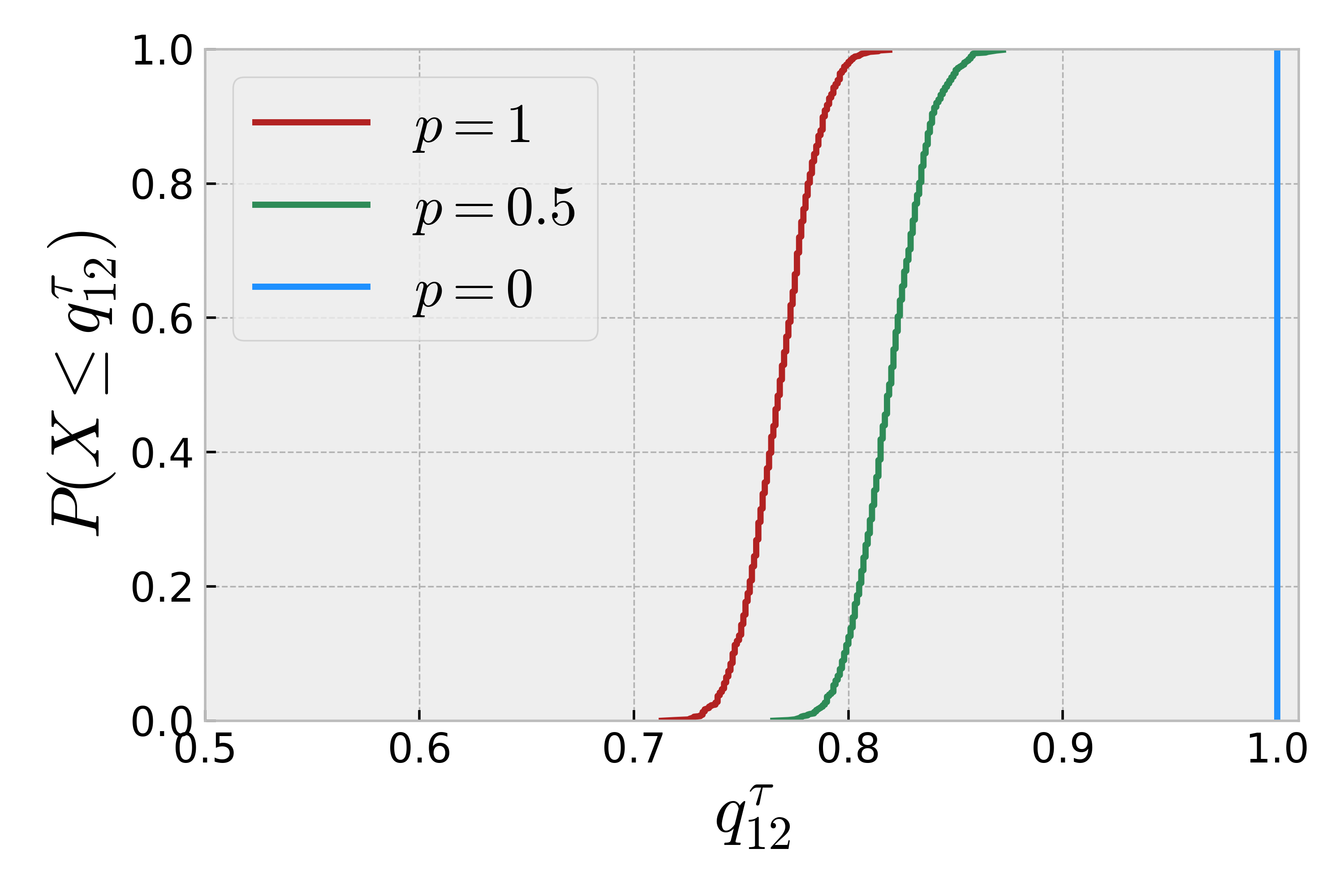} &
    \includegraphics[width=0.33\textwidth]{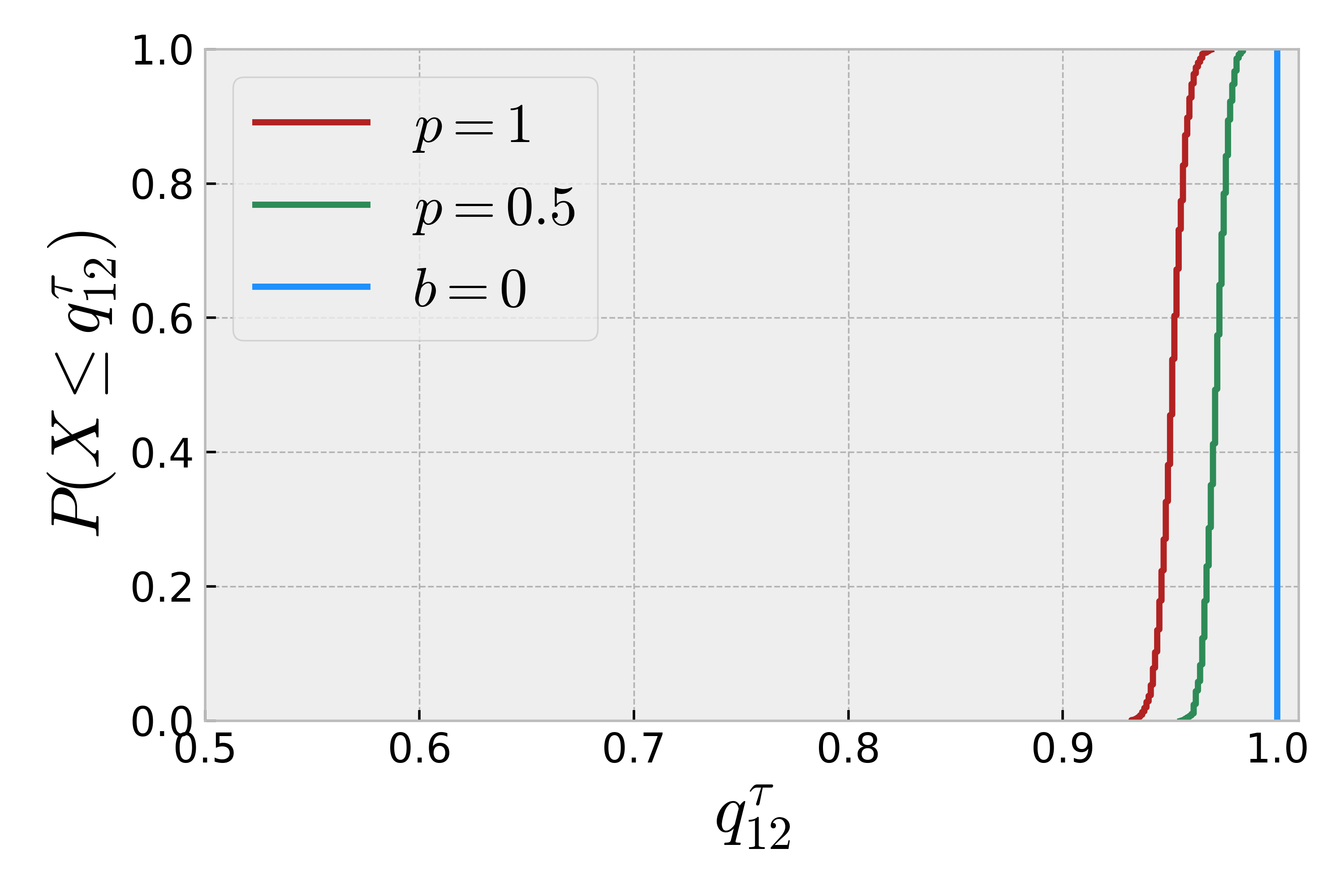} &
  \includegraphics[width=0.33\textwidth]{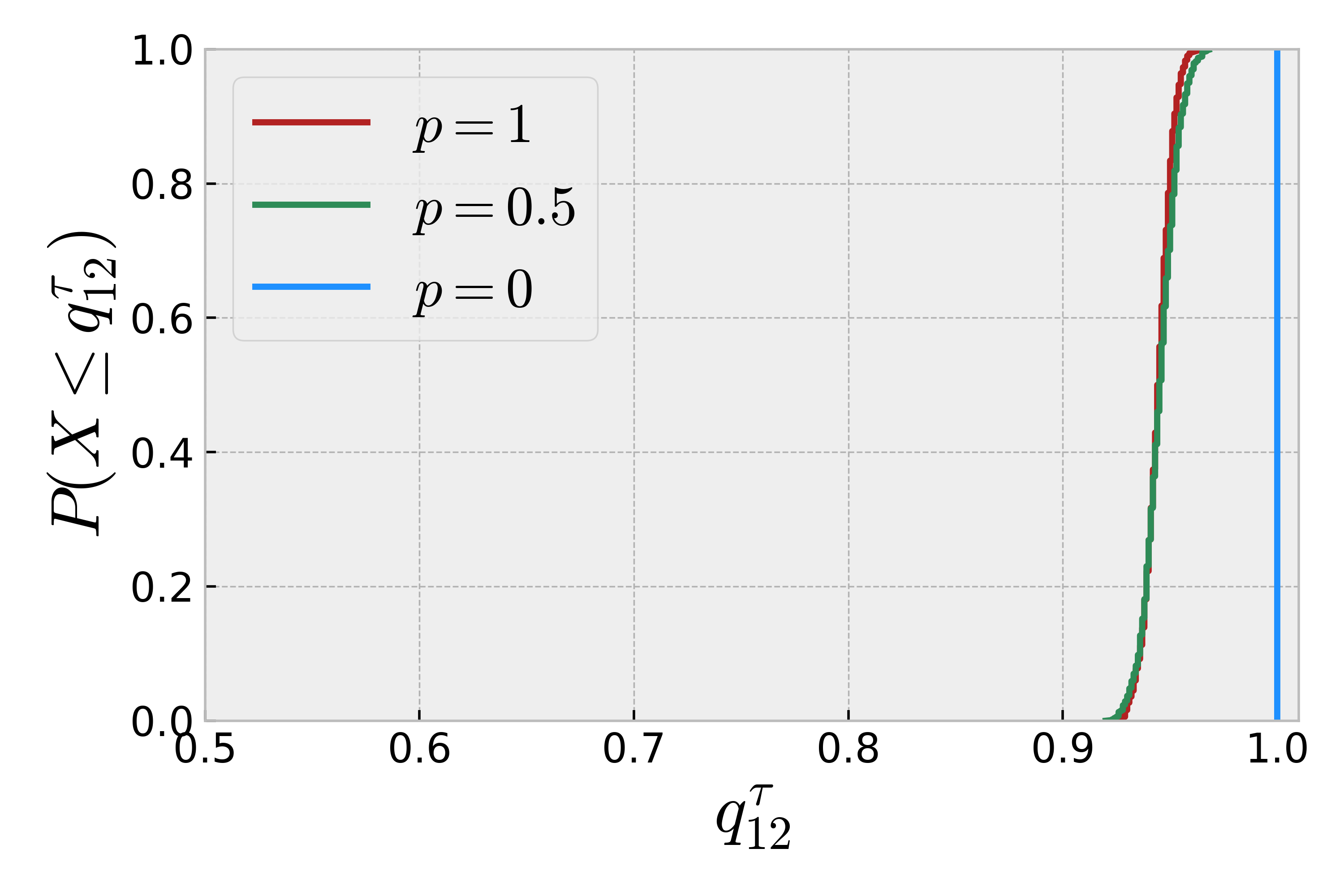} \vspace{-3mm}\\
  \vspace{-2mm}
    \text{d)} & \hspace{-35mm} \text{e)} & \hspace{-35mm} \text{f)}
\end{tabular}
  \caption{ Distribution of overlap of attractors starting from two different initial conditions for genes  $\pr(q_{12}^{s})$ (a)-(c) and TFs $\pr(q_{12}^{\tau})$ (d)-(f). Network parameters are: $N \!=\! 1000$, $c \!=\!1$, $d \!=\!3$, $a \!=\!0$, $b \!=\!1$. Distributions are formed by comparing the attractors reached by 1000  random pairs of initial conditions. Results shown for different levels of bidirectional links indicated by legend. Noise level is $\beta\!=\!\hat{\beta}\!=\!\infty$. Columns, from left to right, indicate cases where $\vartheta_{\mu}\! =\! - c_{\mu} \!+\! \epsilon$,  $\vartheta_{\mu}\! =\! -1 \!-\! \epsilon$ and $\vartheta_{\mu} \! = \! - \epsilon$ with $\epsilon \!=\! 10^{-4}$. We also have $\vartheta_{i} \!=\! \epsilon$.}
  \label{fig: overlap dist s + tau }
\end{figure}

To elucidate the effect that the density of bidirectional links has on the attractors of our model, we study a dynamical variant of the overlap distribution. We exploit the fact that we are at zero temperature to run MC simulations until the system reaches its attractor, store them, and compare the attractors that are reached from different initial conditions. At zero temperature, the system is expected to reach a limit cycle, and so we compute the overlap between each point of the limit-cycles reached from two random initial conditions. The overlap for two replicas, $\bfs^{\rho}$ and $\bfs^{\nu}$, of the system starting from different random initial conditions is defined as
\begin{align}
    q^{s}_{\rho \nu} = \frac{1}{N}\sum_i \sum_{n=1}^{p}\left(2s_i^{n,\rho} -1\right)\left(2 s_i^{n,\nu}-1\right)
\end{align}
where $s_i^{n,\rho}$ is the state of site $i$ at the $n^{\text{th}}$ point in the limit cycle in replica $\rho$. We can similarly define the overlap for the TFs, 
\begin{align}
    q^{\tau}_{\rho \nu} = \frac{1}{N}\sum_\mu \sum_{n=1}^{p}\left(2\tau_\mu^{n,\rho} -1\right)\left(2 \tau_\mu^{n,\nu}-1\right)
\end{align}
The overlap takes values in $q\in[-1,1]$ such that $q=1$ if the configurations are identical, and $q=-1$ if they are oppositely aligned i.e $s_i^{\rho}=1-s_i^{\nu}~\forall~i$. The above definition of the overlap is, however, dependent on the ordering of the points in the cycle. For example, it may be that two initial conditions lead to the same attractor, but enter the attractor at different points in the cycle, and so lead to an overlap $q \neq 1$. For this reason we compute the overlap between the attractors for different permutations of the points in the cycle and take the \textit{maximum} value to be the true value of the overlap. By drawing many pairs of initial conditions and computing the overlap between the attractors according the above, we compute the overlap distribution. Figure \ref{fig: overlap dist s + tau } shows that as the density of bidirectional links decreases, the width of the overlap distribution decreases and moves towards an overlap of $q^{s}_{12},q^{\tau}_{12}=1$, suggesting that the attractors become more similar as bidirectional links are removed, and that systems without bidirectional links will have just one attractor. Additionally, the external field acting on the TFs is shown to have a significant effect. When $\vartheta_\mu = - c_\mu + \epsilon$, such that TFs operate with AND logic, we see the existence of multiple attractors, with relatively low overlap (figures \ref{fig: overlap dist s + tau }(a) and \ref{fig: overlap dist s + tau }(d) ). Conversely, for systems with external field $\vartheta_\mu = -\epsilon$, such that TFs operate with OR logic, we see that while these networks do support multiple attractors, the overlap is high, hence the attractors are very similar (figures \ref{fig: overlap dist s + tau }(c) and \ref{fig: overlap dist s + tau }(f) ).
Interpolating between these extreme cases, we set $\vartheta_\mu = -1 - \epsilon$, to add a little cooperativity to the OR logic, such that a TF requires at least two neighbouring genes to be active in order to be synthesised. In this case, we also see high overlap, even for systems with high densities of bidirectional links (figures \ref{fig: overlap dist s + tau }(b) and \ref{fig: overlap dist s + tau }(e) ).

Our work demonstrates the existence of multiple attractors in sparse, partially bidirectional networks. Previous analytical work has focused on the fully connected and fully asymmetric cases \cite{sompolinsky1988chaos,rieger1989glauber,ma1992suppression}. More recently, numerical studies in small sparse networks showed that the number of attractors decreases to one as the dilution of the network is increased \cite{folli2018effect}. 
Our results show that the presence of bidirectional links is crucial to have a multiplicity of attractors in sparse networks and that 
increasing the fraction of bidirectional interactions decreases the overlap between the attractors of the system. To the best of our knowledge, 
our work is the first to demonstrate that the existence of multiple attractors in sparse networks is conditional on the interactions being at least partially bidirectional.

Additionally, we have seen that the overlap is lower in the nonlinear model, where TFs evolve according to AND logic, than in the linear model, where TFs evolve according to OR logic, suggesting that cooperativity 
in gene regulation promotes diversification of attractors.
Our nonlinear model can be regarded as the Boolean 
equivalent of a mixed p-spin model, where the state of each site depends on a varying number of sites. The regular p-spin model has been shown in equilibrium to exhibit a replica symmetry breaking phase at low noise levels, where ergodicity is broken and multiple attractors are to be expected \cite{franz2001exact,ricci2001simplest,mezard2002alternative}. Our numerical results show that ergodicity is also broken in our mixed p-Boolean model with asymmetric interactions, 
at low levels of noise, where the system exhibits a multiplicity of {\it cyclic} attractors. Interestingly, the overlap distribution is reminiscent of the one observed in equilibrium spin models exhibiting one step of replica symmetry breaking. It can be shown that the width of the distributions in figure \ref{fig: overlap dist s + tau } decreases with system size as $N^{-\frac{1}{2}}$, suggesting that this is a finite size effect and that for networks of infinite size these distributions are delta-peaked at a single value.

\begin{figure}[t]
\begin{tabular}{c @{\quad} c }
  \includegraphics[width=0.49\textwidth]{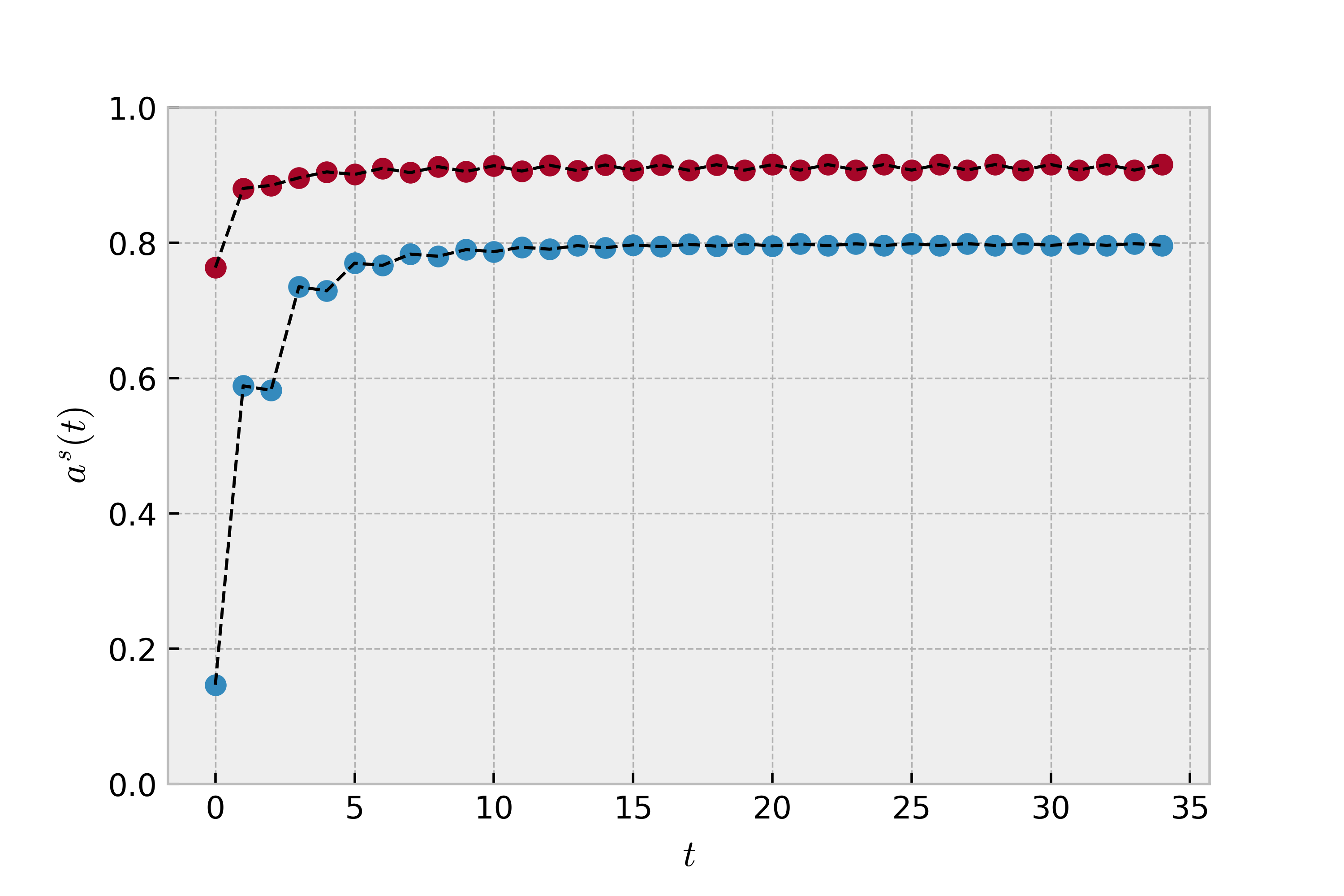} & 
  \includegraphics[width=0.49\textwidth]{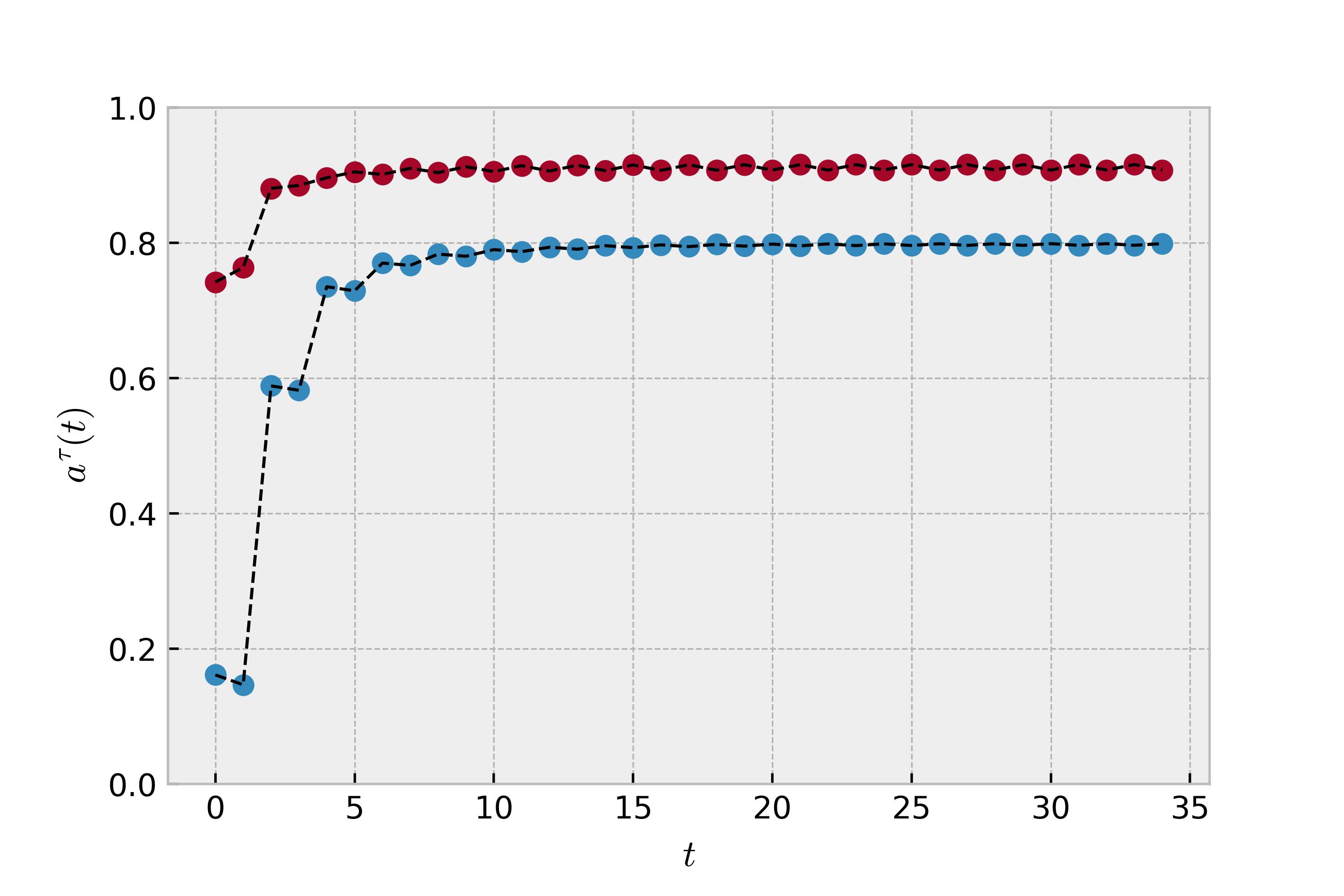} \\
\vspace{-3mm}
\hspace{-60mm} \text{a)} & \hspace{-58mm} \text{b)} 
  \end{tabular}
  \caption{Average activation of (a) genes $a^{s}(t) = N^{-1}\sum_i \langle s_i (t) \rangle $ and (b) TFs $a^{\tau}(t) = (\alpha N)^{-1}\sum_\mu \langle \tau_\mu (t) \rangle $ against time. Symbols indicate result of MC simulations averaged over 100 thermal histories. Network parameters are: $N=1500$, $c=1$, $d=10$, $p=1$, $a=0$ and $b=1$. Noise level for TFs $\hat{\beta}=\infty$ and for genes $\beta=100$. }
  \label{fig: multi attractors finite temp }
\end{figure}

While we have shown the existence of multiple attractors in the absence of noise, we expect that as noise is increased, the system will become ergodic, and the dynamics will converge to a single attractor. We can, however, show in figure \ref{fig: multi attractors finite temp } that if TFs evolve without noise, but genes evolve with low but finite noise, the system still exhibits multiple attractors. 
Here, we have used a network with higher degree than in  figures \ref{fig: multiple attractors }(a) and \ref{fig: multiple attractors }(c)
and see that there is a bigger difference in the average activation, suggesting that as the connectivity of the network is increased, the attractors become more dissimilar. 
Finally, we observe that any low but finite noise in the TFs dynamics leads to a single attractor, even when gene dynamics is noiseless.

\section{OTA in the thermodynamic limit} \label{sec: thermodynamic limit }

In earlier sections we have derived expressions for activation probabilities of individual nodes in single graph instances. We can use these results to obtain 
expressions for typical (or average) activation probabilities by averaging over sites. For large networks, these expressions will depend on the distribution $P(\bfJ)$  (or $P(\bfeta,\bxi)$) of the disorder and not on its realization $\bfJ$ (or $\bfeta, \bxi$), 
hence they give 
information on typical trajectories in the graph ensemble. Similar expressions can also be derived via generating functional analysis (GFA)
\cite{hatchett2004parallel,mimura2009parallel}, however, no equivalent
of the OTA scheme has been formulated within GFA, 
hence the resulting equations exhibit the aforementioned exponential time complexity. Taking averages of the dynamical cavity equations after application of the OTA scheme allows us to obtain equations for typical activation probabilities which do not exhibit such complexity and can therefore be solved explicitly at short and long times.

\begin{figure}[t]
\begin{tabular}{lccccc }
    \includegraphics[width=0.33\textwidth]{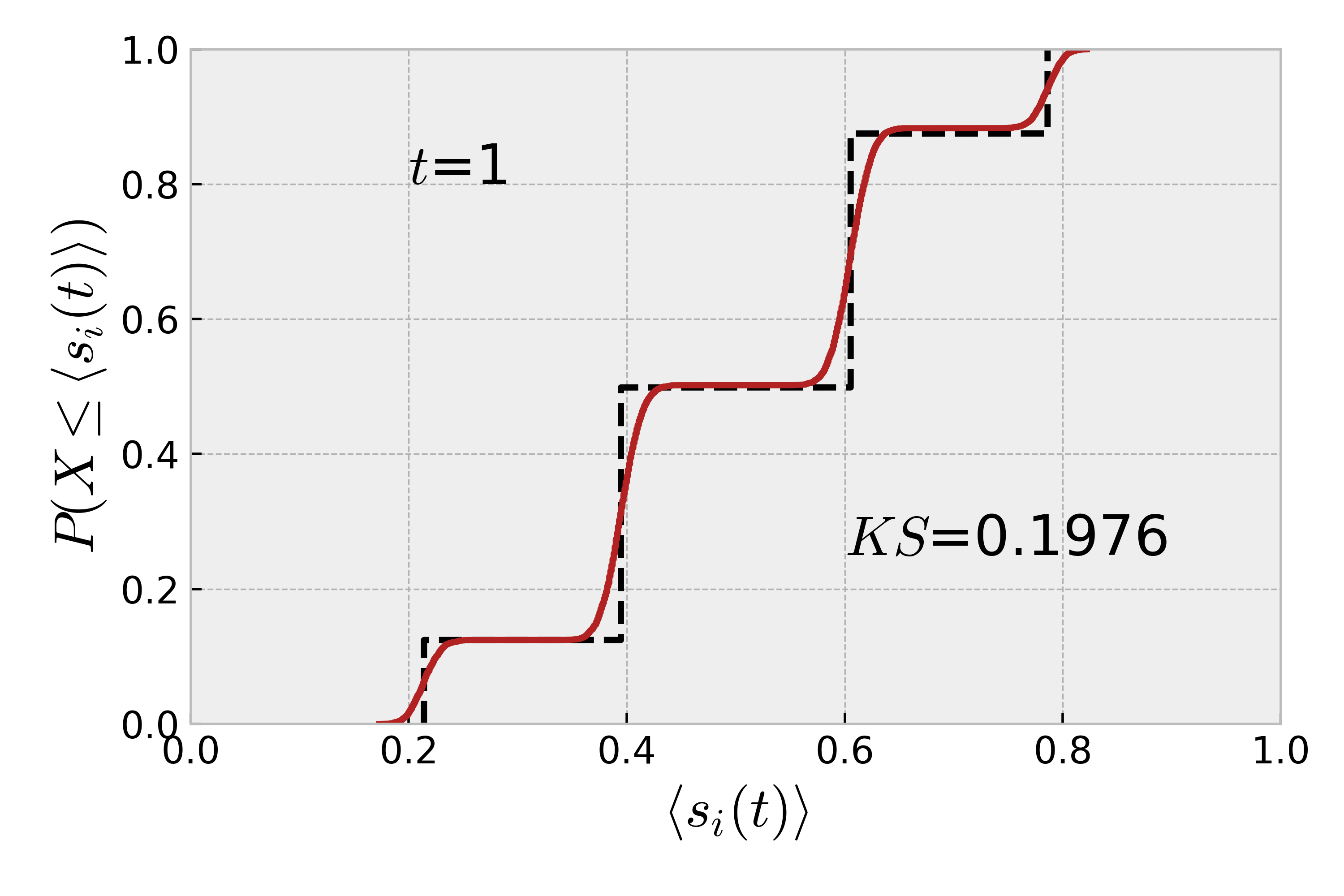} &
    \includegraphics[width=0.33\textwidth]{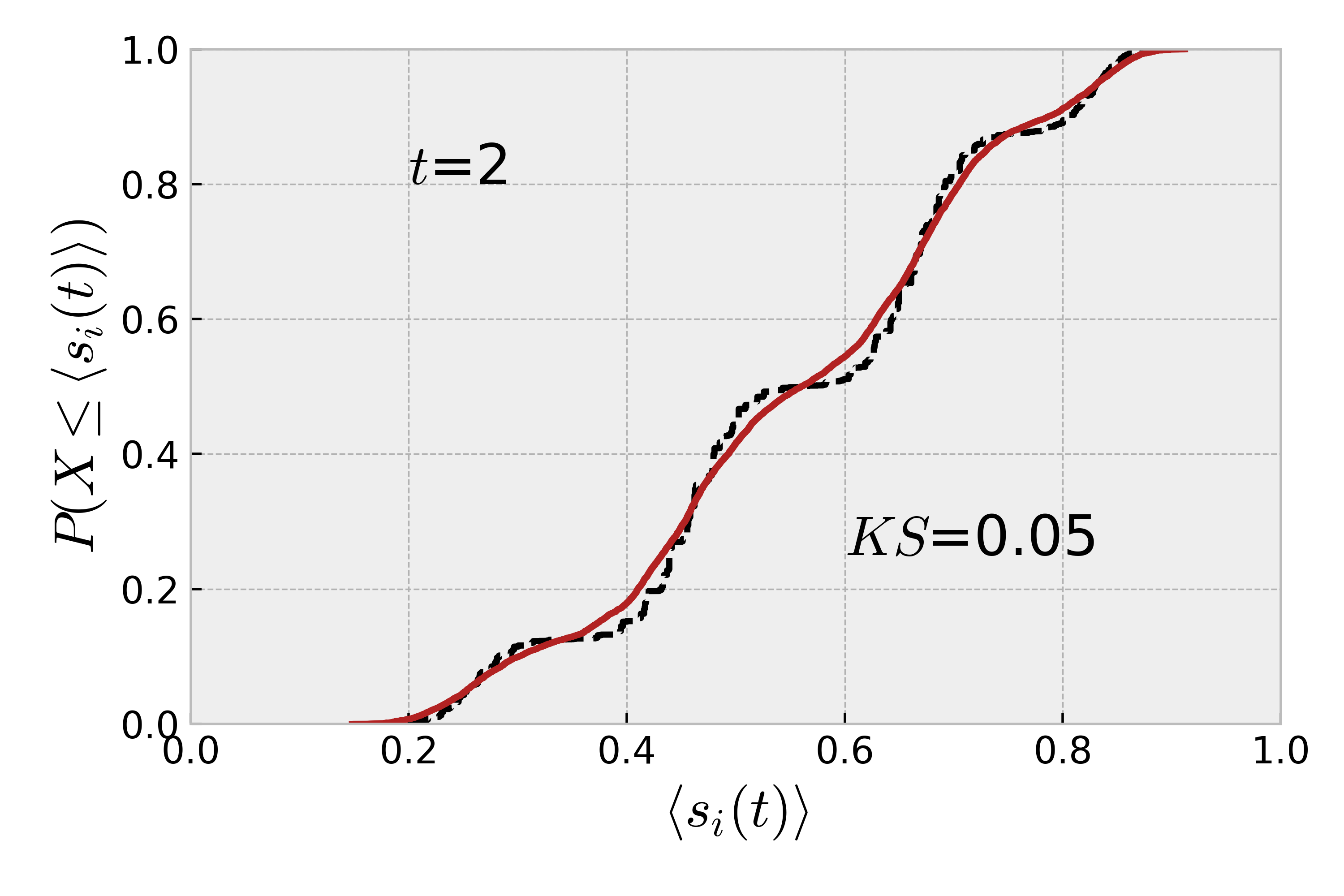} &
    \includegraphics[width=0.33\textwidth]{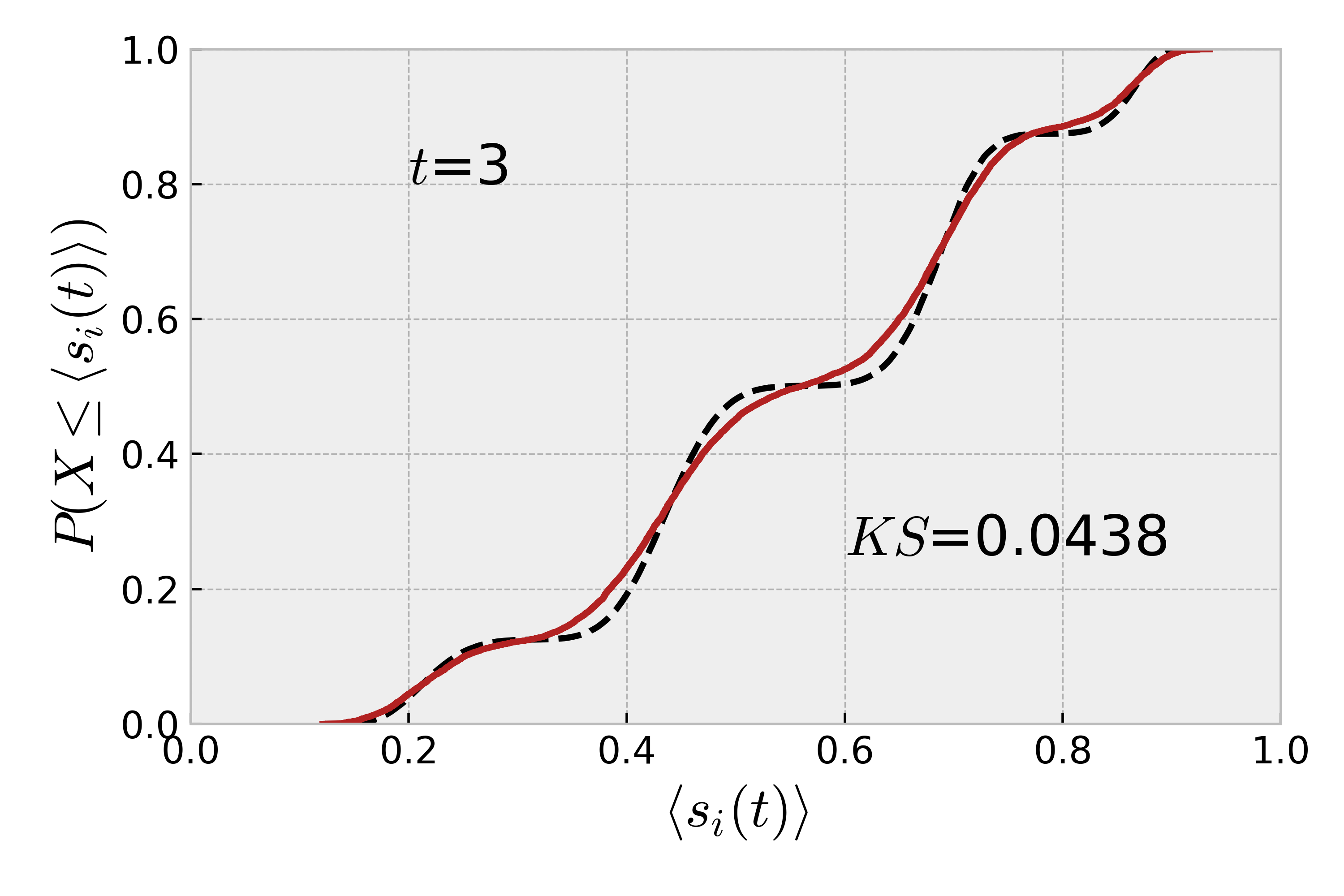} \\
    \vspace{-2mm} \hspace{0.1mm} a) & \hspace{-35mm} b) &  \hspace{-35mm} c) 
\end{tabular}
\caption{CDF of activation probabilities for the Boolean system defined in \eqref{eq: LTMM update rule} at time $t=1,2,3$ (from (a) to (c)). The network is a random regular graph with degree $c=3$. Interactions are fully symmetric, i.e. $J_{ij}=J_{ji}$ and are drawn from the set $\{-1,1\}$ with equal probability. Self-interactions are absent i.e $J_{ii}=0 ~ \forall ~  i$. The inverse temperature is $\beta=1$ and initial conditions are set to $\pr(s_i^{t})=0.5 ~\forall i$. Dashed line indicates solution to equation (\ref{eq: OTA mono thermo maginal dist}) from population dynamics with sample size $S=2.5 \times 10^{5}$. Solid line indicates results from MC simulations for a network of size $N=10^{4}$. Annotations indicate the Kolmogorov–Smirnov (KS) statistic comparing the empirical distributions from the cavity method and MC simulations.  }
  \label{fig: dist site marginal }
\end{figure}

To demonstrate this we consider the linear threshold model (\ref{eq: LTMM update rule}) in the absence of self-interactions, such that $J_{ii}=0$.  In this case the probability of a site to have activation $s_i^{t}$ at some time $t$, under the OTA scheme, is given by, 
\begin{align}
    \mathrm{P}_i(s_i^{t})
    &= \sum_{s_i^{t-2}} \mathrm{P}_i(s_i^{t-2}) \sum_{\bfs_{\partial_i}^{t-1}} \Wr(s_i^{t}| h_i(\bfs_{\partial_i}^{t-1}))  \left[ \prod_{j \in \partial_i}\mathrm{P}^{(i)}_j(s_j^{t-1}| \zeta_j^{(i),t-1})\right], \label{eq: mono site marginal }
    \end{align}
and the probability of a site to have activation $s_i^{t}$ at time $t$, in the cavity graph where site $\ell$ is removed, subject to some external field $\zeta_i^{(\ell),t} = J_{i \ell}s_\ell^{t-1}$, is given by
\begin{align}
    \mathrm{P}^{(\ell)}_i(s_i^{t}| \zeta_i^{(\ell),t})
    &= \sum_{s_i^{t-2}} \mathrm{P}_i(s_i^{t-2}) \sum_{\bfs_{\partial_i \setminus \ell }^{t-1}} \Wr(s_i^{t}| h_i(\bfs_{\partial_i}^{t-1}))  \left[ \prod_{j \in \partial_i \setminus \ell }\mathrm{P}^{(i)}_j(s_j^{t-1}| \zeta_j^{(i),t-1})\right] \label{eq: mono site marginal cav}
\end{align}
as shown in \cite{zhang2012inference}. From this we can define the distribution of site marginals as, 
\begin{align}
    \pi(\{\pr_{t}\}) &= \frac{1}{N} \sum_i\prod_{s^{t}} \delta\left(\pr_{t}(s^{t}) - \pr_i(s^{t})\right). \label{eq: def dist of site probs}
\end{align}
In \ref{app: OTA thermo dist mono} we show that in the limit $N\to\infty$ the distribution of site marginals is found from the following closed set of equations, 
\begin{align}
\begin{split}
    \pi(\{\pr_{t}\}) &= \sum_{k} \pr(k)  \int \left[ \prod_{j=1}^{k} \rmd J_j \rmd \hat{J}_j \{\rmd \hat{\pr}_j\}\pr(J_j)\pr(\hat{J}_j|J_j) \right] \int \{\rmd \pr_{t-2}\}  \\ 
    & \quad \times \pi_{t-2} \left[ \{\pr_{t-2}\} | k, \mathbf{J},\mathbf{\hat{J}}  \right] \prod_{s^{t}} \delta\left( \pr_{t}(s^{t}) - \phi(k, \{ \pr_{t-2}\},\mathbf{J},\mathbf{\hat{J}},\{ \boldsymbol{\hat{\pr}}\} ) \right)\\   
      & \qquad \qquad \qquad \qquad \quad \times \prod_{j=1}^{k} \hat{\pi}_{t-1} \left[ \{\hat{\pr}_j\} | \hat{J}_js^{t-2} \right] \label{eq: OTA mono thermo maginal dist}
\end{split}
\end{align}
with
\begin{align}
    \phi(k, \{ \pr_{t-2}\},\mathbf{J},\mathbf{\hat{J}},\{ \boldsymbol{\hat{\pr}}\} ) &= \sum_{s^{t-2}} \pr_{t-2}(s^{t-2}) \sum_{s_{1}^{t-1},\dots, s^{t-1}_{k}} \Wr(s^{t}| \sum_{j=1}^{k} J_js_j^{t-1}) \prod_{j=1}^{k} \hat{\pr}_j(s_j^{t-1}) \label{eq: phi}
\end{align}
and
\begin{align}
\begin{split}
    \hat{\pi}_{t}(\{\pr_{t}\}| x ) &= \sum_{k} \frac{k \pr(k)  }{ \left<k \right> } \int \left[ \prod_{j=1}^{k-1} \rmd J_j \rmd \hat{J}_j \{\rmd \hat{\pr}_j\}\pr(J_j)\pr(\hat{J}_j|J_j) \right] \int \{\rmd \pr_{t-2}\} \\ 
    & \quad \times  \pi_{t-2} \left[ \{\pr_{t-2}\} | k, \mathbf{J},\mathbf{\hat{J}}  \right] \prod_{s^{t}} \delta\left( \pr_{t}(s^{t}) - \widetilde{\phi}(k, \{ \pr_{t-2}\},\mathbf{J},\mathbf{\hat{J}},\{ \boldsymbol{\hat{\pr}}\},x ) \right) \\
    & \qquad \qquad \qquad \qquad \quad  \times \prod_{j=1}^{k-1} \hat{\pi}_{t-1} \left[ \{\hat{\pr}_j\} | \hat{J}_js^{t-2} \right] \label{eq: OTA mono thermo cav dist}
\end{split}
\end{align}
where we define,
\begin{align}
\begin{split}
    \widetilde{\phi}(k, \{ \pr_{t-2}\},\mathbf{J},\mathbf{\hat{J}},\{ \boldsymbol{\hat{\pr}}\},x ) &= \sum_{s^{t-2}} \pr_{t-2}(s^{t-2}) \\ 
    & \qquad \times \sum_{s_{1}^{t-1},\dots, s^{t-1}_{k}} W(s^{t}| \sum_{j=1}^{k} J_js_j^{t-1} + x) \prod_{j=1}^{k} \hat{\pr}_j(s_j^{t-1}). \label{eq: phi tilde}
\end{split}
\end{align}
Equations (\ref{eq: OTA mono thermo maginal dist}) and (\ref{eq: OTA mono thermo cav dist}) can be solved by a population dynamics procedure \cite{mezard2001bethe,hurry2022vaccination}. In figure \ref{fig: dist site marginal } we show the cumulative
distribution function
of site activation probabilities $\pr(\langle s(t) \rangle) = \pi(\pr(s^{t}=1))$ at different times, 
computed solving 
the above equations via population dynamics, 
for a random regular graph with fully symmetric interactions. We find reasonable agreement with MC simulations on a single instance of a network with size $N=10^{4}$. We however, see small deviations due to the cavity equations in this instance being averaged over the disorder $\pr(\bfJ)$. As expected from earlier analysis in Sec. \ref{sec: linear } the time-dependent distribution of activation probabilities reaches a multi-modal steady-state.

\section{Discussion}

In our work we have studied  systems of sparsely connected Boolean variables with multi-node and self-interactions. Previous work has shown that self-interactions complicate the analysis of dynamics. However, by mapping to an equivalent bipartite system, we find that the dynamical cavity method, within a OTA scheme, provides an efficient numerical framework to study the dynamics of systems with arbitrary bidirectionality, multi-node and self-interactions.
We have shown for such systems that the OTA scheme predicts the activation probability of each site in the transient and non-equilibrium steady-state, down to relatively low temperature, where the system is ergodic. As temperature is lowered further, ergodicity is eventually broken, and we have found that the error increases with the bidirectionality of the interactions.
At zero temperature, however, 
if the dynamical cavity is given the same initial configuration as MC simulations, the OTA scheme is in excellent agreement with simulations, showing that OTA accurately describes the dynamics within any given ergodic sector. We have also derived equations for the distribution (and mean) of site activations at a given time, in the thermodynamic limit, within a OTA scheme, which is numerically more efficient than previous approaches involving sums over the history of a trajectory \cite{neri2009cavity}.  We comment that these equations, when compared to their analogue derived by generating functionals \cite{hatchett2004parallel,mimura2009parallel} provide physical insight into the effect of truncating memory terms in the GFA. In principle, the dynamical cavity equations are also solveable for networks with strong degree correlations, as shown in previous work \cite{hurry2022vaccination}. By providing closed expressions for the distribution of site marginals, the dynamical cavity method can provide insight into the heterogeneity of site trajectories for a typical network drawn from some ensemble. This is unlike GFA which only provides information on the trajectory of a typical site.

From our work, it is clear that sparse networks with partially bidirectional interactions support multiple (cyclic) attractors, at low 
noise. Additionally, multi-node interactions are found to decrease the overlap between attractors, suggesting that cooperativity increases diversification of attractors. The existence of a multiplicity of attractors in partially bidirectional sparse systems is an important feature for models of GRNs to sustain  multi-cellular life. It means that if a GRN is in an attractor, it can be pushed out of this attractor by some event, for example the sudden change in external field, and can move from one attractor to another. This could describe how cells move from one cell type to another. Both cooperative effects and gene self-regulation (i.e. a gene coding for a TF that regulates the gene itself) are commonly observed in GRNs and our work suggests that it is these features in combination which allow GRNs to support a diverse set of stable gene expression profiles, corresponding to different cell types, with self-regulation playing an important role in determining the nature of attractors.

Our work poses interesting questions for future work. One point of focus would be to study how a GRN may move from one attractor to another, either by the application of time-dependent external fields, or dynamic variations in the interaction network itself. One could also focus on adding prescribed degree sequences to impose structure that is a more realistic description of GRNs. Of course while we can deduce that a multiplicity of (cyclic) attractors exist, it remains an open question how the attractors relate to the specific realisation of the bipartite network. An analytical framework to deduce the number of attractors that are supported by a network also remains an open point of investigation. Our work also suggests that there is an ergodicity breaking phase at low temperatures characterised by an overlap distribution which resembles the one observed in equilibrium spin systems (with symmetric interactions and fixed-point attractors) exhibiting one step of replica symmetry breaking. 
Due to the arbitrary symmetry of the interactions, however, this would have to be formally ascertained by as yet unknown dynamical methods. 
Such methods 
will be necessary to advance our understanding of 
sparse systems with multi-node interactions out of equilibrium
and may give further insight into the behaviour of GRNs.

\ack
CJH is supported by the EPSRC Centre for Doctoral Training in Cross-Disciplinary Approaches to Non- Equilibrium Systems (CANES, EP/L015854/1). AA wishes to thank the Chimera group at the University of Rome, Sapienza, for the kind hospitality.

\appendix

\section{Dynamical cavity approach to bipartite systems} \label{app: dyn cav bipartite}

We show here how the dynamical cavity method may be used to analyse the dynamics of the general bipartite system defined in (\ref{eq: GLTBM update rule s}) and (\ref{eq: GLTBM update rule tau}), from which the linear model with self-interactions (\ref{eq: LTMM update rule}) and the nonlinear model with multi-node interactions (\ref{eq: NLTM update rule}) may be recovered by suitable choice of parameters. 
Starting with equation 
\eqref{eq:double-trajectory},
we wish to derive a closed set of equations for single site quantities at a given time, like $s_i^t$ and $\tau_\mu^t$. 
To do so, we first consider the trajectory of a single gene $i$, 
\begin{align}
\mathrm{P}_i(s_i^{0...t_m}) &= \sum_{\bfs^{0...t_m} \setminus s_i^{0...t_m} ,\btau^{0...t_m}} \mathrm{P}(\bfs^{0...t_m},\btau^{0...t_m}).
\end{align}
As is typical of the dynamical cavity formalism, we assume that the bipartite network has the topological structure of a tree and that initial conditions factorise over the sites of the tree
$P(\bfs^0,\btau^0)=\prod_{i\mu}P(s_i^0,\tau_\mu^0)$. In doing so we can rewrite the equation above as follows, 
\begin{eqnarray}
\hspace{-1.5cm}\mathrm{P}_i(s_i^{0\ldots t_m}) 
&=&\sum_{\bfs^{0\ldots t_m} 
\backslash s_i^{0\ldots t_m} ,\btau^{0\ldots t_m}} \mathrm{P}(\bfs^{0},\btau^{0}) 
\left[\prod_{t=1}^{t_m} \Wr(s_i^{t}| h_i(\btau_{\partial_i}^{t-1})) \prod_{j \in T_i \backslash i} \Wr(s_j^{t}| h_j(\btau_{\partial_j}^{t-1})) \right] 
\nonumber\\
\hspace*{-1.5cm}
&&\times \prod_{\mu \in \partial_i} \left[\prod_{t=1}^{t_m} \widetilde{\Wr}(\tau_\mu^{t}| h_\mu(\bfs_{\partial_\mu}^{t-1})) \prod_{\nu \in T_\mu \backslash \mu} \widetilde{\Wr}(\tau_\nu^{t}| h_\nu(\bfs_{\partial_\nu}^{t-1})) \right] 
\nn\\
\hspace{-1.5cm}
&=&\sum_{\bfs^{0...t_m} \backslash s_i^{0...t_m} ,\btau^{0...t_m}} \mathrm{P}(\bfs^{0}, \btau^{0}) \left[\prod_{t=1}^{t_m} \Wr(s_i^{t}| h_i(\btau_{\partial_i}^{t-1})) \right] 
\nn\\
\hspace{-1.5cm}
&&\times \prod_{\mu \in \partial_i}
\left[\prod_{t=1}^{t_m} \widetilde{\Wr}(\tau_\mu^{t}| h_\mu(\bfs_{\partial_\mu}^{t-1})) \prod_{\nu,j \in T_\mu \backslash \mu,i} \widetilde{\Wr}(\tau_\nu^{t}| h_\nu(\bfs_{\partial_\nu}^{t-1}))\Wr(s_j^{t}| h_j(\btau_{\partial_j}^{t-1})) \right]
\nn\\
\hspace*{-1.5cm}
&=&\mathrm{P}(s_i^{0})\sum_{\btau_{\partial_i}^{0...t_m}} \Big[\prod_{t=1}^{t_m} \Wr(s_i^{t}| h_i(\btau_{\partial_i}^{t-1})) \Big] 
\prod_{\mu \in \partial_i}\Bigg\{ \sum_{\bfs_{T_\mu\backslash i}^{0...t_m} ,\btau_{T_\mu\backslash \mu}^{0...t_m}} \mathrm{P}(\bfs_{T_\mu}^{0}\backslash s_i^{0}) \mathrm{P}(\btau_{T_\mu}^{0})  
\nn\\
\hspace*{-1.5cm}
&&\hspace*{-1.5mm}\times \Big[\prod_{t=1}^{t_m} \widetilde{\Wr}(\tau_\mu^{t}| h_\mu(\bfs_{\partial_\mu}^{t-1})) \prod_{\nu,j \in T_\mu \backslash \mu,i} \widetilde{\Wr}(\tau_\nu^{t}| h_\nu(\bfs_{\partial_\nu}^{t-1}))\Wr(s_j^{t}| h_j(\btau_{\partial_j}^{t-1})) \Big]\Bigg\} 
\end{eqnarray}
where we have defined 
$\bfs_{T_\mu} = \left\{s_i, i \in T_\mu\right\}$ and $\btau_{T_\mu} = \left\{\tau_\nu, \nu \in T_\mu\right\}$
where $T_\mu$ and $T_i$ are trees rooted at nodes  $\mu$ and $i$.
We may then take the external sum over $\btau_{\partial_i}^{t_m}$ and the internal sums over $\btau_{T_\mu}^{t_m}$ and $\bfs_{T_\mu}^{t_m}$,
\begin{align}
\begin{split}
\mathrm{P}_i(s_i^{0...t_m})
&=\mathrm{P}(s_i^{0})\sum_{\btau_{\pa_i}^{0...t_m-1}} \left[\prod_{t=1}^{t_m} \Wr(s_i^{t}| h_i(\btau_{\pa_i}^{t-1})) \right] \prod_{\mu \in \pa_i}\Bigg\{ \sum_{\bfs_{T_\mu\backslash i}^{0...t_m-1} ,\btau_{T_\mu\backslash \mu}^{0...t_m-1}} \mathrm{P}(\bfs_{T_\mu}^{0}\backslash s_i^{0}) \\
& \quad\times \mathrm{P}(\btau_{T_\mu}^{0})\Big[\prod_{t=1}^{t_m-1} \widetilde{\Wr}(\tau_\mu^{t}| h_\mu(\bfs_{\partial_\mu}^{t-1})) \prod_{\nu,j \in T_\mu \backslash \mu,i} \widetilde{\Wr}(\tau_\nu^{t}| h_\nu(\bfs_{\partial_\nu}^{t-1}))\Wr(s_j^{t}| h_j(\btau_{\partial_j}^{t-1})) \Big]\Bigg\}. \label{app eq: bipartite cav curly brackets}
\end{split}
\end{align}
We note that the local field that acts on TF $\mu$ can be written as, 
\begin{align}
    h_\mu(\bfs_{\partial_\mu}^{t-1}) &= \sum_{j \in \pa_\mu \setminus i}\eta_j^\mu s_j^{t-1} + \vartheta_\mu +  \eta_i^\mu s_i^{t-1} =  h_\mu^{(i)}(\bfs_{\partial_\mu}^{t-1}) + \zeta_\mu^{(i),t}  \label{app eq: h_mu}
\end{align}
where we have defined the local field in a new copy of the network where site $i$ is removed, referred to as the \textit{cavity graph}, $h_\mu^{(i)}(\bfs_{\partial_\mu}^{t-1})=\sum_{j \in \pa_\mu \setminus i}\eta_j^\mu s_j^{t-1} + \vartheta_\mu $, and the time dependent external field $\zeta_\mu^{(i),t} = \eta_i^\mu s_i^{t-1}$, which is the effect of the removed gene $i$ on TF $\mu$ in the cavity graph. Similarly, we can write the local field that acts on gene $i$ as, 
\begin{align}
    h_i(\btau_{\partial_i}^{t-1}) &= \sum_{\mu \in \pa_i \setminus \nu}\xi_i^\mu\tau_\mu^{t-1} +\vartheta_i + \xi_i^{\nu}\tau_\nu^{t-1} =  h_i^{(\nu)}(\btau_{\partial_i}^{t-1}) + \zeta_i^{(\nu),t} 
\end{align}
denoting the local field in the cavity graph where TF $\nu$ is removed, $h_i^{(\nu)}(\btau_{\partial_i}^{t-1})=\sum_{\mu \in \pa_i \setminus \nu}\xi_i^\mu\tau_\mu^{t-1} +\vartheta_i$, and external field $\zeta_i^{(\nu),t} = \eta_i^\mu s_i^{t-1}$.
We can then define the contents of the curly brackets in equation (\ref{app eq: bipartite cav curly brackets}) as $\mathrm{P}_\mu^{(i)}(\tau_\mu^{0\dots t_m-1}| \zeta_\mu^{(i),1 \dots t_m-1})$, which is the probability to observe the trajectory $\tau_\mu^{0\dots T-1}$ in the cavity graph where gene $i$ is removed, given that the time dependent external field, $\zeta_\mu^{(i),1,\dots,t_m}$, acts on TF $\mu$. With this definition in place, equation (\ref{app eq: bipartite cav curly brackets}) is now equivalent to
\begin{align}
\mathrm{P}_i(s_i^{0...t_m}) &=\mathrm{P}(s_i^{0})\sum_{\btau_{\pa_i}^{0 \dots t_m-1}} \left[\prod_{t=1}^{t_m} \Wr(s_i^{t}| h_i(\btau_{\pa_i}^{t-1})) \right] \prod_{\mu \in \pa_i} \mathrm{P}_\mu^{(i)}(\tau_\mu^{0\dots t_m-1}|\zeta_\mu^{(i),1 \dots t_m-1}).\label{app eq:p_s}
\end{align}
Following the same reasoning, we can derive an analogous expression for the trajectory of a single TF $\mu$,
\begin{align}
\mathrm{P}_\mu(\tau_\mu^{0...t_m}) &=\mathrm{P}(\tau_\mu^{0})\sum_{\bfs_{\pa_\mu}^{0 \dots t_m-1}} \left[\prod_{t=1}^{t_m} \widetilde{\Wr}(\tau_\mu^{t}| h_\mu(\bfs_{\pa_\mu}^{t-1})) \right] \prod_{j \in \pa_\mu} \mathrm{P}_j^{(\mu)}(s_j^{0\dots t_m-1}|\zeta_j^{(\mu),1 \dots t_m-1}) \label{app eq:p_tau}
\end{align}
where $\mathrm{P}_j^{(\mu)}(s_j^{0\dots t_m-1}|\zeta_j^{(\mu),1 \dots t_m-1})$ is the probability to observe the trajectory $s_j^{0\dots t_m-1}$ in the cavity graph where site $\mu$ is removed, given that the time-dependent external field $\zeta_j^{(\mu),1 \dots t_m-1} = \xi_j^\mu \tau_\mu^{1 \dots t_m-1}$, acts on site $j$. Equations for the probability to observe the trajectory of a single gene $i$ in the \textit{cavity graph} are found by considering the effect of removing TF $\nu$ from equation (\ref{app eq:p_s}),
\begin{align}
\mathrm{P}^{(\nu)}_i(s_i^{0...t_m}) &=\mathrm{P}(s_i^{0})\sum_{\btau_{\pa_i \backslash \nu}^{0...t_m-1}} \left[\prod_{t=1}^{t_m} \Wr(s_i^{t}| h_i^{(\nu)}(\btau_{\pa_i }^{t-1})) \right] \prod_{\mu \in \pa_i\backslash \nu} \mathrm{P}_\mu^{(i)}(\tau_\mu^{0\dots t_m-1}|\zeta_\mu^{(i),1 \dots t_m-1}),\label{eq:p_s_cav}
\end{align}
which depends upon the cavity field $h_i^{(\nu)}(\btau_{\pa_i }^{t-1})$. Similarly, by removing gene $\ell$ from equation (\ref{app eq:p_tau}) we find, 
\begin{align}
\mathrm{P}^{(\ell)}_\mu(\tau_\mu^{0...t_m}) &=\mathrm{P}(\tau_\mu^{0})\sum_{\bfs_{\pa_\mu \backslash \ell}^{0...t_m-1}} \left[\prod_{t=1}^{t_m} \widetilde{\Wr}(\tau_\mu^{t}| h^{(\ell)}_\mu(\bfs_{\pa_\mu}^{t-1})) \right] \prod_{j \in \pa_\mu \backslash \ell} \mathrm{P}_j^{(\mu)}(s_j^{0\dots t_m-1}|\zeta_j^{(\mu),1 \dots t_m-1}), \label{eq:p_tau_cav}
\end{align}
which depends upon the cavity field $h^{(\ell)}_\mu(\bfs_{\pa_\mu}^{t-1})$. If we now consider an external field $\zeta_i^{(\nu),1\dots t_m}$ acting on gene $i$ in the cavity graph with TF $\nu$ removed, the cavity field in equation (\ref{eq:p_s_cav}) will become $h_i^{(\nu)}(\btau_{\pa_i }^{t-1}) + \zeta_i^{(\nu),t} = h_i(\btau_{\pa_i }^{t-1})$, from which we deduce equation (\ref{eq:p_s_cav_exfield}). Similarly, if an external field $\zeta_\mu^{(\ell),1\dots t_m}$ acts on TF $\mu$ in the cavity graph where gene $\ell$ is removed, from equation (\ref{eq:p_s_cav}), we can deduce equation (\ref{eq:p_tau_cav_exfield}). We have derived a closed system of equations (\ref{app eq:p_s}), (\ref{app eq:p_tau}), (\ref{eq:p_s_cav_exfield}) and (\ref{eq:p_tau_cav_exfield}) which in principle can be solved recursively for the trajectory of the system. As noted in the main text, their computational complexity grows exponentially with the length of the trajectory.

If we assume that the system has unidirectional interactions, however, the cavity equations simplify. In this case,  $\xi_i^\mu \neq 0$ implies $\eta_i^\mu =0$, such that the external fields in equation (\ref{app eq:p_s}) vanish, $\zeta_\mu^{(i),t}=0$, and the RHS of equations \eqref{app eq:p_tau} and \eqref{eq:p_tau_cav_exfield} become identical, as $\ell \not\in \partial_\mu$ when $\mu \in \partial_\ell$ and given \eqref{app eq: h_mu} we have
$\mathrm{P}^{(i)}_\mu(\tau_\mu^{0...t_m} | 0) = \mathrm{P}_\mu(\tau_\mu^{0...t_m})$. This simplifies equation (\ref{app eq:p_s}) to 
\begin{align}
\begin{split}
    \mathrm{P}_i(s_i^{0...t_m}) &=\mathrm{P}(s_i^{0})\sum_{\btau_{\pa_i }^{0...t_m-1}} \left[\prod_{t=1}^{t_m} \Wr(s_i^{t}| h_i(\btau_{\pa_i}^{t-1})) \right] 
\prod_{\mu \in \pa_i} \mathrm{P}_\mu(\tau_\mu^{0\dots t_m-1})\label{eq:p_s_unidirectional}
\end{split}
\end{align}
and by the same argument one can also derive the trajectory for the TFs, 
\begin{align}
\begin{split}
    \mathrm{P}_\mu(\tau_\mu^{0...t_m}) &=\mathrm{P}(\tau_\mu^{0})\sum_{\bfs_{\pa_\mu}^{0...t_m-1}} \left[\prod_{t=1}^{t_m} \widetilde{\Wr}(\tau_\mu^{t}| h_\mu(\bfs_{\pa_\mu}^{t-1})) \right] \prod_{j \in \pa_\mu} \mathrm{P}_j(s_j^{0\dots t_m-1}). \label{eq:p_tau_unidirectional}
\end{split}
\end{align}
We may now marginalise equation (\ref{eq:p_s_unidirectional}) over $s_i^{0,\dots,t_m-1}$ and  equation (\ref{eq:p_tau_unidirectional}) over $\tau_\mu^{0,\dots,t_m-1}$ and find,
\begin{align}
\begin{split}
    \mathrm{P}_i(s_i^{t_m}) &=\sum_{\btau_{\pa_i}^{t_m-1}} \left[ \Wr(s_i^{t_m}| h_i(\btau_{\pa_i}^{t_m-1})) \right] 
\prod_{\mu \in \pa_i} \mathrm{P}_\mu(\tau_\mu^{t_m-1})\label{eq:p_s_unidirectiona_onestep}
\end{split}\\
\begin{split}
    \mathrm{P}_\mu(\tau_\mu^{t_m}) &=\sum_{\bfs_{\pa_\mu}^{t_m-1}} \left[ \widetilde{\Wr}(\tau_\mu^{t_m}| h_\mu(\bfs_{\pa_\mu}^{t_m-1})) \right] \prod_{j \in \pa_\mu } \mathrm{P}_j(s_j^{ t_m-1}). \label{eq:p_tau_unidirectional_onestep}
\end{split}
\end{align}
Equations (\ref{eq:p_s_unidirectiona_onestep}) and (\ref{eq:p_tau_unidirectional_onestep}) are a closed set of equations which may be solved by simple iteration. They reveal that in systems with unidirectional interactions, $P(\bfs_{\partial_\mu}^t)=\prod_{j\in\partial_\mu}P_j(s_j^t)$.

As we note in Sec. \ref{sec: cavity method}, for systems with arbitrarily bidirectional interactions, we may use the OTA scheme to reduce the time complexity of the cavity equations, which in the case of our bipartite system means to assume (\ref{app eq: OTA assumption s}) and (\ref{app eq: OTA assumption tau}).  If we insert (\ref{app eq: OTA assumption tau}) into (\ref{eq:p_s_cav_exfield}) and isolate the terms which contain $s_i^{T}$ and $\btau_{\partial_i}^{T-1}$, we can rewrite the RHS in terms of the cavity distribution 
of the $s_i$ trajectory up until time $t_m-1$,
\begin{eqnarray}
\hspace*{-2.5cm}\mathrm{P}^{(\nu)}_i(s_i^{0...t_m} | \zeta_i^{(\nu),1\dots t_m})&=&\mathrm{P}(s_i^{0})\sum_{\btau_{\pa_i \backslash \nu }^{0...t_m-1}} \left[\prod_{t=1}^{t_m} \Wr(s_i^{t}| h_i(\btau_{\pa_i}^{t-1})) \right] \prod_{\mu \in \pa_i\backslash \nu} \mathrm{P}_\mu(\tau_\mu^{0})\prod_{t=1}^{t_m-1}  \mathrm{P}^{(i)}_\mu(\tau_\mu^{t}| \zeta_\mu^{(i),t})
\nonumber\\
&=&\sum_{\btau_{\pa_i \backslash \nu}^{t_m-1}} \Wr(s_i^{t_m}| h_i(\btau_{\pa_i}^{t_m-1}))
\left[\prod_{\mu \in \pa_i\backslash \nu}\mathrm{P}^{(i)}_\mu(\tau_\mu^{t_m-1}| \zeta_\mu^{(i),t_m-1})\right] \mathrm{P}(s_i^{0})
\nonumber \\
&&\times \sum_{\btau_{\pa_i}^{0...t_m-2}} \left[\prod_{t=1}^{t_m-1} \Wr(s_i^{t}| h_i(\btau_{\pa_i}^{t-1})) \right] \prod_{\mu \in \pa_i\backslash \nu} \mathrm{P}_\mu(\tau_\mu^{0})\prod_{t=1}^{t_m-2}  \mathrm{P}^{(i)}_\mu(\tau_\mu^{t}| \zeta_\mu^{(i),t}) \nonumber\\
&=& \sum_{\btau_{\pa_i \backslash \nu}^{t_m-1}} \Wr(s_i^{t_m}| h_i(\btau_{\pa_i}^{t_m-1})) \left[ \prod_{\mu \in \pa_i\backslash \nu}\mathrm{P}^{(i)}_\mu(\tau_\mu^{t_m-1}| \zeta_\mu^{(i),t_m-1})\right] 
\nn\\
&&\times \mathrm{P}^{(\nu)}_i(s_i^{0...t_m-1}|\zeta_i^{(\nu),1\dots t_m-1}).
\end{eqnarray}
At this stage it is possible to marginalise the above over $s_i^{0,\dots,t_m-1}$,
\begin{eqnarray}
\hspace*{-2cm} \mathrm{P}_i(s_i^{t_m}| \zeta_i^{(\nu),1\dots t_m}) 
    &=& \sum_{s_i^{0 \dots t_m-1}} \sum_{\btau_{\pa_i \backslash \nu}^{t_m-1}} \Wr(s_i^{t_m}| h_i(\btau_{\pa_i}^{t_m-1})) \left[ \prod_{\mu \in \pa_i\backslash \nu}\mathrm{P}^{(i)}_\mu(\tau_\mu^{t_m-1}| \zeta_\mu^{(i),t_m-1})\right]\nonumber \\
    &&\times  \mathrm{P}^{(\nu)}_i(s_i^{0...t_m-1} | \zeta_i^{(\nu),1\dots t_m-1})
\nn\\
    &=& \sum_{s_i^{0 \dots t_m-1}} \sum_{\btau_{\pa_i\backslash \nu}^{t_m-1}} \Wr(s_i^{t_m}| h_i(\btau_{\pa_i}^{t_m-1})) \left[ \prod_{\mu \in \pa_i\backslash \nu}\mathrm{P}^{(i)}_\mu(\tau_\mu^{t_m-1}| \zeta_\mu^{(i),t_m-1})\right]\nonumber \\ 
     &&\times \mathrm{P}_i(s_i^{0}) \prod_{t=1}^{t_m-1} \mathrm{P}^{(\nu)}_i(s_i^{t} | \zeta_i^{(\nu),t}) \nn\\
    &=& \sum_{s_i^{t_m-2}} \sum_{\btau_{\pa_i\backslash \nu}^{t_m-1}} \Wr(s_i^{t_m}| h_i^{(\nu)}(\btau_{\pa_i}^{t_m-1}) + \zeta_i^{(\nu),t_m} ) 
    \nn\\
    &&\times \left[ \prod_{\mu \in \pa_i\backslash \nu}\mathrm{P}^{(i)}_\mu(\tau_\mu^{t_m-1}| \zeta_\mu^{(i),t_m-1})\right] 
 \mathrm{P}^{(\nu)}_i(s_i^{t_m-2} | \zeta_i^{(\nu),t_m-2}).
\end{eqnarray}
It is now apparent that the LHS only depends on $\zeta_i^{(\nu),t_m-2}$ and $\zeta_i^{(\nu),t_m}$ i.e $\mathrm{P}_i(s_i^{t_m}| \zeta_i^{(\nu),1\dots t_m}) = \mathrm{P}_i(s_i^{t_m}| \zeta_i^{(\nu),t_m-2,t_m})$, such that we can write
\begin{align}
\begin{split}
    \mathrm{P}_i^{(\nu)}(s_i^{t_m}| \zeta_i^{(\nu),t_m-2, t_m})
    &= \sum_{s_i^{t_m-2}} \sum_{\btau_{\pa_i\backslash \nu}^{t_m-1}} \Wr(s_i^{t_m}| h_i^{(\nu)}(\btau_{\pa_i}^{t_m-1}) + \zeta_i^{(\nu),t_m}) \\
    & \qquad \times \left[ \prod_{\mu \in \pa_i\backslash \nu}\mathrm{P}^{(i)}_\mu(\tau_\mu^{t_m-1}| \zeta_\mu^{(i),t_m-1})\right] \mathrm{P}^{(\nu)}_i(s_i^{t_m-2} | \zeta_i^{(\nu),t_m-2}). \label{app eq: p i no nu OTA step 1 }
\end{split}
\end{align}
In principle, this is in contradiction with the Markovian assumption made for the cavity distribution, which 
would require $ \mathrm{P}_i^{(\nu)}(s_i^{t_m}| \zeta_i^{(\nu),t_m-2, t_m})= \mathrm{P}_i^{(\nu)}(s_i^{t_m}| \zeta_i^{(\nu), t_m})$. Such inconsistency arises from treating a non-Markovian process as Markovian.  
In order to get a closed set of equations for the one-time cavity marginal, further closure assumptions are thus required. As stated in Sec. \ref{sec: cavity method}, we assume that the cavity distribution, with external field from the removed site, may be approximated by its non-cavity counterpart i.e  $\mathrm{P}^{(\nu)}_i(s_i^{t_m-2} | \zeta_i^{(\nu),t_m-2}) \approx\mathrm{P}_i(s_i^{t_m-2})$, following  \cite{zhang2012inference}. Under this approximation, from (\ref{app eq: p i no nu OTA step 1 }) we retrieve equation (\ref{eq: p i without nu}), 
and by considering adding the removed TF $\nu$ back in and setting external fields to zero, we find, for the non-cavity distribution, the expression given in (\ref{eq: p i}).
By the same reasoning, we can derive analogous equations for the trajectory of a TF $\mu$. In summary, the dynamics of the bipartite system are fully described, under the OTA scheme, by the closed, coupled set of equations (\ref{eq: p i without nu}),(\ref{eq: p mu without l }),(\ref{eq: p i}) and (\ref{eq: p mu}), which can be solved by iteration given some initial conditions $\mathrm{P}_i(s_i^{0})$ and $\mathrm{P}_\mu(\tau^{0}_\mu)$.


\section{Equilibrium analysis of (0,1) spins with parallel update and self-interactions} \label{app: equib analysis}

In this appendix we use the cavity method to calculate the equilibrium value 
of site marginals for the linear threshold model 
(\ref{eq: LTMM update rule}) with self-interactions.  Assuming that $z_i(t)$ is a random variable with c.d.f. (\ref{eq:thermal}), we can write the evolution of the state probability as a Markov chain
\begin{equation}
    \mathrm{P}_{t+1}(\bfs)=\sum_{\bfs'} W(\bfs|\bfs')\mathrm{P}_t(\bfs')
    \label{eq:mono parallel_dynamics}
\end{equation}
with transition probability
\begin{align}
    W(\bfs|\bfs')=\prod_{i=1}^N \Wr(s_i|h_i(\bfs'_{\pa_i},s'_i))=\prod_{i=1}^N\frac{e^{\frac{\beta}{2} (2 s_i - 1) h_i(\bfs'_{\pa_i},s'_i)}}{2\cosh \frac{\beta}{2} h_i(\bfs'_{\pa_i},s'_i)} \label{eq: mono transition prob},
\end{align}
where $h_i(\bfs_{\pa_i},s_i) = \sum_{j \in \pa_i} J_{ij} s_j + J_{ii}s_i + \vartheta_i$. If the interactions are fully symmetric, $J_{ij} = J_{ji}$, 
the transition probabilities \eqref{eq: mono transition prob} satisfy detailed balance 
\begin{align}
    p_{\text{eq}}(\bfs) W(\bfs'| \bfs) &= p_{\text{eq}}(\bfs') W(\bfs| \bfs') \label{eq: detailed balance}
    \end{align}
with the equilibrium distribution 
\begin{align}
    p_{\text{eq}}(\bfs) &= \frac{1}{Z} \rme^{ - \beta H_{\beta}(\bfs)} 
\end{align}
where $H_{\beta}(\bfs) =  - \frac{1}{\beta} \sum_i \ln 2 \cosh \frac{\beta}{2}h_i(\bfs_{\pa_i},s_i) - \frac{1}{2}\sum_i h_i(\bfs_{\pa_i},s_i) - \sum_i s_i\vartheta_i$ 
We can rewrite this distribution in a form more amenable to analysis if we consider a fictitious set of variables $\btau = \{0,1\}^{N}$,
\begin{align}
    p_{\text{eq}}(\bfs) &= \frac{1}{Z} \prod_{i=1}^{N } 2 \cosh \frac{\beta}{2}h_i(\bfs_{\pa_i},s_i) \rme^{ \frac{\beta}{2} h_i(\bfs_{\pa_i},s_i)}\rme^{\beta \vartheta_i s_i} \\  
    &= \frac{1}{Z} \prod_{i=1}^{N } \sum_{\tau_i} \rme^{\frac{\beta}{2}(2 \tau_i-1) h_i(\bfs_{\pa_i},s_i)} \rme^{ \frac{\beta}{2} h_i(\bfs_{\pa_i},s_i)} \rme^{\beta \vartheta_i s_i} 
\end{align}
where we have used $2 \cosh(x) = \sum_{\tau \in \{0,1\}} e^{(2 \tau -1 ) x }$ as in \cite{fontanari1988information,castillo2004little}. The equilibrium distribution can then be written as
\begin{align}
    p_{\text{eq}}(\bfs) &= \frac{1}{Z} \sum_{\btau} \rme^{\beta \sum_{i \neq j} \tau_i J_{ij} s_j + \beta \sum_i \tau_i J_{ii} s_i  + \beta \sum_i \vartheta_i (s_i + \tau_i) }. 
\end{align}
the marginalisation of the joint distribution of real and fictitious variables (\ref{eq: statics joint dist}). 

Now we use the cavity method to find a closed set of equations for single site quantities. To do so we marginalise (\ref{eq: statics joint dist}) over all sites except $i$, 
\begin{align}
  p_i(s_i,\tau_i) &= \frac{1}{Z} \sum_{\bfs \backslash s_i} \sum_{\btau \backslash \tau_i}\rme^{-\beta \mathcal{H}(\bfs,\btau)} 
\end{align}
and note that the Hamiltonian can be written in the form $\mathcal{H}(\bfs,\btau) = -s_ih_i(\btau_{\pa_i}) - \tau_ih_i(\bfs_{\pa_i}) - s_iJ_{ii}\tau_i -\vartheta_i (\tau_i + s_i) + \mathcal{H}^{(i)}(\bfs,\btau)$ where we have defined the Hamiltonian of the system where site $i$ has been removed $\mathcal{H}^{(i)}(\bfs,\btau) = -\sum_{\ell \neq (i,j)} s_{\ell} J_{\ell j} \tau_j  -\sum_{\ell \neq i} s_{\ell} J_{\ell } \tau_{\ell}-\sum_{\ell \neq i} \vartheta_{\ell} ( s_{\ell} + \tau_{\ell})$. With this definition the site marginal may be written in the following form, 
\begin{align}
  p_i(s_i,\tau_i) &= \frac{1}{Z_i} 
  \sum_{\btau_{\pa_i}}\rme^{\beta \left(s_i h_i(\btau_{\pa_i}) + \tau_ih_i(\bfs_{\pa_i}) + s_iJ_{ii} \tau_i + \vartheta_i(s_i + \tau_i)\right)} p^{(i)}(\bfs_{\pa_i}, \btau_{\pa_i})
\end{align}
where $ p^{(i)}(\bfs_{\pa_i}, \btau_{\pa_i})$ is the equilibrium distribution of the neighbours of site $i$, 
in the cavity graph where $i$ is removed, $Z^{(i)} = \sum_{\bfs \backslash s_i} \sum_{\btau \backslash \tau_i} \rme^{- \beta \mathcal{H}^{(i)}(\bfs,\btau) }$ is its corresponding partition function and $Z_i=Z/Z^{(i)}$. The cavity approach assumes that cavity fields are independent of each other, such that $ p^{(i)}(\bfs_{\pa_i}, \btau_{\pa_i}) = \prod_{j\in \pa_i} p_j^{(i)}(s_j,\tau_j)$, which is exact on trees and sparse graphs which are locally tree-like in the limit $N \to \infty$. Under this assumption the site marginals and cavity site marginals can be found from 
the closed set of equations 
(\ref{eq: site marginal }) and (\ref{eq: cavity site marginal }). 
In order to solve these equations, it is convenient to 
parameterise the cavity marginals in terms of effective fields 
\begin{eqnarray}
  p_i^{(\ell)}(s_i,\tau_i) &=& \frac{1}{Z_i^{(\ell)}} \rme^{\frac{\beta}{2}\left( \left(2s_i-1\right)h_{i\ell}^{s} + \left(2\tau_i-1\right)h_{i\ell}^{\tau} + \left(2s_i-1\right)\left(2\tau_i-1\right)h_{i\ell}^{s \tau}  \right)}
\end{eqnarray}
and use (\ref{eq: cavity site marginal }) to obtain a closed set of equations for the cavity fields, 
\begin{align}
    h_{i\ell}^{s} &= \frac{1}{2 \beta} \sum_{j \in \pa_i \setminus \ell } \ln\left(\frac{f_{ij}(1,1)f_{ij}(1,0)  }{f_{ij}(0,1) f_{ij}(0,0) }\right) + \frac{1}{2} J_{ii} + \vartheta_i \nonumber \\
    h_{i\ell}^{\tau} &= \frac{1}{2 \beta} \sum_{j \in \pa_i \setminus \ell } \ln\left(\frac{f_{ij}(1,1)f_{ij}(0,1)  }{f_{ij}(1,0) f_{ij}(0,0) }\right) + \frac{1}{2} J_{ii} + \vartheta_i\nonumber \\
    h_{i\ell}^{s \tau} &= \frac{1}{2 \beta} \sum_{j \in \pa_i \setminus \ell } \ln\left(\frac{f_{ij}(1,1)f_{ij}(0,0)  }{f_{ij}(0,1) f_{ij}(1,0) }\right) + \frac{1}{2} J_{ii} \label{app: equib stau cav field},
\end{align}
where we define, 
\begin{align}
    f_{ij}(s_i,\tau_i) &= \sum_{s_j} \sum_{\tau_j}\rme^{\beta \left(s_i J_{ij}\tau_j + \tau_iJ_{ij}s_j\right)} p^{(i)}_j(s_j, \tau_j). 
\end{align}
These equations may be solved by simple iteration from some random initial condition. We can similarly parameterise the single site marginal as
\begin{eqnarray}
  p_i(s_i,\tau_i) &=& \frac{1}{Z_i} \rme^{\frac{\beta}{2}\left( \left(2s_i-1\right)h_i^{s} + \left(2\tau_i-1\right)h_i^{\tau} + \left(2s_i-1\right)\left(2\tau_i-1\right)h_i^{s \tau}  \right)}
\end{eqnarray}
from which one finds, 
\begin{align}
    h_i^{s} &= \frac{1}{2 \beta} \sum_{j \in \pa_i  } \ln\left(\frac{f_{ij}(1,1)f_{ij}(1,0)  }{f_{ij}(0,1) f_{ij}(0,0) }\right) + \frac{1}{2} J_{ii}  +\vartheta_i \nonumber\\
    h_i^{\tau} &= \frac{1}{2 \beta} \sum_{j \in \pa_i  } \ln\left(\frac{f_{ij}(1,1)f_{ij}(0,1)  }{f_{ij}(1,0) f_{ij}(0,0) }\right) + \frac{1}{2} J_{ii} +\vartheta_i \nonumber\\
    h_i^{s \tau} &= \frac{1}{2 \beta} \sum_{j \in \pa_i  } \ln\left(\frac{f_{ij}(1,1)f_{ij}(0,0)  }{f_{ij}(0,1) f_{ij}(1,0) }\right) + \frac{1}{2} J_{ii}. \label{app: equib stau field}
\end{align}
One must solve equations (\ref{app: equib stau cav field}) by iteration, and then substitute this solution into equations (\ref{app: equib stau field}) to find the fields $h_i^{s}$, $h_i^{\tau}$ and $h_i^{s \tau}$
from which one can find the average activation probability, $\left<s_i\right> = \sum_{s_i,\tau_i}s_ip_i(s_i,\tau_i)$, from the following expression
\begin{align}
    \left<s_i\right> &= \frac{1}{Z_i} \rme^{\frac{\beta}{2}h_i^{s}}\cosh\frac{\beta}{2}\left(h_i^{\tau} + h_i^{s\tau}\right),
\end{align}
where we have defined the normalisation constant
\begin{align}
 Z_i &= \sum_{s_i,\tau_i} \rme^{\frac{\beta}{2}\left( \left(2s_i-1\right)h_i^{s} + \left(2\tau_i-1\right)h_i^{\tau} + \left(2s_i-1\right)\left(2\tau_i-1\right)h_i^{s \tau}  \right)}.
\end{align}

\section{Dynamical cavity approach to systems  with self-interactions} \label{app: dyn cav self-interactions}
Here we detail the dynamical cavity approach for a %
system of $N$ Boolean variables 
with pairwise as well as self- interactions, which evolves in time according to equation (\ref{eq:mono parallel_dynamics}). As we shall show, it is possible to derive a closed set of equations for such systems but, even for systems with unidirectional interactions, these equations are exponential in complexity, and can not be simplified using OTA schemes. To begin, we consider the trajectory of the system from time $t=0$ to $t=t_m$, $\bfs^{0} \to \bfs^{1} \to \dots \to \bfs^{t_m}$, which we denote $\bfs^{0 \dots t_m}$. From equation (\ref{eq:mono parallel_dynamics}) the trajectory of the system follows, 
\begin{align}
    \pr(\bfs^{0\dots t_m}) &= \pr_{0}(\bfs^{0}) \prod_{t=1}^{t_m} W( \bfs^{t} | \bfs^{t-1} ).
\end{align}
We now assume that the interactions, due to their sparsity, are represented by a network with the topology of a tree. 
By following steps in \ref{app: dyn cav bipartite}, now applied to a monopartite system, 
 we may write the probability to observe a single site trajectory as,
\begin{align}
    \pr_i(s_i^{0\dots t_m}) &= \pr_i(s_i^{0}) \sum_{\bfs_{\partial_i}^{0 \dots t_m -1 }} \left[ \prod_{t=1}^{t_m}\Wr( s_i^{t} | h_i(\bfs^{t-1}_{\partial_i},s_i^{t-1} ) ) \right] \nonumber\\
    &\times
    \prod_{j \in \partial_i } \pr_j^{(i)}(s_j^{0 \dots t_m-1} | \zeta_j^{(i),1 \dots t_m-1}) \label{app eq: mono site marginal}
\end{align}
where $\pr_j^{(i)}(s_j^{0 \dots t_m-1} | \zeta^{(i),1,\dots t_m-1}_j)$ is the probability to observe the trajectory $s_j^{0,\dots t_m-1}$ in the cavity graph where site $i$ has been removed, given that site $j$ feels a time dependent external field $\zeta_j^{(i),t} = J_{ji} s_i^{t-1}$.
Similarly, the probability of a trajectory in the cavity graph is found to be given by,
\begin{align}
\begin{split}
    \pr_i^{(\ell)}(s_i^{0\dots t_m} | \zeta_i^{(\ell),1,\dots, t_m})
    &= \pr_i(s_i^{0}) \sum_{\bfs_{\partial_i\setminus \ell}^{0 \dots t_m -1 }} \left[ \prod_{t=1}^{t_m}\Wr( s_i^{t} | h_i(\bfs^{t-1}_{\partial_i},s^{t-1}_i )  \right] \\
    & \qquad \times \prod_{j \in \partial_i \setminus \ell } \pr_j^{(i)}(s_j^{0 \dots t_m-1} | \zeta_j^{(i),1 \dots t_m-1}). \label{app eq: mono cav site marginal recursive}
\end{split}
\end{align} 
We are interested in the site marginals at a given time, however, in the presence of self-interactions, we cannot perform the sum over 
$s_i^{0\ldots t_m-1}$ explicitly, even in the simplest scenario where interactions are unidirectional, as we will show explicitly below. As noted earlier, for unidirectional interactions, cavity and non-cavity distributions are equal, so we can write
\begin{align}
    \pr_i(s_i^{0\dots t_m}) &= \pr_i(s_i^{0}) \sum_{\bfs_{\partial_i}^{0 \dots t_m -1 }} \left[ \prod_{t=1}^{t_m}\Wr( s_i^{t} | h_i(\bfs^{t-1}_{\partial_i},s_i^{t-1} ) ) \right] \nonumber\\
    &\times
    \prod_{j \in \partial_i } \pr_j(s_j^{0 \dots t_m-1}). \label{app eq: site trajectory self int before ota}
\end{align}
These equations cannot, however, be simplified by assuming suitably
factorised forms of the trajectory distribution. 
Even if we were to assume 
a fully factorised distribution 
$$
\pr_j(s_j^{0 \dots t_m-1})=
\prod_{t=0}^{t_m-1}\pr_j(s_j^t)
$$
we would be left with 
\begin{align}
    \pr_i(s_i^{t_m}) &= \sum_{s_i^{0\ldots t_m-1}}
    \pr_i(s_i^{0})\prod_{t=1}^{t_m} \sum_{\bfs_{\partial_i}^{t -1 }} \Wr( s_i^{t} | h_i(\bfs^{t-1}_{\partial_i},s_i^{t-1} ) )  \prod_{j \in \partial_i } \pr_j(s_j^{ t-1}) 
\end{align}
which requires summing over a number $2^{t_m (1+|\partial_i|)}$ of 
variables which grows exponentially in time. We would thus need to face the full complexity of equations (\ref{app eq: mono site marginal}) and (\ref{app eq: mono cav site marginal recursive}).

For unidirectional systems \textit{without} self-interactions, as shown in \ref{app: dyn cav bipartite}, it is possible to reduce the exponential complexity of the dynamical cavity equations, by deriving an expression for the single site marginal $\pr_i(s_i^{t})$. However, here we find that one can not marginalise (\ref{app eq: site trajectory self int before ota}) over $s_i^{0 \dots t_m-1}$ due to the presence of self-interactions, and one is forced to consider other methods to reduce the complexity of the cavity equations. Previous works have applied the OTA scheme to systems without self-interactions, but with %
bidirectional interactions, where strong memory effects are present. The analogous OTA for unidirectional systems with self-interactions is applied by assuming the cavity trajectories factorise in the following way,
\begin{align}
    \pr_i^{(\ell)}(s_i^{0\dots t_m-1}) &=\pr_i(s_i^{0}) \prod_{t=1}^{t_m-1} \pr_i^{(\ell)}(s_i^{t} | s_i^{t-1})
\end{align}
where we have used $P_i^{(\ell)}(s_i^{0}) = \pr_i(s_i^{0})$.
Inserting this into (\ref{app eq: mono cav site marginal recursive}) we find, 
\begin{align}
\begin{split}
      \pr_i^{(\ell)}(s_i^{0\dots t_m} ) &= \pr_i(s_i^{0}) \sum_{\bfs_{\partial_i\setminus \ell}^{0 \dots t_m -1 }} \left[ \prod_{t=1}^{t_m}\Wr( s_i^{t} | h_i(\bfs^{t-1}_{\partial_i},s_i^{t-1} )  \right] \\ 
      & \qquad \quad \times \prod_{j \in \partial_i \setminus \ell } \pr_j(s_j^{0}) \prod_{t=1}^{t_m-1} \pr_j^{(i)}(s_j^{t} | s_j^{t-1}). \label{app eq: mono OTA before}
\end{split}
\end{align}
However, we find that it is still \textit{not} possible to marginalise (\ref{app eq: mono OTA before}) over $s_i^{0\dots t_m-1}$ without making \textit{uncontrolled} approximations. Hence, a different procedure is necessary to solve for the dynamics of systems with self-interactions, as we detail in Sec. \ref{sec: linear dynamics}.
%

\section{One time approximation in the thermodynamic limit}\label{app: OTA thermo dist mono}

We now show how the distribution of site marginals for monopartite systems without self-interactions, $\pi(\{\pr_{t}\})$, defined in equation (\ref{eq: def dist of site probs}), may be computed from the cavity method in the limit $N\to\infty$. Starting from equation (\ref{eq: def dist of site probs}) we insert unity of the form, 
\begin{align}
\begin{split}
    1 &= \sum_{k} \delta_{k, |\partial_i|} \int \{\rmd \pr_{t-2}\} \prod_{s^{t-2}} \delta\left(\pr_{t-2}(s^{t-2}) - \pr_i(s^{t-2})\right)  \\
    & \quad \times \prod_{j \in \partial_i}\Bigg\{ \int \rmd J_j \rmd \hat{J}_j \delta\left(J_j - J_{ij}\right) \delta\left(\hat{J}_j - J_{ji}\right) \\
    & \qquad \qquad  \times \int \{\rmd \hat{\pr}_j\} \prod_{s_j^{t-1}} \delta\left(\hat{\pr}_j(s_j^{t-1}) - \pr_j^{(i)}(s_j^{t-1}| \hat{J}_js^{t-2})\right) \Bigg\}
\end{split}
\end{align}
which yields,
\begin{align}
    \pi(\{\pr_{t}\}) &= \sum_{k}  \int \{\rmd \pr_{t-2}\} \int \left[ \prod_{j=1}^{k} \rmd J_j \rmd \hat{J}_j \{\rmd \hat{\pr}_j\}  \right] \mathcal{P}\left[k,\{ \pr_{t-2}\}, \mathbf{J},\mathbf{\hat{J}},  \{ \boldsymbol{\hat{\pr}}\} \right] \nonumber \\
    & \qquad \times \prod_{s^{t}} \delta\left( \pr_{t}(s^{t}) - \phi(k, \{ \pr_{t-2}\},\mathbf{J},\mathbf{\hat{J}},\{ \boldsymbol{\hat{\pr}}\} ) \right) \label{app eq: pi dist before }
\end{align}
where we have defined,
\begin{align}
\begin{split}
     &\mathcal{P}\left[k,\{ \pr_{t-2}\}, \mathbf{J},\mathbf{\hat{J}},  \{ \boldsymbol{\hat{\pr}}\} \right] \\
     &\quad= \frac{1}{N}\sum_i \delta_{k, |\partial_i|} \prod_{s^{t-2}} \delta\left(\pr_{t-2}(s^{t-2}) - \pr_i(s^{t-2})\right)  \\
    & \quad \quad \times \prod_{j \in \partial_i}\left \{ \delta\left(J_j - J_{ij}\right) \delta\left(\hat{J}_j - J_{ji}\right) \prod_{s_j^{t-1}} \delta\left(\hat{\pr}_j(s_j^{t-1}) - \pr_j^{(i)}(s_j^{t-1}| \hat{J}_js^{t-2})\right) \right\}
    \end{split}
\end{align}
which is the probability that a site drawn at random has degree $k$, site marginal at two earlier time steps $\pr(s^{t-2})$, where each of the $k$ neighbours acts on the site with interactions $\bfJ = (J_{1},\dots,J_{k})$, and where the site acts on each neighbour with interactions $\boldsymbol{\hat{J}} = (\hat{J}_{1},\dots,\hat{J}_{k})$, and where each neighbour in the cavity graph with site $i$ removed is described by the cavity site marginals at one earlier time step $\{ \boldsymbol{\hat{\pr}}\} = ( \hat{\pr}(s^{t-1}_{1}),\dots,\hat{\pr}(s^{t-1}_{k}) )$. The function $\phi(k, \{ \pr_{t-2}\},\mathbf{J},\mathbf{\hat{J}},\{ \boldsymbol{\hat{\pr}}\} )$ is defined in equation (\ref{eq: phi}). 
To proceed we assume that all interactions $J_{ij}$ are drawn independently from some distribution $\pr(J_{ij})$, and also assume that cavity fields are independent. Under these assumptions, we may use Bayes theorem to write,
\begin{align}
\begin{split}
    &\mathcal{P}\left[k,\{ \pr_{t-2}\}, \mathbf{J},\mathbf{\hat{J}},  \{ \boldsymbol{\hat{\pr}}\} \right] \\ 
    & \qquad = \pr(k) \pi_{t-2} \left[ \{\pr_{t-2}\} | k, \mathbf{J},\mathbf{\hat{J}} \right] \prod_{j=1}^{k}\pr(J_j)\pr(\hat{J}_j|J_j) \hat{\pi}_{t-1} \left[ \{\hat{\pr}_j\} | \hat{J}_js^{t-2}\right] \label{app eq: Prob k, J, pr}
\end{split}
\end{align}
where we have defined $\hat{\pi}_{t-1} \left[ \{ \hat{\pr}_j \} | x \right]$, the distribution of cavity site marginals, $\hat{\pr}_{j}(s_{j}^{t-1})$, given that the site removed acts on the cavity graph with external field $x$. We note that by definition in absence of external field the cavity distribution is equal to its non-cavity counterpart, $\hat{\pi}_{t-1} \left[ \{ \hat{\pr}_{j}\} | 0 \right]  =\pi_{t-1} \left[ \{ \hat{\pr}_{j} \} \right] $, since if there is no external field, this implies that the cavity distribution is uninfluenced by the removed site, such that $\hat{J}=0$ i.e $\pr_{j}^{(i)}(s_{j}^{t-1} | 0)=\pr_{j}(s_j^{t-1}) $. We then insert (\ref{app eq: Prob k, J, pr}) into (\ref{app eq: pi dist before }) from which we find equation (\ref{eq: OTA mono thermo maginal dist}).

 What remains is to find an expression for the distribution of cavity marginals $\hat{\pi}_{t-1} \left[ \{\hat{\pr}_j\} | x \right]$. By definition this object is, in the limit $N\to \infty$, equivalent to,
\begin{align}
    \hat{\pi}_{t} \left[ \{\hat{\pr}_i\} | x \right] = \frac{1}{N} \sum_{i\ell} \frac{A_{i\ell}}{k_i}\prod_{s_i^{t}}\delta\left( \hat{\pr}_i(s_i^{t}) -  \pr_i^{(\ell)}(s_i^{t}| x) \right). \label{app eq: def cav dist of dists}
\end{align}
We insert into (\ref{app eq: def cav dist of dists}) unity of the form, 
\begin{align}
\begin{split}
    1 &= \sum_{k} \delta_{k, |\partial_i|} \sum_{k'}\delta_{k', |\partial_{\ell}|} \int \{\rmd \pr_{t-2}\} \prod_{s^{t-2}} \delta\left(\pr_{t-2}(s^{t-2}) - \pr_i(s^{t-2})\right) \\
    & \qquad \times \prod_{j \in \partial_i\setminus \ell}\Bigg \{ \int \rmd J_j \rmd \hat{J}_j \delta\left(J_j - J_{ij}\right) \delta\left(\hat{J}_j - J_{ji}\right) \\
    & \qquad \qquad \qquad \times \int \{\rmd \hat{\pr}_j\} \prod_{s_j^{t-1}} \delta\left(\hat{\pr}_j(s_j^{t-1}) - \pr_j^{(i)}(s_j^{t-1}| \hat{J}_js^{t-2})\right) \Bigg\}
\end{split}
\end{align}
and define, 
\begin{align}
\begin{split}
    &\mathcal{W}\left[k,k',\{ \pr_{t-2}\}, \mathbf{J},\mathbf{\hat{J}},  \{ \boldsymbol{\hat{\pr}}\} \right] \\
    &= \frac{1}{N \left< k \right>} \sum_{i \ell} A_{i \ell} \delta_{k, |\partial_i|} \delta_{k', |\partial_{\ell}|} \prod_{s^{t-2}} \delta\left(\pr_{t-2}(s^{t-2}) - \pr_i(s^{t-2})\right)  \\
    & \quad \times \prod_{j \in \partial_i\setminus \ell}\left \{  \delta\left(J_j - J_{ij}\right) \delta\left(J_j - J_{ji}\right) \prod_{s_j^{t-1}} \delta\left(\hat{\pr}_j(s_j^{t-1}) - \pr_j^{(i)}(s_j^{t-1}| \hat{J}_js^{t-2})\right) \right\}
\end{split}
\end{align}
which, assuming that interactions are drawn independently, and that cavity fields are independent, can be written using Bayes theorem, 
\begin{align}
\begin{split}
    &\mathcal{W}\left[k,k',\{ \pr_{t-2}\}, \mathbf{J},\mathbf{\hat{J}},  \{ \boldsymbol{\hat{\pr}}\} \right] \\
    &\qquad =    W(k,k') \pi_{t-2}\left[\{ \pr_{t-2}\}| k, \mathbf{J},\mathbf{\hat{J}} \right] \prod_{j=1}^{k-1}\pr(J_j)\pr(\hat{J}_j|J_j) \hat{\pi}_{t-1}\left[ \{ \hat{\pr}_j\} | \hat{J}_j s^{t-2}\right]
\end{split}
\end{align}
where $\mathcal{W}\left[k,k',\{ \pr_{t-2}\}, \mathbf{J},\mathbf{\hat{J}}, \{ \boldsymbol{\hat{\pr}}\} \right]$ is the probability to draw a link at random connecting site $i$ with degree $k'$ and site $j$ with degree $k$, where $j$ has neighbours (excluding $i$) which act on site $j$ with interactions $\bfJ  = (J_{1},\dots,J_{k-1})$, and where $j$ acts on its neighbours with interactions $\boldsymbol{\hat{J}} = (\hat{J}_{1},\dots,\hat{J}_{k-1})$, and where the activation of each neighbour is described by the cavity site marginals $\{ \boldsymbol{\hat{\pr}}\} = ( \hat{\pr}(s^{t-1}_{1}),\dots,\hat{\pr}(s^{t-1}_{k-1}) )$. We have also defined the degree correlation function $W(k,k')= (N \langle k \rangle)^{-1}\sum_{ij}A_{ij}\delta_{k',|\pa_i|}\delta_{k,|\pa_j|}$ which is the probability to draw a link at random with a node of degree $k$ at one end and $k'$ at the other. By writing the cavity fields as independent, we have implicitly assumed that the network lacks strong degree correlations, and so we now explicitly assume that degrees are uncorrelated such that  $W(k,k') = W(k) W(k') $ where $W(k) = \frac{k \pr(k)}{\langle k \rangle}$ is the probability to draw a link at random that is attached to a node of degree $k$. Substituting this form of unity into the definition for the cavity marginal (\ref{app eq: def cav dist of dists}) we find equation (\ref{eq: OTA mono thermo cav dist}).
Equations (\ref{eq: OTA mono thermo maginal dist}) and (\ref{eq: OTA mono thermo cav dist}) are a closed set of equations which may be solved via a population dynamics procedure. Note, this a dynamical variant of the population dynamics procedure, and can be used to compute the distribution of site marginals at each point in time.

From the distributional equations  (\ref{eq: OTA mono thermo maginal dist}) and (\ref{eq: OTA mono thermo cav dist}), one can also derive a closed set of equations for the \textit{average} site marginal, which by definition is given by, 
\begin{align}
    \overline{\pr}_{t}(s^{t}) &= \int \{\rmd \pr_{t}\}\pi(\{\pr_{t}\})\{\pr_{t}\}. \label{app eq: av site marginal thermo}
\end{align}
If we substitute equation (\ref{eq: OTA mono thermo maginal dist}) into (\ref{app eq: av site marginal thermo}) we find, 
\begin{align}
    \overline{\pr}_{t}(s^{t}) &= \sum_{k} \pr(k)  \int \left[ \prod_{j=1}^{k} \rmd J_j \rmd \hat{J}_j \{\rmd \hat{\pr}_j\}\pr(J_j)\pr(\hat{J}_j|J_j) \right] \nonumber  \\
    & \qquad \times \sum_{s^{t-2}} \overline{\pr}_{t-2}(s^{t-2}|k,\bfJ, \boldsymbol{\hat{J}}) \sum_{s_{1}^{t-1},\dots,s_{k}^{t-1}} \Wr(s^{t}| \sum_{j=1}^{k}J_js_j^{t-1})  \label{app eq: av site marginal eq} \\
    & \qquad \quad \times \prod_{j=1}^{k}\left[(1-\delta_{\hat{J}_j,0})\overline{Q}_{t-1}(s_j^{t-1}|\hat{J}_js^{t-2}) +\delta_{\hat{J}_j,0} \overline{\pr}_{t-1}(s_j^{t-1}) \right] \nonumber
\end{align}
where we have defined, 
\begin{align}
    \overline{Q}_{t}(s^{t}|x) &= \int \{\rmd \hat{\pr}\}\hat{\pi}(\{\hat{\pr}_{t}\}|x)\{\hat{\pr_{t}}\}. \label{app eq: av cav site marginal thermo}
\end{align}
If we then insert equation (\ref{eq: OTA mono thermo cav dist}) into (\ref{app eq: av cav site marginal thermo}), we find, 
\begin{align}
    \overline{Q}_{t}(s^{t}|x) &= \sum_{k} \frac{k \pr(k)}{\langle k \rangle}  \int \left[ \prod_{j=1}^{k-1} \rmd J_j \rmd \hat{J}_j \{\rmd \hat{\pr}_j\}\pr(J_j)\pr(\hat{J}_j|J_j) \right] \nonumber \\
    & \qquad \times \sum_{s^{t-2}} \overline{\pr}_{t-2}(s^{t-2}|k,\bfJ, \boldsymbol{\hat{J}}) \sum_{s_{1}^{t-1},\dots,s_{k-1}^{t-1}} \Wr(s^{t}| \sum_{j=1}^{k-1}J_js_j^{t-1} + x) \label{app eq: av cav site marginal equation} \\
    & \qquad \quad \times \prod_{j=1}^{k-1}\left[(1-\delta_{\hat{J}_j,0})\overline{Q}_{t-1}(s_j^{t-1}|\hat{J}_js^{t-2}) +\delta_{\hat{J}_j,0} \overline{\pr}_{t-1}(s_j^{t-1}) \right] \nonumber .
\end{align}
Equations (\ref{app eq: av site marginal eq}) and (\ref{app eq: av cav site marginal equation}) are a closed set of equations that may be solved via iteration. These equations may also be derived via generating functional analysis, if one makes appropriate assumptions to deal with memory effects \cite{hurry2022thesis}.

\vspace{5mm}

\section*{References}
\bibliographystyle{iopart-num}
\bibliography{SelfInteractions}

\end{document}